\documentclass[%
superscriptaddress,
showpacs,
showkeys,
preprintnumbers,
nofootinbib,T
 amsmath,amssymb,
 aps,
prd, twocolumn, 
floatfix,
usenames,
dvipsnames
]{revtex4-1}

\usepackage{epsfig}
\usepackage{graphicx}
\usepackage{dcolumn}
\usepackage{bm}
\usepackage[colorlinks=true,linkcolor=blue,citecolor=blue]{hyperref}
\usepackage{natbib}
\usepackage{comment}
\usepackage{xcolor}
\usepackage{enumerate}
\usepackage{makecell, boldline}
\usepackage{gensymb}
\usepackage[normalem]{ulem}
\usepackage{subfigure}
\usepackage{multirow}
\usepackage{booktabs}
\usepackage{tabularx}
\usepackage{amsmath}
\usepackage[capitalize]{cleveref}

\newcommand{\am}[1]{{\textcolor{brown}{\sf{[AM: #1]}}}}

\newcommand{\yr}{\, \text{yrs}}

\newcommand{\msun}{{\, \rm M}_\odot}

\newcommand{\uncer}{uncertainties }  %do NOT change this command 

\newcommand{\lcdm}{$\Lambda$CDM }
\newcommand{\xgw}{x_{\rm gw}}
\newcommand{\xem}{x_{\rm em}}

\newcommand{\hi}{\mathcal{H} \mathcal{I}}
\newcommand{\thetabin}{\hat{\theta}_{\rm bin}}
\newcommand{\thetaenv}{\hat{\theta}_{\rm env}}
\newcommand{\zem}{z_{\rm em}}
\newcommand{\colorref}{black}

\newcommand\mnras{Monthly Notices of the Royal Astronomical Society}
\newcommand\apjl{The Astrophysical Journal Letters}
\newcommand\apjs{The Astrophysical Journal Supplement Series}

\newcommand\jcap{Journal of Cosmology and Astroparticle Physics}

\newcommand\aap{Astronomy and Astrophysics}
\newcommand\nar{New Astronomy Reviews}
\newcommand\baas{Bulletin of the American Astronomical Society}

\definecolor{mycolormodel}{HTML}{E79819}  %orange
\definecolor{mycolormodel}{HTML}{EC13A8}   %magenta
\begin{document}

\preprint{APS/123-QED}

\title{Massive black hole binaries in LISA: constraining cosmological parameters at high redshifts}

\author{Alberto Mangiagli}
\email{mangiagli@apc.in2p3.fr}
\affiliation{Universit\'e Paris Cit\'e, CNRS, Astroparticule et Cosmologie, F-75013 Paris, France}

\author{Chiara Caprini}
\affiliation{Universit\'e de Gen\`eve, D\'epartement de Physique Th\'eorique and Centre for Astroparticle Physics,
24 quai Ernest-Ansermet, CH-1211 Gen\`eve 4, Switzerland}
\affiliation{CERN, Theoretical Physics Department, 1 Esplanade des Particules, CH-1211 Gen\'eve 23, Switzerland}

\author{Sylvain Marsat}
\affiliation{Laboratoire des 2 Infinis - Toulouse (L2IT-IN2P3), Universit\'e de Toulouse, CNRS, UPS, F-31062 Toulouse Cedex 9, France}

\author{Lorenzo Speri}
\affiliation{Max Planck Institute for Gravitational Physics (Albert Einstein Institute) Am M\"uhlenberg 1, 14476 Potsdam, Germany}

\author{Robert R. Caldwell}
\affiliation{Department of Physics and Astronomy, Dartmouth College, 6127 Wilder Laboratory, Hanover, NH 03755}

\author{Nicola Tamanini}
\affiliation{Laboratoire des 2 Infinis - Toulouse (L2IT-IN2P3), Universit\'e de Toulouse, CNRS, UPS, F-31062 Toulouse Cedex 9, France}

% \author{Alexandre Toubiana}
% \affiliation{Max Planck Institute for Gravitational Physics (Albert Einstein Institute) Am M\"uhlenberg 1, 14476 Potsdam, Germany}

% \author{Marta Volonteri}
% \affiliation{Institut d'Astrophysique de Paris, CNRS \& Sorbonne Universit\'e, UMR 7095, 98 bis bd Arago, F-75014 Paris, France}

\date{\today}% It is always \today, today,
             %  but any date may be explicitly specified

\begin{abstract}

One of the scientific objectives of the Laser Interferometer Space Antenna (LISA) is to probe the expansion of the Universe using gravitational wave observations. Indeed, as gravitational waves from the coalescence of a massive black hole binary (MBHB) carry direct information of its luminosity distance, an accompanying electromagnetic (EM) counterpart can be used to determine its redshift.
This method of \emph{bright sirens}, when applied to LISA, enables one to build a gravitational Hubble diagram to high redshift.
In this work, we forecast the ability of LISA-detected MBHB bright sirens to constrain cosmological models. 
\textcolor{\colorref}{The expected EM emission from MBHBs can be detected up to redshift $z\sim 7$ with future astronomical facilities, and the distribution of MBHBs with detectable counterpart peaks at $z\sim 2-3$. Therefore, 
%in this paper, 
we propose several methods to leverage the ability of LISA to constrain the expansion of the Universe at $z\sim 2-3$, a poorly charted epoch in cosmography.}
\textcolor{\colorref}{ 
We find that the most promising method consists in
using a model-independent approach based on a spline interpolation of the luminosity distance-redshift relation:
in this case, LISA can constrain the Hubble parameter at $z\sim2-3$ with a relative precision of at least $10\%$.}

\end{abstract}


\maketitle

%\tableofcontents

\section{\label{sec:intro}Introduction}
In the next decade, the Laser Interferometer Space Antenna (LISA) \cite{Seoane17} will observe  gravitational waves (GWs) from the coalescence of massive black hole binaries (MBHBs) of mass $10^4-10^7 \msun$ at redshifts up to $z \sim 20$.

For nearly 40 years, \textcolor{\colorref}{\emph{standard sirens} have been proposed as a method to shed light}  on the cosmic expansion history of the Universe \cite{schutz1986determining, 2022arXiv220405434A}. Indeed, a coalescing binary system can be considered as a standard siren because the GW emission carries direct information on the luminosity distance of the source. The GW signal is degenerate in the redshift, however, meaning that additional information is required to infer the distance-redshift relation.

If an electromagnetic (EM) counterpart is associated with the coalescence, the redshift may be determined with spectroscopic or photometric follow-up observations \cite{Tamanini16,holz2005using,PhysRevD.74.063006,Nissanke_2010,PhysRevD.86.043011, 2017Natur.551...85A, Mangiagli22, Orazio23}. This approach to constructing a standard-siren distance-redshift relation is usually referred to in the literature as the method of \emph{bright sirens}. In the absence of an EM counterpart, the redshift may be estimated probabilistically by cross-correlating the region hosting the GW merger with galaxy catalogs \cite{2011ApJ...732...82P,2018MNRAS.475.3485D,PhysRevD.95.083525, Muttoni22,Laghi21,2022PhRvR...4a3247Z, Dalang23}. This approach is referred to as the method of \emph{dark sirens}. Finally, the redshift may also be inferred from the detected mass distribution of the source. Due to cosmic expansion, the gravitational waveform determines the product $M(1+z)$ where $M$ is the intrinsic, rest-frame mass parameter.
%Under some assumptions about
Assuming a parametrized functional form for the intrinsic mass distribution, the cosmological parameters can be constrained together with the parameters describing the mass distribution \cite{2021arXiv211103604T, Ezquiaga22}. This approach has been recently named the method of \emph{spectral sirens}.
The dark and spectral siren approaches can be combined in a single inference methodology where both information on the intrinsic distribution of sources and cross-correlation with galaxy catalogs provide more stringent cosmological constraints \cite{Gray:2023wgj,Mastrogiovanni:2023emh,2023arXiv231112093L}.
Further approaches have been proposed to test cosmology with GWs, for example, by exploiting the cross-correlation of the weak lensing of both GWs and galaxy fields \cite{2022arXiv221006398B, 2021MNRAS.502.1136M}, prior knowledge either of the equation of state of neutron stars  \cite{2012PhRvL.108i1101M, 2021PhRvD.104h3528C} or of the merger rate evolution of GW sources \cite{Ding:2018zrk,Ye:2021klk}.

A complementary method to determine the distance-redshift relation would provide valuable information about the nature of our Universe. The standard $\Lambda$CDM model provides a good fit to the bulk of cosmological data. However, in recent years, some tensions between late- and early-Universe measurements have arisen. The most famous one is the Hubble constant $H_0$ tension wherein early-time measurements from the Cosmic Microwave Background (CMB) report a value of $H_0 \sim 67 \, \rm km/s/Mpc$ \cite{Planck18,DESI24} and late-time measurements from supernovae (SNe) obtain $H_0 \sim 72 \, \rm km/s/Mpc$ \cite{SNe21}. (See also Ref.~\cite{Divalentino21,abdalla22,Perivolaropoulos22} and references therein for recent reviews on the topic.) Whether these tensions are due to systematics in the different measurements and datasets or are the hints of new physics is a pressing, open question.

% (RRC: is this sentence necessary?) In this context, numerous early- and late-time dark energy models have been introduced in the literature in order to resolve the tension (\cite{Divalentino21,Huterer18, 2023arXiv230209032P,2022PhR...984....1S} for recent reviews) and new cosmological probes have emerged \cite{Moresco22}.

\begin{figure*}    \subfigure{\includegraphics[width=1\textwidth]{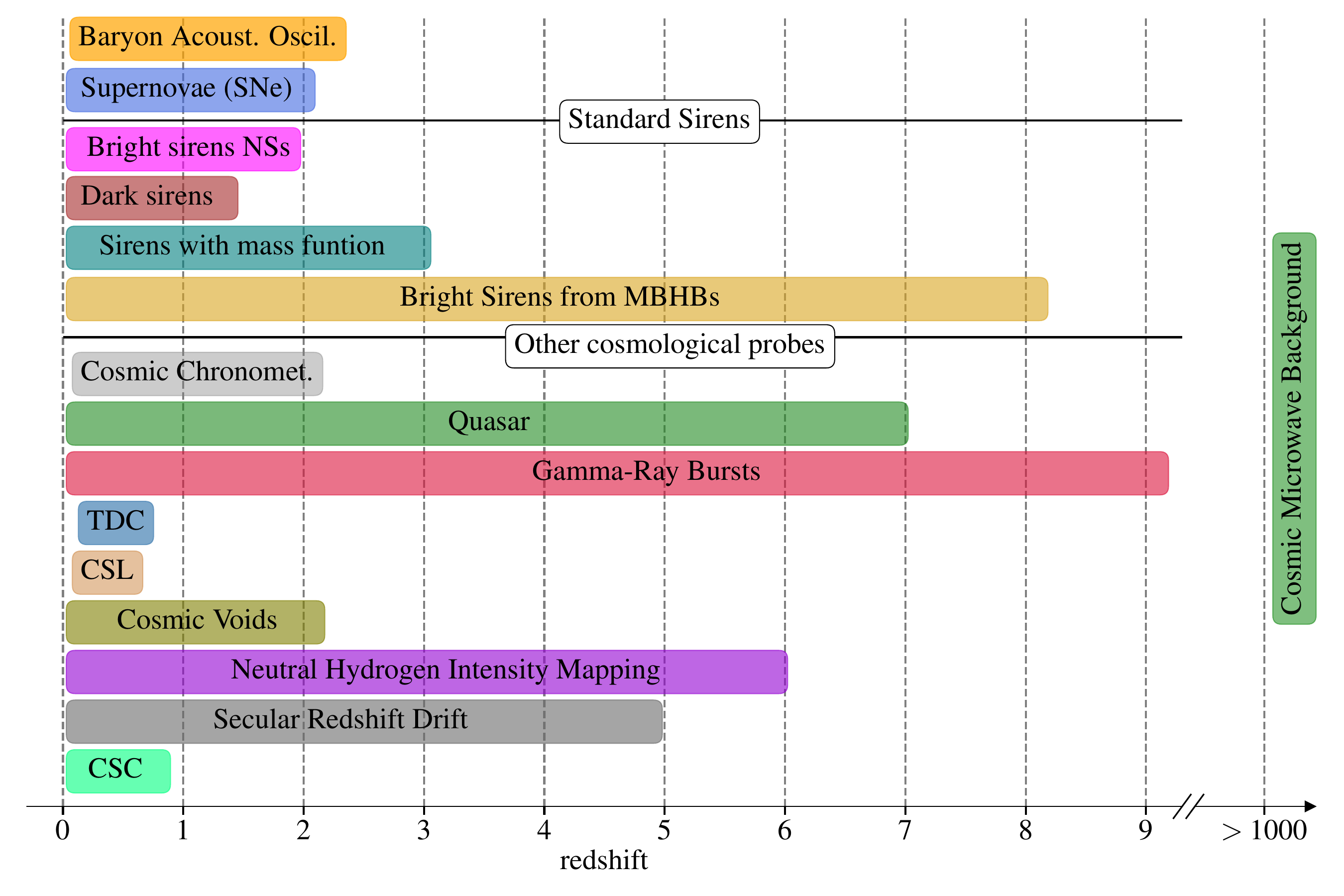}} 
    \caption{Redshift range covered by several cosmological probes. On the right and on the top, the three main cosmological probes, i.e. CMB, BAO and SNe. In the middle, the standard sirens forecasts with different approaches for future 3G detectors and LISA. In the bottom part, other emerging cosmological probes. The acronyms stand for cosmic chronometers (`Cosmic Chronomet.'), time delay cosmography (TDC), cluster strong lensing (CSL) and clustering of standard candles (CSC). 
    Bright sirens from MBHBs can probe the expansion of the Universe up to $z\sim8$ where few other cosmological probes are available. We refer to \cite{Moresco22} for more details on each of the techniques listed in the plot.
    %\nt{I would lower the spectral siren method to $z<3$ (see text). Also it's important to specify that all sirens forecasts here are from 3G detectors and will come after LISA (ET and CE will not be fully operational before $\sim$2040). Also aren't quasars and GRB ranges too optimistic? \am{in principle you can detect them up to the reported redshift, but, for example, for quasars there are not many sources at $z>4$. In a sense, the problem of this plot is that it does not convey `uncertainties' but I think it's still useful. I would not write explicitly that 3G detectors forecasts will arrive after LISA because, you know, maybe they speed up the construction at some point or LISA might be a bit late.}\nt{You wanna bet?}}
    }
    \label{fig:cosmo_probes}
\end{figure*}

In Fig.~\ref{fig:cosmo_probes} we report the redshift range covered by different cosmological probes (see also Fig. 52 in \cite{Moresco22} for a similar plot). 
\textcolor{\colorref}{In order to clarify the complementarity of LISA MBHB bright sirens, we proceed to describe the probes shown in this figure.}
The three main measurements techniques %in the literature 
of the Universe expansion
use the CMB at early-times and SNe and BAO at late-times. While the former is at $z>1000$, the latter methods can test the expansion up to $z \sim 2$ \cite{Riess18} and $z \sim 2.5$ \cite{2018ApJ...853..119A}, respectively.

In the case of standard sirens, current ground-based detectors are expected to observe NS-NS and NS-BH mergers up to $z\sim 0.1$ \cite{Colombo22}. Third-generation ground-based detectors as Einstein Telescope (ET) \cite{et} and Cosmic Explorer (CE) \cite{cosmic_explorer}, together with EM facilities as \emph{Fermi} \cite{fermi} and \emph{Theseus} \cite{theseus}, should be able to detect the GW signal and the EM emission up to $z\sim 2.5$ in X-ray and up to $z\sim 3$ in $\gamma$-band \cite{Ronchini22}. 
The dark sirens approach can be, in principle, applied to any type of standard sirens, i.e.~compact object binaries in ground-based interferometers or in the high-frequency portion of LISA band \cite{2018MNRAS.475.3485D,Kyutoku:2016zxn,Muttoni22} and extreme-mass ratio inspiral (EMRIs) \cite{MacLeod:2007jd,Laghi21,Liu:2023onj}. However, independently of the type of GW source, this technique is limited to $z\sim1$-1.5 by the completeness of the galaxy catalogs. 
Concerning spectral sirens, current detectors are limited to $z\sim 1.5$ at design sensitivity \cite{Taylor:2011fs,Leyde22,Mancarella22, Leandro:2021qlc}, but ET and CE can potentially expand this approach up to $z\sim3$. The peak of the star formation rate is at $z\sim2-3$, meaning that the number of BHs and NSs will decrease quickly at higher redshifts  \cite{Santoliquido21} and BHBs may not provide strong constraints on $H(z)$ at $z\gtrsim2$ (see Fig.~1 in \cite{Ezquiaga22}).
% $z\sim10$.
% \am{to be checked with someone, update the figure and add a ref}
% \nt{I don't think we will ever be able to use spectral sirens above say $z\sim3$. The peak of the star formation history is around $z\sim2$, meaning that the number of BHs and NSs will decrease quickly after that redshift (see e.g. Fig.~10 in \cite{Taylor:2012db} for NSs). Jose and Dan found that after $z\sim 2$ even BBHs will not provide great constraints on $H(z)$ (see Fig.~1 in \cite{Ezquiaga22}). Other papers on spectral siren with 3G we may want to cite: \cite{Taylor:2011fs,Leandro:2021qlc}}
% However the mergers distribution is expected to peak at $z\sim2$-3, since these sources follow the evolution of the star formation rate \cite{Santoliquido21}. 
MBHs can also be used as spectral sirens even if, at the moment,  consistent studies are still missing. Finally, the detection of the GW signal from a MBHB merger together with the identification of the host galaxy, might probe the expansion of the Universe up to $z\sim8$ \cite{Tamanini16,Mangiagli22,Speri21}, depending on the astrophysical model assumed. 
\textcolor{\colorref}{Since our analysis focuses on these sources, we discuss them in detail later}.

%However, although MBHBs may become interesting sources to test alternative cosmological models at high-redshift \cite{2016JCAP...10..006C,2017JCAP...05..031C,2022PhRvD.105f4061C,Corman:2020pyr,Belgacem19,2022PhRvD.105l3531F}, there are large uncertainties on the expected rates of MBHB mergers (see Sec. 2.4.2 in \cite{astrowg-lisa-paper}), on the modeling of the EM counterpart and on the possibility to identify the host galaxy at high redshift.

In the lower part of the plot, we show other cosmological probes that have been exploited in  recent years. We describe briefly the probes at $z>3$ and refer the interested reader to Ref.~\cite{Moresco22} for a complete review.

Quasars have been proposed as standardizable candles, exploiting the nonlinear relation between the X-ray and UV luminosities (\cite{Risaliti19} and reference therein). This relation should represent a universal mechanism taking place in quasars and their emission can be detected up to $z\sim7$. However the selection of the samples is affected by observational issues, leading to an intrinsic dispersion of $0.2$ dex (even if this value reduces to $0.12$ with high-quality data \cite{Sacchi22}), and the majority of quasars are located at $z<4$.

Similarly, gamma-ray bursts (GRBs) can be promoted to cosmological probes exploiting correlations between rest- and observer-frame quantities, as the relation between the intrinsic peak energy and the total radiated energy \cite{amati02}. GRBs can be detected up to very high redshift $z\sim9$-10 and, emitting in $\gamma$ and hard X-ray band, they are not affected by dust absorption. However the correlations cannot be calibrated in a cosmology-independent way due to the lack of low-redshift events. Therefore the parameters describing the empirical relations have to be fitted together with the cosmological parameters \cite{amati13} or calibrated with lower redshift standard candles, as SNe \cite{demianski17}.

Neutral hydrogen intensity mapping consists in exploiting the $21 \, \rm cm$ emission to map the large-scale structure \cite{kovetz20}. Even if single galaxies are not resolved, the neutral hydrogen follows the matter density fluctuations, providing information on the  evolution of the Universe. This emission can be detected up to very high redshift but there might be contamination from other sources. 

The secular redshift drift consists in measuring the variation in redshift due to the expanding Universe \cite{loeb-redshift-drift}. Any type of redshift indicator can be used (absorption/emission lines and feature in the spectrum) and the approach is completely cosmological model-independent. However it requires a long observing time and ultimately it might not be as accurate as other cosmological probes.

%\nt{Maybe here add one  or two sentences explain the scope of our paper? The transition from explaining the figure to explaining what we do here seems too abrupt IMO. We do not say what is the objective of our investigation.}
 %\am{I moved here the paragraph that as before `In Fig.~\ref{fig:cosmo_probes} \dots}
 %Distinct from other methods, the GW signals carry direct information of the luminosity distance of the source, depending solely on gravitational physics. 
% Moreover, the GW signal depends only on gravitational physics so no luminosity distance scatter is present, as in SNe, and GW measurements are completely independent on early and late-time observations, as CMB, SNe and Baryon Acoustic Oscillations (BAO).   
\textcolor{\colorref}{Standard sirens represent a new cosmological probe to test the expansion of the Universe across a broad range of redshift, as illustrated in Fig.~\ref{fig:cosmo_probes}. In this work we examine the potential of MBHBs as bright sirens to constrain the cosmic evolution of the Universe at intermediate redshifts.}
 In particular, current knowledge of the properties of MBHBs suggests that we may be able to probe the expansion across redshifts $2\lesssim z \lesssim 8$, a territory that is still poorly explored in modern cosmology (see \cite{DESI:2024lzq} for the recent DESI estimates at $z\sim 2$). \textcolor{\colorref}{For this reason, MBHBs are also interesting sources to test alternative cosmological models at high-redshift \cite{2016JCAP...10..006C,2017JCAP...05..031C,2022PhRvD.105f4061C,Corman:2020pyr,Belgacem19,2022PhRvD.105l3531F}. However, 
it is important to remark that there are large uncertainties on the expected rates of MBHB mergers (see Sec. 2.4.2 in \cite{astrowg-lisa-paper}), on the modeling of the EM counterpart and on the possibility to identify the host galaxy at high redshift.}

In this context, we build this paper on the previous work done in \citealt{Mangiagli22} (hereafter `M22') where we explored the number of MBHBs mergers emitting a detectable EM counterpart under different astrophysical models and EM emission channels.
\textcolor{\colorref}{In particular, in M22 we identified a population of MBHB mergers that we called \emph{EMcps} (from EM counterparts, see Sec.~\ref{sec:general_methods}) that satisfy the following conditions}:
%Here we focus on the subset of EM counterparts (EMcps), i.e. systems with :
\begin{enumerate}
    \item GW signal-to-noise ratio (SNR) above 10 in LISA;
    \item Detectable EM emission;
    \item Sufficiently accurate sky-localization, depending on the EM telescopes considered (see \cite{Mangiagli22} for more details).
\end{enumerate}
For this population, we assume that we are able to identify the host galaxy and to get an independent measurement of the redshift. In other words, EMcps are \emph{standard sirens}, i.e. systems that can be used to test the expansion of the Universe. 

The cosmological inference in the present analysis is based on mock EMcps catalogues.
For different astrophysical populations, we build 100 realisations of Universes with EMcps, and we perform the cosmological tests. We divide the analysis in two parts: the first focuses on local Universe quantities such as the local Hubble constant, $\Omega_m$ or late-time dark energy models, while the second explores LISA capabilities to constrain $H(z)$ at $z\gtrsim 2$ using various strategies, both model-dependent and -independent.

The paper is organised as follows. In Sec.~\ref{sec:general_methods} we review the results of M22. In Sec.~\ref{sec:cosmo_intro} we introduce some useful notions in cosmology and we present the models we tested in this work. The catalogues of MBHBs are constructed in Sec.~\ref{sec:cat_construction} and the likelihood is formulated in Sec.~\ref{sec:data_analysis}. In Sec.~\ref{sec:inference} we present the analysis setup and discuss some caveats. In Sec.~\ref{sec:results} we report our main results. In Sec.~\ref{sec:discu_and_concl} we conclude with some final remarks and comments. In Appendix~\ref{sec:app_convergence_realisation} we check that the number of realisations is sufficient to provide solid results. In Appendix~\ref{sec:gaussianity_dl} we assess the Gaussianity of the luminosity distance posterior distributions. In Appendix~\ref{sec:js_convergence} we discuss a test we adopted to determine informative realisations. In Appendix~\ref{sec:pivot_parameters} we derive a redshift where the correlation between the Hubble parameter and $\Omega_m$ is minimum. \textcolor{\colorref}{In Appendix~\ref{sec:accuracy_spline} we report the accuracy for the model-independent approach adopted in this study.}

\begin{comment}
    
General structure of the paper:
\begin{itemize}
    \item General Introduction
    \item Reference to the stsi paper with table of the models that we are gonna use for this paper
    \item Theory section to introduce fom (the fom concept does not depend on a particular cosmological model because its just $H/\sigma_H$
    \item Section to introduce  the cosmological models: standard analysis with $H_0$ and $\Omega_m$ (curvature? we might be competitive with supernovae but we have to check), no $\omega_0$ and $\omega_a$ for simplicity
    \item Analysis with methodology 1 and several pivot redshift and study the most efficient binning
    \item Analysis with methodology 2 without the systems at $z<1$. The scope here is to say that we will be able to say if the universe is matter dominated at pivot redshift
    \item theory section on the $d_L$ error from lensing and peculiar velocity
    \item Bayesian inference and selection effect
    \item Results
\end{itemize}
\end{comment}

\section{\label{sec:general_methods}Review of M22}
In this section we briefly summarise the main results of M22. We have built our methodology on the previous work done in Ref.~\cite{Tamanini16}, with some major improvements.
Since this work is a follow-up of M22, here we limit our discussion to a summary of the most important results and refer the interested reader to the original paper.

There are still large uncertainties in the populations of  MBHBs that LISA will observe, mostly due to the lack of observational evidence. Therefore, we have to rely on simulations.
In the past years, semi-analytical models (SAMs) have been established as one of the possible methods to predict the population of merging MBHBs. In M22, and consequently in this work, we adopted the SAM developed in \cite{Barausse12} (with contributions from \cite{ Sesana14_spin_evolution,Antonini15_1, Antonini15_2}) to track the evolution of MBHs across cosmic time.
We consider three different astrophysical models:
\begin{enumerate}
    \item Pop3: a model where MBHs grow from light seed BHs that are the remnant of young metal-poor Pop3 stars. This model takes into account the delay between the galaxy merger and the MBHB merger;
    \item Q3d: in this case, MBHs originate from the collapse of proto-galactic disks at $10^4-10^6 \msun$. The time-delay between mergers is also included;
    \item Q3nd: a heavy-seed scenario, similar to Q3d, but without mergers time-delays, causing an increase of the number of MBHB mergers.
\end{enumerate}

The three population models above yield three qualitatively different catalogs of MBHB mergers across cosmic time. Among these, we are interested in the MBHB mergers producing an EM counterpart.

Even larger uncertainties affect the EM emission from MBHBs. A few binary-AGN (Active Galactic Nucleus) candidates at sub-parsec separation have been reported (see Part II in \cite{ROSA2020101525} for a recent review). However, these systems are still far from merger in the LISA frequency band, and more massive than the ones LISA will be able to detect. The behaviour of gas in a rapidly changing space-time is yet unclear so we have to rely again on simulations.
During the inspiral phase, General Relativity MagnetoHydroDynamic (GRMHD) simulations show that the binary excavates a cavity in the circumbinary disk and streams of gas flow from the inner edge to form minidisks around each BH \cite{Bowen_2018,Gold14, 2021ApJ...922..175N, 2022ApJ...928..187C, Franchini_2022, Cattorini22}. While UV photons are produced by the inner edge of the circumbinary disk, a large amount of X-ray radiation is emitted by the minidisks  \cite{10.1093/mnras/sty423,d_Ascoli_2018}.
The motion of the binary is expected to imprint a modulation in the EM emission \cite{DalCanton19}. If the binary is already `on' years before the merger, the modulation might appear in optical and, possibly, it can be detected with surveys as LSST \cite{LSST}. In the last phase of the inspiral, the modulation can be instead detected in X-ray with future telescopes, such as the Advanced Telescope for High ENergy Astrophysics  (Athena) \cite{nandra13, 10.1093/mnras/stad659}. 

During and/or after the merger, flare or jet emissions are expected at different wavelengths and on timescales of weeks or months \cite{Milosavljevic:2004cg, Fontecilla17, Yuan20}. Additional transient features might be produced by a re-brightening of the accretion disk or by internal shocks in the gas, adjusting to the new gravitational potential \cite{2010MNRAS.401.2021R}.

To test the expansion of the Universe with bright MBHBs sirens,  the luminosity distance estimate is obtained from the GW signal and the redshift estimate from the EM counterpart. In M22 we proposed the following strategy: if the source is sufficiently bright in optical, its redshift can be determined directly, e.g.~with the Vera C. Rubin Observatory \cite{LSST, about_Rubin}. Otherwise, the MBHB can be first localised in radio with the Square Kilometre Array (SKA) \cite{ska} telescope or in X-ray with Athena \cite{ Athena-synergies}; its redshift can be subsequently determined with photometric or spectroscopic observations of the host galaxy performed, for example, with the Extremely Large Telescope (ELT) \cite{elt}.

In M22, we defined a \emph{GW event with EM counterpart} (EMcp for the rest of the paper)  
as a system whose EM counterpart can be detected by any of the proposed strategies, and whose sky localization is sufficiently accurate to fall inside the aforementioned telescope's fields of view (FOV). %(see the original work for more details).
Therefore, these sources represent  the subset of MBHB systems for which we have both the luminosity distance and the redshift measurements.

The rate of EMcps changes significantly depending on the processes responsible for the production of the EM counterpart (i.e.~the accretion rate for the UV/X-ray emission or the jet opening angle for the radio emission), on the 
%assumptions of the 
environment surrounding the MBHBs (i.e.~the AGN obscuration) and on the sky localization provided by LISA. In order to simplify the presentation of results, in M22 we considered two models, labelled  `maximising' and `minimising' following the number of EMcps that they predict.
The two main differences between these models are that in the former there is no AGN obscuration and the radio flare emission is isotropic, while, in the latter, we include the AGN obscuration and the radio flare emission is collimated with an opening angle of $\sim 30^\circ$.
In Tab.~\ref{tab:number_of_emcps} we report  the average number of EMcps for each model, assuming 
that the LISA mission lasts for 4yrs, its nominal duration. 
%4yrs of observations. 
The `maximising' model predicts on average between $\sim7$ and $\sim 20$ EMcps in 4 yrs, depending on the astrophysical population, while the `minimising' one predicts $\sim 2-3$ EMcps. The AGN obscuration and the collimated jet emission are the two main factors that drastically reduce the number of EMcps. 

Standard siren cosmology is one of the LISA science objectives that strongly depend on the number of sources and on the mission time \cite{2022arXiv220405434A}. In this study, we focus only on 
the `maximising' case and drop the `minimising' one, due to its limited number of EMcps. 
If the `minimising' case is the one closer to reality, the LISA science case of probing the expansion of the Universe with bright MBHBs will be undermined, if LISA will operate for only 4 yrs. 
However, in 10yrs, which is the maximal possible duration of the mission, the heavy models (Q3d and Q3nd) in the `minimising case' predict $\sim 3.3 \times 10/4 \sim 8.2$ EMcps: a number comparable to the one of
%which are close to the number of 
EMcps predicted by the Pop3 model, in the `maximising' case, and in 4yrs.

\begin{comment}
\begin{table}
\begin{tabular}{ m{1.4cm} | m{1.4cm} | m{2.5cm}  } 
\hline \hline
    Light  & Heavy & Heavy-no-delays  \\
\Xhline{0.1pt}
   16.0  & 37.0 & 51.7  \\
   \hline \hline

%\Xhline{0.1pt} 
\end{tabular}
\end{table}
\end{comment}

\begin{table}
\caption{\label{tab:number_of_emcps} Average number of EMcps in 4yrs in the `maximising' and `minimising' models.}
\begin{ruledtabular}
\begin{tabular}{c | m{2.4cm} | m{3.4cm}  } 
 (in 4yr)  & Maximising  & Minimising  \\
\Xhline{0.1pt}
Pop3   & 6.4  & 1.6  \\
Q3d  & 14.8 & 3.3  \\
Q3nd  &  20.7 & 3.5  \\
%\Xhline{0.1pt} 
\end{tabular}
\end{ruledtabular}
\end{table}

\section{\label{sec:cosmo_intro}Cosmological models}
The Friedmann-Lema\^{i}tre-Robertson-Walker (FLRW) metric is \cite{Dodelson2003, 2018arXiv180300070P}
\begin{equation}
    ds^2 = -c^2dt^2 + a^2(t) \left( \frac{dr}{1-Kr^2} + r^2d\Omega   \right)
\end{equation}
with $K=0, -1, +1$, $a(t)$ the scale factor and $c$ the light speed. 
From the FLRW metric, one derives the Friedmann equations
\begin{align}
H^2 = \frac{8\pi G}{3}\rho - \frac{Kc^2}{a^2} \label{eq:fried_eq1}\\
\frac{\ddot{a}}{a} = \frac{-4\pi G}{3} \left( \rho + \frac{3P}{c^2}  \right)
\label{eq:fried_eq2}
\end{align} 
where $H = (da/dt) / a = \dot{a}/a$ is the Hubble rate, $G$ the gravitational constant and $\rho$ and $P$ are the sum of the energy densities of all the components of the universe, i.e. $(\rho, P) = (\rho_r+\rho_m+\rho_\Lambda, P_r+P_m+P_\Lambda)$, where the subscripts `$r$', `$m$' and `$\Lambda$' refer to radiation, matter and cosmological constant, respectively.
Each component satisfies the continuity equation 
\begin{equation}
\label{eq:continuity_eq}
    \dot{\rho_i} + 3H \left( \rho_i + \frac{P_i}{c^2}\right) = 0
\end{equation}
where $i$ runs over the individual components.
Eq.~\ref{eq:continuity_eq} can be easily solved if we assume $P_i = \omega_i \rho_i c^2$ where $\omega_i$ is the equation of state for the i-component. For example, in the standard $\Lambda$CDM model, the equations of state are $w_i = (1/3, 0, -1)$ for radiation, matter and cosmological constant, respectively.
Plugging these values in Eqs.~(\ref{eq:fried_eq1}-\ref{eq:continuity_eq}), we obtain the Hubble rate as (from this moment we neglect the contribution from radiation, i.e. $\rho_r=0$, and we assume a flat universe, i.e. $K=0$)
\begin{equation}
    \label{eq:hubble_rate_h0_omegam}
    %\frac{H(z)}{H_0} = \sqrt{ \Omega_m(1+z)^3 + (1-\Omega_m-\Omega_\Lambda)(1+z)^2 + \Omega_\Lambda  }
    H(z) = H_0\sqrt{ \Omega_m(1+z)^3 + (1-\Omega_m)}
\end{equation}
where $H_0 = h \times 100 \, \rm km/(s \cdot  Mpc)$ is the Hubble constant %at the present day, 
$\Omega_m = 8 \pi G \rho_{m,0}/(3 H^2_0 )$ is the matter relative energy density today  and $z = 1/a-1$ is the redshift. For the fiducial cosmological model, we adopted $h = 0.6774 \,(H_0 = 67.74 \, \rm km/(s \cdot  Mpc))$ and $\Omega_m = 0.3075$ \cite{Planck15}; we note that in this case, $\Omega_\Lambda = \Lambda c^2/3H_0^2$ is fixed by the condition $\sum_i \Omega_i = 1$ (from Eq.~\eqref{eq:fried_eq1}), i.e. $\Omega_\Lambda = 1- \Omega_m$.

Assuming that the Universe is flat, we can define the luminosity distance $d_L(z)$  and the comoving distance $d_C(z)$, respectively as 
\begin{align}
    d_L(z) = c(1+z) \int_0^z \frac{dz'}{H(z')} \label{eq:dl} \\
    d_C(z) = \frac{d_L(z)}{1+z} 
    = c \int_0^z \frac{dz'}{H(z')} 
    \label{eq:dc} .
\end{align}

A %final 
useful remark for the future discussions is that, under the assumption of a flat Universe, the comoving distance $d_C$ is related to $H(z)$ as in Eq.~\ref{eq:dc} and we can express $H(z)$ as the inverse of the derivative in $z$ of the comoving distance, i.e.
\begin{equation}
\label{eq:H_of_z}
    H(z) = c \left( \frac{d}{dz} d_C \right)^{-1} \,.
\end{equation}
Given the broad span of redshift covered by MBHB standard sirens with LISA, in this work we assess their potential to test both the local and the higher redshift universe.

%$\Omega_\Lambda = 8 \pi G \rho_{\Lambda,0}/(3 H^2_0 ) = \Lambda c^2 /(3 H^2_0 ) $ is the relative energy density of the cosmological constant $\Lambda$ today

\subsection{\label{subsec:local_universe_models}Local Universe models}

Concerning the local universe,
we test the standard cosmological model and two additional beyond $\Lambda$CDM models. %For all the models presented here, we perform the analysis on $(h, \Omega_m)$ \emph{and} some additional parameters.
In particular, we analyse the following models:
\begin{enumerate}
    \item \label{itm:lcdm} $\boldsymbol{(h, \Omega_m)}$ :  Standard $\Lambda$CDM model. This is a two-parameter model where we fit for $(h, \Omega_m)$ using Eq.~\ref{eq:hubble_rate_h0_omegam}.
    
\item \label{itm:cpl} $\boldsymbol{(h, \Omega_m, \omega_0, \omega_a)}$: one of the most adopted parametrizations of the dark energy equation of state in the literature is the Chevallier-Polarski-Linder (CPL) formalism \cite{Linder03, Chevallier01} where one defines 
\begin{equation}
    \label{eq:cpl_dark_energy}
    \omega (z) = \omega_0 +\omega_a (1-a) =  \omega_0 +\omega_a \frac{z}{z +1}.
\end{equation}
With this equation of the state, the Hubble rate becomes 
\begin{equation}
    \label{eq:hubble_rate_cpl}
    \begin{split}
    \frac{H(z)}{H_0} & = \bigg( \Omega_m(1+z)^3  +  (1-\Omega_m)  \\
     & \times \exp{\left[ -\frac{3\omega_a z }{1+z}\right]}  (1+z)^{3(1+\omega_0+\omega_a)}  \bigg)^{1/2}.
    \end{split}
\end{equation}
Here we fit for $(h, \Omega_m, \omega_0, \omega_a)$ assuming $\omega_0 = -1$ and $\omega_a=0$ as fiducial values.

\begin{comment}
\item \label{itm:bull20} $\boldsymbol{(h, \Omega_m, \omega_0, \Delta \omega, z_c, \Delta z)}$ : Following \cite{Bull21}, we test a phenomenological dark energy model with a smooth transition of the equation of state of the form
\begin{equation}
    \omega(z) = \omega_0 + \frac{1}{2}\Delta \omega \left( 1 + \tanh\left( \frac{z-z_c}{\Delta z}\right)   \right)
\end{equation}
where $\omega_0=-1$ and $\Delta \omega = 0$ corresponds to the \lcdm Universe. We note that, in this case, $z_c$ and $\Delta z$ might assume any value.
Depending on the value of $z_c$ and $\Delta z$, the transition might happen in the post-reionisation matter-dominated era $(1<z<6)$ where we expect to detect the EM counterpart from MBHB mergers. 
\end{comment}

%\item \label{itm:sign-switching} \am{add the sign-switching model proposed by Nicola in arxiv:2307.10899}

\item \label{itm:belgacem19} $\boldsymbol{(h, \Omega_m, \omega_0, \Xi_0)}$: Alternative gravity theories \textcolor{\colorref}{leading to accelerating expansion predict that the luminosity distance measured by GWs may differ from the one measured by EM observations.}
%\sout{with non trivial dark energy models predict that GWs do not scale as $1/d_L$ even if they travel at light speed.
%In such theories the luminosity distance measured by GWs assuming the usual $1/d_L$ scaling may differ from the luminosity distance measured by EM observations.}
One phenomenological parametrization that encompass several of these models is \cite{Belgacem18, Belgacem19}
\begin{equation}
    \frac{d_L^{gw}(z)}{d_L^{em}(z)} = \Xi_0 + \frac{1-\Xi_0}{(1+z)^n} \,,
\end{equation}
where $d_L^{gw}$ and $d_L^{em}$ are the luminosity distances as measured by GW and EM observations, respectively.
In this case the Hubble rate is expressed as in Eq.~\ref{eq:hubble_rate_cpl}. This is a 4-parameter model where we fit for $(h, \Omega_m, \omega_0, \Xi_0)$ assuming $\omega_0=-1$ and $\Xi_0=1$ as fiducial values (these values correspond to $\Lambda$CDM) and  fixing $n=2.5$ and $\omega_a = 0$ in the inference process. 
\end{enumerate}

\subsection{\label{subsec:high-redshift-models}High-redshift Universe approaches}
\begin{figure}
    \includegraphics[width=0.5\textwidth]{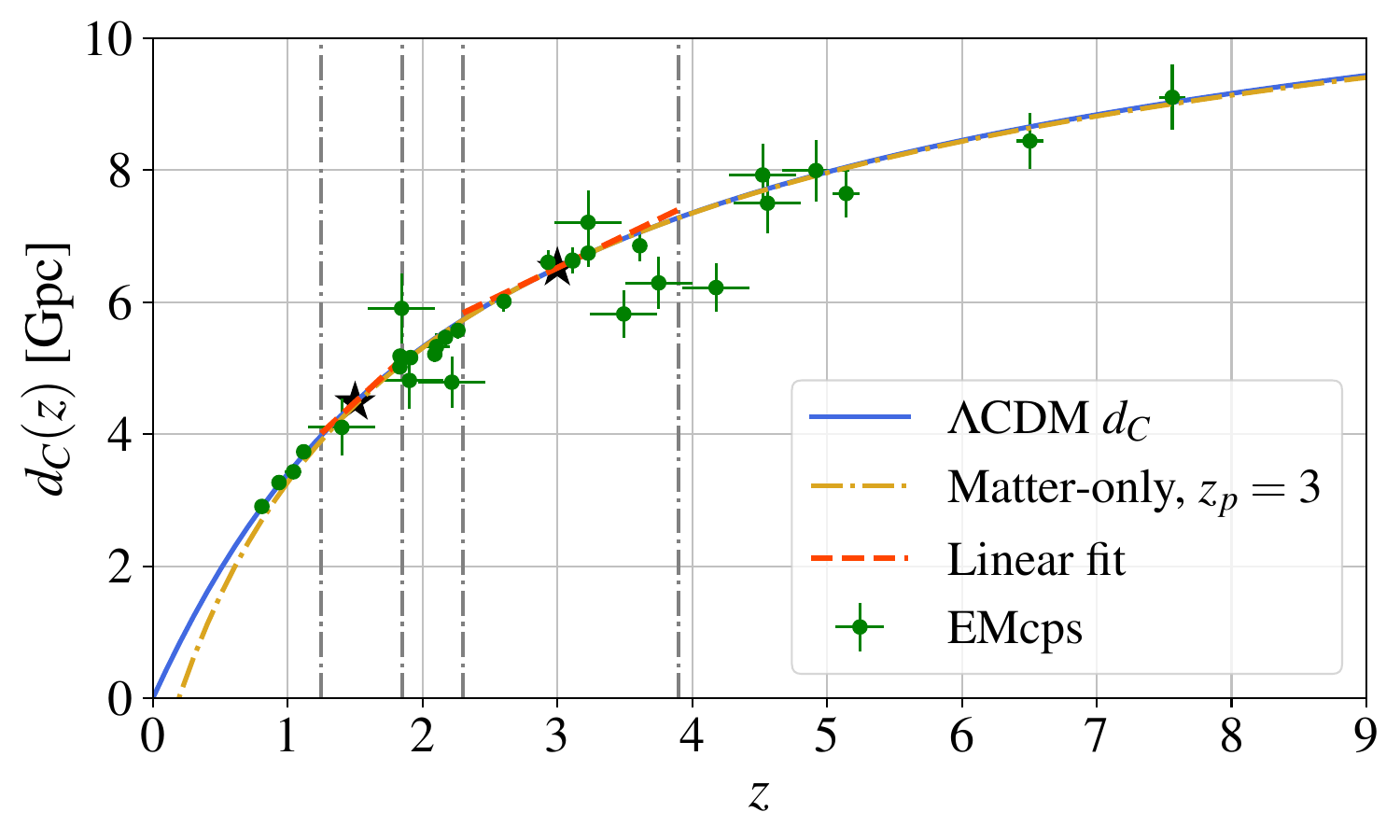} 
    \caption{Representation of the \hyperref[itm:bins]{\color{mycolormodel}redshift bins} and \hyperref[itm:matteronly]{\color{mycolormodel}matter-only} %approximation 
    models.
    %, according to the legend. 
    The blue, solid line corresponds to the comoving distance in the standard \lcdm Universe; the yellow, dotted-dashed line represents the matter-only approximation from Eq.~\ref{eq:matter-only} with $z_p=3$; the two red, dashed lines denote the redshift bins approach in Eq.~\ref{eq:linear_fit} for two redshift bins with $z_p=1.5$ and $z_p=3$ (black stars) \textcolor{\colorref}{and the vertical dotted-dashed grey lines visualise the boundaries of the corresponding redshift bins}. The green points show the MBHBs from a random Universe realisation, \textcolor{\colorref}{based on the astrophysical model Q3d, assuming 10 yrs of data (and therefore correspondingly of LISA mission duration),} %for 10 yr of data and 
    with the respective \textcolor{\colorref}{1$\sigma$} errors bars on the redshift and comoving distance, accounting also for lensing and peculiar velocities errors (as described in Sec.~\ref{sec:cat_construction}). For the low-redshift events, the errors are smaller than the size of the dot. }
    \label{fig:dc_matteronly_bins}
\end{figure}

By the time LISA will be operational, EM telescopes as Euclid \cite{Euclid}, LSST \cite{LSST} will have provided accurate measurements of the expansion of the Universe up to $z\lesssim2$.
However, thanks to the fact that we expect to detect the EM counterpart of MBHB mergers up to $z\sim 8$, we may wonder how we can use these systems to test the high redshift portion of the Universe (up to the redshift where we have both the EM and GW signal).  

In contrast with the previous ones, here we present methods that can be used to test the expansion of the Universe at $z>1$. The first one introduces a possible deviation to the matter equation of state; the second and third methods have in common that the cosmological inference is performed over two parameters $(H(z_p), d_C(z_p))$ corresponding to the Hubble parameter and the comoving distance at a given pivot redshift $z_p$; the last one is a model-independent approach based on a spline interpolation of the luminosity distance. More in detail, these approaches are:

\begin{enumerate}
     \item \label{itm:lcdm_beta}  $\boldsymbol{(h, \Omega_m, \beta)}$: we introduce a deviation to the matter equation of state of the form $\omega_m=\beta$. In this case,  Eq.~\ref{eq:hubble_rate_h0_omegam} becomes
     \begin{equation}     \label{eq:hubble_rate_h0_omegam_and_beta}
        H(z) = H_0\sqrt{ \Omega_m(1+z)^{3(1+\beta)} + (1-\Omega_m)}.
     \end{equation}
     This is a 3-parameter model where we fit for $(h, \Omega_m, \beta)$, assuming $\beta=0$ as the fiducial value. The scope of this model is to test if LISA can put constraints on the cold dark matter equation of state if it deviates from zero at high redshift; for this reason we decide to place this model in the `high-redshift' part even if we still have $h$ and $\Omega_m$ in the inference.
     Note that this is a simple phenomenological model which can be applied only to the late-time universe. Very strong constraints on $\beta$ would apply if CMB is taken into account. We must thus implicitly assume that ordinary %$\Lambda$CDM 
     evolution happens (i.e.~that $\beta\rightarrow 0$) at %say 
     $z\gtrsim 10$, outside the range of LISA MBHB multi-messenger data. \textcolor{\colorref}{Such a model arises in the context of dark matter models where the mass of the dark matter particles depends on the acceleration of the Universe \cite{2007PhRvD..75h3506A}}.

    \item \label{itm:matteronly} \textbf{Matter-only approximation}:
    Since we aim at constraints at high redshift, one reasonable assumption is that the Universe is matter-dominated, i.e. $H(z)
    \simeq H_0 \sqrt{\Omega_m} (1+z)^{3/2}$. In this case, the comoving distance can be written as 
    \begin{align}
       d_C(z) & = d_C(z_p) + \label{eq:matter-only}\\ & 2(1+z_p) H^{-1}(z_p) \bigg(  1-\frac{\sqrt{1+z_p}}{\sqrt{1+z}}\bigg). \nonumber
    \end{align}
    This is a 2-parameter model and we infer $(h(z_p), d_C(z_p))$ where $ H(z_p) =  h(z_p) \times 100 \, \rm km/(s \cdot  Mpc) $.
    For both parameters, we assume the \lcdm values as the fiducial ones.
    %\am{@Nicola/Robert can you send me the calculations on how you derived this expression?}\nt{Sent.}

    \item \label{itm:bins}\textbf{Redshift bins}: According to Eq.~\ref{eq:H_of_z}, $H(z)$ is the slope of the comoving distance relation. If we consider a small redshift interval around a pivot redshift $z_p$, we can  approximate the $d_c-z$ relation as a Taylor expansion at $z_p$ as
    \begin{equation}
        d_C(z) = d_C(z_p) + \frac{c}{H(z_p)}(z-z_p).
        \label{eq:linear_fit}
    \end{equation}
    This is also a 2-parameter model and we fit for $(h(z_p), d_C(z_p))$ where $ H(z_p) =  h(z_p) \times 100 \, \rm km/(s \cdot  Mpc) $ and $z_p$ corresponds to the pivot redshift. 
    Note that we can recover Eq.~\ref{eq:linear_fit} by expanding in $z\rightarrow z_p$ at first order Eq.~\ref{eq:matter-only}; however, within this model we do not assume that the Universe is matter dominated:
    %the same parameters as in the matter-only model, though $H(z_p)$ does not have the same exact meaning of the corresponding parameter in the previous model, but they coincide to first order in the limit $z\rightarrow z_p$.
    the only assumption behind Eq.~\ref{eq:linear_fit}  is that the Universe is flat.
     %This approach is independent on the chosen cosmological model \textcolor{\colorref}{ and the only assumption is that the Universe is flat}. % (except for the flatness assumption).

    \item \label{itm:splines} \textbf{Spline interpolation}: This model consists in interpolating the luminosity distance at several knot redshifts with cubic polynomials. The final product of the inference is the multi-dimensional posterior distribution on the $d_L$ at the knots. For the spline, we adopt the implementation in `\texttt{InterpolatedUnivariateSpline}' from \texttt{SciPy} \cite{scipy}. %\ls{what is the default boundary conditions? sometimes results change depending on this and it might be worth specifying this} \am{I didn't modify them so I guess it's the default ones. If someone is interested, they can just look at the documentation}.
\end{enumerate}
%\hyperref[itm:bins]{model}

For clarity, in Fig.~\ref{fig:dc_matteronly_bins} we show (i) an example of the redshift bins model for two bins with pivot redshift fixed at $z_p=1.5$ and $z_p=3$ respectively, and (ii) the matter-only approximation with pivot redshift $z_p=3$.
For all the models, we report more details on the technical implementation and some caveats in  Sec.~\ref{sec:inference}.

\section{\label{sec:cat_construction}Catalogues construction}

The cosmological inference on the models presented above is performed starting from catalogues of EMcps. 
In this work we use the catalogues produced in M22. However, for consistency, we provide below a summary of the catalogue production method.

For each astrophysical model, we have \textcolor{\colorref}{simulated 90 years of data, assuming $\Lambda$CDM. 
From these 90 years of data,}%and
we want to construct different Universe realisations, depending on the LISA mission time of observation ($t_m$).
We proceed in the following way:
\begin{enumerate}
    \item We compute the average intrinsic number of mergers per year, $\Lambda_{i}$, multiplying the total number of mergers in the 90 years of data by  $1/90$. The average intrinsic number of events during a certain time mission $\lambda_{i}$  is obtained as $ \lambda_{i} = t_m\cdot \Lambda_{i} $. For example, assuming $t_m=4 \yr$, the average number of events %in 4 years 
    is 691, 31 and 475 respectively for Pop3, Q3d and Q3nd, as reported in the first column of Tab.~(III) in M22;
    \item Since each realisation is independent on the others, we extract the intrinsic number of events $n$ in each realisation according to a Poisson distribution with mean $\lambda = \lambda_{i}$. Each realisation is then constructed drawing $n$ random events from the 90 years of data;
    \item In each realisation, we select only the events that are EMcps, i.e.~satisfy the requirements of SNR, detectability of the EM emission and sky localization accuracy, explained in \cref{sec:general_methods}.
    %\ls{Per each GW event how do you assign the probability that it has a counterpart?}\am{It's not a probability: for all the binaries, I compute the EM counterpart, with the information on the surrounding gas, spins and so on. So for each event, I know if it has and EM counterpart or not}
    \footnote{It may happen that the same EMcp is extracted twice during the second step. In this case, we remove the second one and extract a new EMcp.}
\end{enumerate}
For the purpose of the cosmological inference, we need the luminosity distance and redshift of the EMcps, and the corresponding errors on these quantities. 

\subsection{Error on the luminosity distance}

%While the former is provided directly from the catalogues, the latter requires some considerations.
%\am{Concerning the luminosity distance errors, we start from the} 
%For the luminosity distance, we consider the 
The luminosity distance measurement is provided by the GW emission; therefore, the error on the luminosity distance of a MBHB is provided by its detection with LISA. 
In this work, we adopt the
marginalized posterior distributions obtained with the LISA data analysis process, which we performed in M22. 
However, other sources of error are also expected to affect the measurement and should be added to the $d_L$ posterior distribution from LISA.
%Moreover, 

Weak lensing from the inhomogeneous distribution of matter between the source and the observer perturb the propagation of the GW signal, and potentially affect the recovered parameters \cite{Canevarolo23}. 
We model %this source of 
the weak lensing error as \cite{Cusin21}
    \begin{equation}
        \label{eq:lensing_popIII}
        \frac{\sigma_{\rm lens}}{d_L} = \begin{cases}
        \frac{0.061}{2}\left( \frac{1-(z+1)^{-0.264}}{0.264} \right)^{1.89} & \mbox{for } z \le 9.35 \\
        0.034+0.015z & \mbox{for } z > 9.35 \\
        \end{cases}
    \end{equation}
for Pop3 and as
    \begin{equation}
        \label{eq:lensing_Q3}
        \frac{\sigma_{\rm lens}}{d_L} = \frac{0.096}{2}\left( \frac{1-(z+1)^{-0.62}}{0.62} \right)^{2.36}
    \end{equation}
for the two massive astrophysical models. 
The total error on the luminosity distance must account for the contribution of weak lensing.
Since the weak-lensing depends on the amount of matter along the line-of-sight, it plays a significant role at high redshift. 
We find that weak lensing dominates the error budget of the luminosity distance at $z\gtrsim0.6$.

\textcolor{\colorref}{Eq.~\ref{eq:lensing_popIII}-\ref{eq:lensing_Q3} describe the amount of (de)magnification that the GW signal experiences while travelling from the source to the detector, due to the stochastic distribution of matter.} Similarly to \cite{Speri21, 2010MNRAS.404..858S}, we also take into account the possibility that specific observations along the line of sight of the GW event, estimating the amount of matter along the line of sight, can be used to reduce the weak lensing error scatter. We estimate this possible delensing factor as \cite{Speri21}
\begin{equation}
    F_{\rm delens} = 1-\frac{0.6}{\pi} \arctan \left( \frac{z}{0.073} \right).
    \label{eq:delens}
\end{equation}
The final lensing uncertainty accounting for delensing is then 
\begin{equation}
\label{eq:lensing_error}
    \sigma_{\rm delens} = F_{\rm delens} \sigma_{\rm lens}.
\end{equation}

The peculiar motion of the host galaxy also %add an additional source of uncertainty, 
affects the luminosity distance measurement,
especially at low redshift. We model the error on the luminosity distance measurement due to the peculiar velocity associated to the MBHB as \cite{Kocsis_2006}
\begin{equation}
    \label{eq:pv_error}
    \frac{\sigma_{\rm pv}}{d_L} = \bigg[   1 + \frac{c(1+z)^2}{H(z)D_L(z)} \bigg]
    \frac{\sqrt{\langle v^2 \rangle}}{c},
\end{equation}
where we fix $\langle v^2 \rangle = 500 \, \rm km/s$, in agreement with the value observed in galaxy surveys.
This error must also be accounted for in the total error budget.

\begin{figure}
    \includegraphics[width=0.5\textwidth]{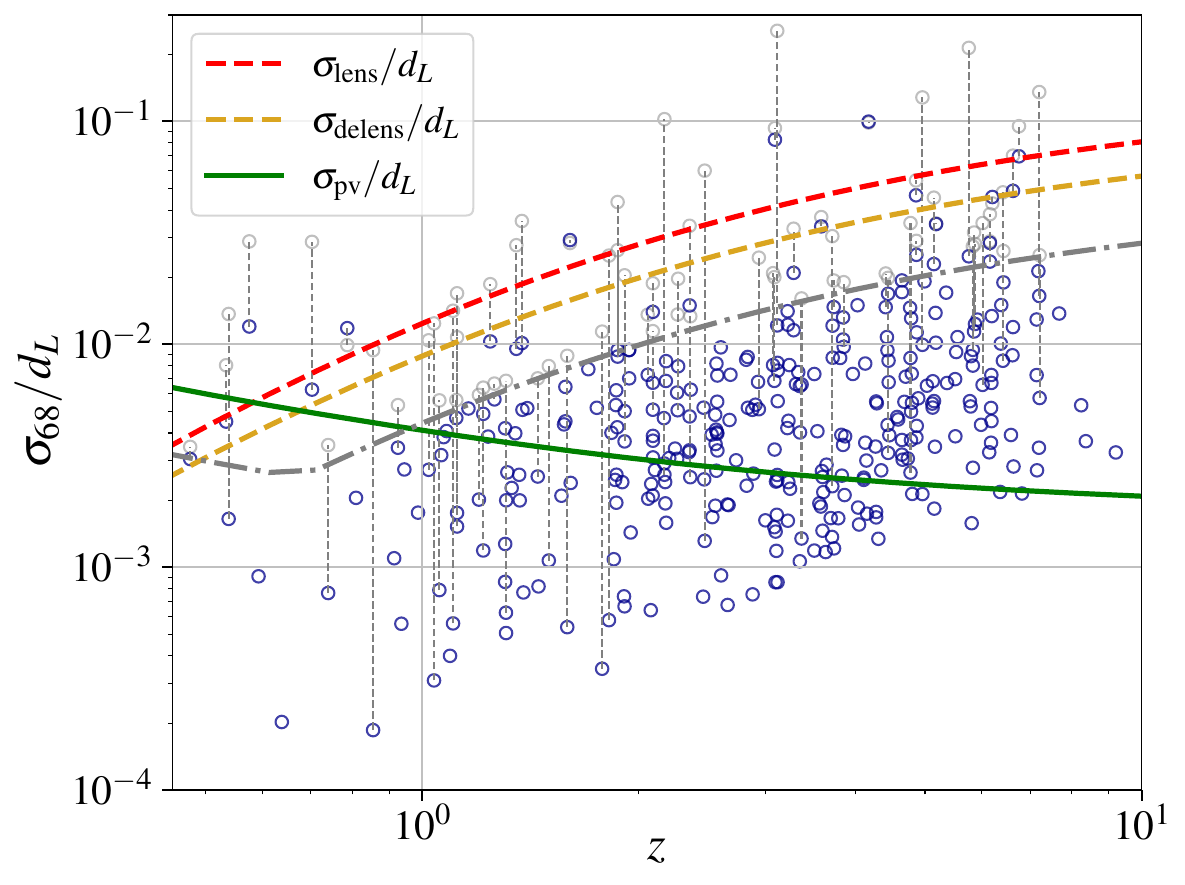} 
    \caption{Scatter plot of the luminosity distance uncertainty at 1$\sigma$ from LISA parameter estimation, as a function of redshift. The blue points correspond to all the MBHBs we have simulated within the Q3d model, i.e.~the 90 years of data. \textcolor{\colorref}{We convolve %these values 
    the luminosity distance uncertainty from the GW detection
    with the lensing and peculiar velocity errors as described in Sec.~\ref{sec:cat_construction}, to get the luminosity distance uncertainties adopted in the cosmological analysis}. The green solid line represents the errors from peculiar velocities as in Eq.~\ref{eq:pv_error}. The red (yellow) dashed line corresponds to the lensing error as in Eq.~\ref{eq:lensing_Q3} without (with) the delensing correction.  \textcolor{\colorref}{The grey dotted-dashed line corresponds to the arbitrary cut-off we impose on the $1\sigma$  error on $d_L$ in order to choose which systems ro rerun, assuming their sky position is known, i.e. $\sigma_{68,d_L} > 0.5 \, \rm \sigma_{\rm delens}$ or $0.5\sigma_{\rm pv}$.
   The grey points above the grey dotted-dashed correspond to the subset of systems for which we rerun the parameter estimation assuming perfect localisation, leading to the corresponding blue points (connected with a vertical thin dashed grey line)}. For points below the grey dotted-dashed line, the error on $d_L$ is already dominated by lensing or peculiar velocities.
    }
    \label{fig:dl_vs_z_lensing}
\end{figure}

\begin{figure}
    \includegraphics[width=0.5\textwidth]{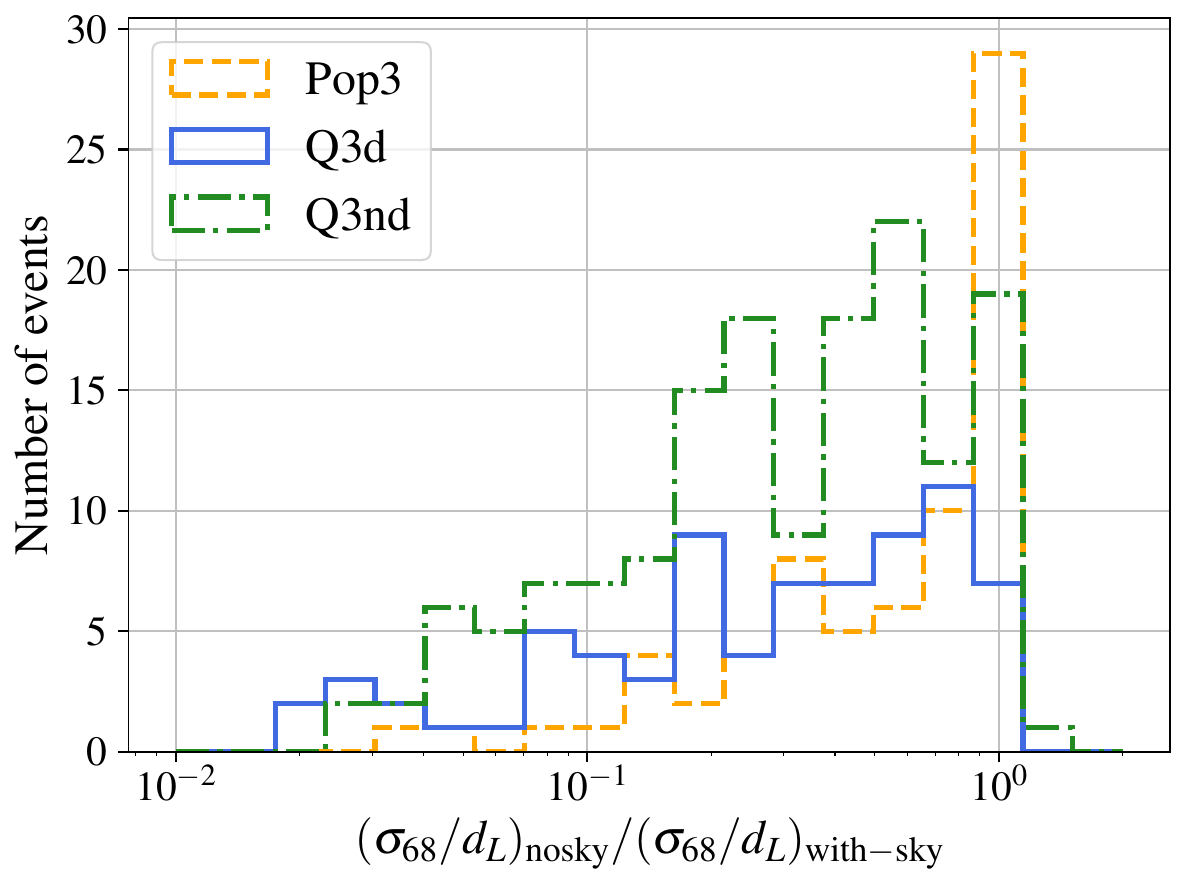} 
    \caption{Ratio of the luminosity distance error %without 
    fixing the sky position in the parameter inference %over luminosity distance error from 
     over the one obtained in the full analysis, i.e. including as parameters the two sky position coordinates. Both errors are at 90\% level. Colors and line styles correspond to the three astrophysical models, as reported in the legend.}
    \label{fig:ratio_dl_with_and_without_skypos}
\end{figure}

Fig.~\ref{fig:dl_vs_z_lensing} shows the relative error, as a function of redshift, on the luminosity distance measurement of all the MBHBs simulated within the Q3d model and it presents our strategy to deal with this error.  
The grey points represent the luminosity distance uncertainty from the GW parameter estimation process
% for the MBHBs simulated in the Q3d model as a function of redshift 
\textcolor{\colorref}{only, i.e. 
%the uncertainties reported do not include, for the moment, 
before including the lensing and peculiar velocity errors.
These latter are shown respectively by the dashed and solid lines (the yellow dashed line takes into account the possibility of delensing, see \cref{eq:delens} and \eqref{eq:lensing_error}).}
Lensing dominates the error budget at $z>0.6$ for the majority of sources, while peculiar velocities are relevant only at $z<0.6$. 
\textcolor{\colorref}{For the majority of events, the weak lensing uncertainty dominates the one from the GW detection: for these events, improving the error on the luminosity distance from the LISA data analysis would not change the total error budget.}
However, %it is clear that 
there is a sub-population of events for which the uncertainty on the luminosity distance from the GW detection is larger than the lensing or peculiar velocities errors. 
\textcolor{\colorref}{For these events, improving the luminosity distance measurement from the LISA data analysis can have an impact on the error budget.}

\textcolor{\colorref}{A common practice when an EM counterpart is present is to assume that the sky position of the binary is known with perfect accuracy 
%to remove this uncertainty from the analysis 
\cite{2017Natur.551...85A,Tamanini16}. This approach is motivated by the fact that the sky position accuracy from EM observations is significantly better than what one can obtain with GW detectors. If one %marginalise over the sky position,
considers that the sky position is known and fixes the binary position, i.e.~eliminates it from the MBHB binary parameter estimation,
%we expect that 
the uncertainties on the other binary parameters, including the luminosity distance, naturally improve.} 
Therefore, we decided to rerun the MBHB parameter estimation with LISA for all the systems with $\sigma_{68,d_L} > 0.5 \, \rm \sigma_{\rm delens}$ or $0.5\sigma_{\rm pv}$ (the grey dotted-dashed line in \cref{fig:dl_vs_z_lensing}), assuming that the detection of the EM counterpart allows to localise precisely the galaxy hosting the merger. 
In this way, we can remove the two parameters describing the binary sky position from the inference of the GW signal, and obtain better estimates on the luminosity distance error. 
\textcolor{\colorref}{The improved luminosity distance errors obtained after rerunning this subset of systems is represented in \cref{fig:dl_vs_z_lensing} by the blue points that are connected to the grey ones with grey dashed lines. 
Note that our specific threshold to select systems to rerun is motivated by the presence of the lensing error.}
%computational cost associated with analysing each GW signal. 
For systems below the grey dotted-dashed line in Fig.~\ref{fig:dl_vs_z_lensing}, the luminosity distance error budget is already dominated by the lensing error. Consequently, 
%even if we were to rerun the parameter estimation and achieve a smaller luminosity distance uncertainty for the GW signal, 
there is no need to rerun those, since the lensing error would still prevail as the dominant factor.
%\cc{I COMMENTED OUT THE PART BELOW, BECAUSE THIS IS JUST A MISUNDERSTANDING OF THE REFEREE: I DON'T THINK WE NEED TO SPECIFY THIS IN THE TEXT BUT JUST IN THE ANSWER.}
%Furthermore, we stress two points: (i) we are only considering the case of \emph{bright sirens} in which we assume that host galaxies are always identified (we comment on this choice at the end of Sec.~\ref{sec:discu_and_concl}); (ii) the identification of the host galaxy does not imply that the redshift is known with perfect accuracy and we have to fold this uncertainty in the analysis (as discussed later in this same section and in the likelihood construction in Sec.~\ref{sec:data_analysis}).
%
%

In Fig.~\ref{fig:ratio_dl_with_and_without_skypos}, we show the ratio of the luminosity distance error from the GW measurement with and without fixing the sky position, for the subset of systems whose $d_L$ \uncer was %originally 
above the grey dotted-dashed line in Fig.~\ref{fig:dl_vs_z_lensing} after the first binary parameter inference. Overall, we find an improvement in the estimate of the luminosity distance of almost two orders of magnitude for the two massive models Q3d and Q3nd. However, in Pop3 the gain is slightly less pronounced, due to the intrinsic low mass of these systems that make the the parameter estimation more complicated. We have a single MBHB merger in the Q3nd model for which the ratio is larger than unity, due to the stochastic behaviour of the MCMC chains.

From this point on, and unless otherwise specified, we will always 
\textcolor{\colorref}{consider as the luminosity distance error coming from the GW detection the one obtained after rerunning the LISA parameter estimation by fixing the sky position, for the subset of systems whose error was big to start with (`big' following the criterion explained above). This error, for the Q3d events, is
%refer to the luminosity distance error on the GW signal 
represented by the blue points in Fig.~\ref{fig:dl_vs_z_lensing}; we proceed similarly for the other populations (see \cref{fig:ratio_dl_with_and_without_skypos})}.

Since the catalogues have been constructed assuming $\Lambda$CDM, the intrinsic luminosity distance and redshift correspond to their $\Lambda$CDM values.
However, contrary to the simulated case, in reality the luminosity distance posterior distribution for each EMcp is not expected to be centred around the true \lcdm value, because of the aforementioned sources of errors.
Therefore, the weak lensing and peculiar velocity errors are expected not only to increase the spread of the luminosity distance posterior, but also to shift it randomly. Furthermore, besides the weak lensing and peculiar velocity, another source of error perturbing the luminosity distance estimate must be accounted for.
The LISA sensitivity \cite{Babak21} is in fact expected to fluctuate around an average value, due to the orbital motion of the spacecraft and the instrumentation. We model this noise by shifting
%This noise is expected to shift 
the posterior distribution of the luminosity distance by a factor drawn from a Gaussian distribution with the same dispersion of the posterior.
%
%different sources of errors
%are expected to affect \am{the measurement}.

In summary, to take into account all the different sources of errors contributing to the luminosity distance and mimic a realistic measurement, 
%the aforementioned sources of errors in luminosity distance for each MBHB event, 
we proceed in the following way:
%\nt{I wonder if there is a simple figure we can make to sketch what the following three shifts do to the posterior. Maybe show what happens to a single sample point?}\am{In principle I could plot a posterior, then how it is shifted for LISA noise and then the effects in points 3-4 but I don't think it's really useful and I would like to limit the number of plot}\nt{Fair enough}
\begin{enumerate}
    \item We compute the $1\sigma$ dispersion  of the luminosity distance posterior, $\sigma_{68, d_L}$, and shift all the $d_L$ samples by a random value extracted from $\mathcal{N}(0, \sigma_{68, d_L})$ where $\mathcal{N}$ is a Gaussian distribution;
    \item To take into account the lensing and peculiar velocities errors, we scatter all the $d_L$ samples $x_i$ as 
    \begin{equation}
        x_{i, \mbox{new}} = x_i + x_{\mathcal{N}(0, \sigma_{\rm delens})} + x_{\mathcal{N}(0, \sigma_{\rm pv})}
    \end{equation}
    where $x_{\mathcal{N}(0, \sigma_{\rm delens})}$ and $x_{\mathcal{N}(0, \sigma_{\rm pv})}$ are random numbers extracted from a Gaussian distribution with zero mean and the corresponding standard deviation; 
    \item Finally, we shift all the samples by a random value extracted from  $\mathcal{N}(0, \sqrt{\sigma^2_{\rm delens} + \sigma^2_{\rm pv}})$.
\end{enumerate}
Step (1) models the effect of LISA noise realisations while steps (2) and (3) represent the fact that lensing and peculiar velocities are expected to spread our luminosity distance posteriors \emph{and} to shift them with respect to the true value. 
At the end of this procedure, for each event, we have a new $d_L$ posterior distribution that is wider than the original one and not centered on the $d_L$ value from the assumed cosmology.

\subsection{Error on the redshift}

The redshift measurement is provided by the detection of the
%Moving to the redshift, it can be obtained by the 
EM counterpart. We assume that such measurement provides an estimate of the redshift uncertainty $\sigma_z$. 
Because of the construction of the EMcps catalogues, the redshift measurement will be centred around the true source redshift $z_\mathrm{true}$ corresponding to the \lcdm value.
%and its 

%\ls{We also assume that the redshift measurement is normally distributed which is justified by the fact.... Do we have any justification we can make here?} \am{as far as I know, no}
The value of $\sigma_z$ depends on the technique adopted to detect the EM emission and on the magnitude of the source. 
For the EMcps for which the EM emission is detected with LSST,  the redshift of the source can be measured photometrically with an error $\sigma_z = 0.031(1+z)$ \cite{Laigle19}. 
For ELT, if the source is sufficiently bright, the redshift can be spectroscopically estimated: in this case, we assume $\Delta z = 10^{-3}$. Otherwise, the redshift can be also measured photometrically with the Lyman-$\alpha$ ($\Delta z= 0.2$) or the Balmer break ($\Delta z= 0.5$) as summarised in Tab.~I of M22 (see also the discussion at the end of Sec.~IVD).
%We also stress that 
While the spectroscopic error depends only on the spectral resolution of the instruments, the redshift uncertainties for the photometric measurements are more uncertain: our assumptions can be considered as conservative.
%and it has to be considered as conservative.
These errors correspond to the 90\% confidence interval. 
%Therefore we can simply assume that, if the system is detected 
To summarise, for the EMcps whose redshift is measured with ELT, we set $\sigma_z \simeq \Delta z/2$ where $\Delta z$ might be $10^{-3}, \, 0.2, \, 0.5$. %\ls{"it might be": It is not clear what the actual error is. Do you make different analyses with different errors, if so, say so.}
%\am{Different sources have different errors because their EM counterpart is detected spectroscopically or photometrically so there are indeed different errors.}

Similarly to the posteriors for $d_L$, we also %scatter
%our observations in 
shift the posteriors for the redshift: for each EMcp, we extract a new redshift to perform the cosmological inference $z_{\rm em}$ as
\begin{equation}
    \label{eq:z_infer}
    z_{\rm em} = z_{\rm true} + \mathcal{N}_{\rm trunc}(0, \sigma_z)
\end{equation}
where $\mathcal{N}_{\rm trunc}$ is the truncated normal distribution, to avoid the extraction of negative redshift values at small redshift.

%\ls{Question out of curiosity: Are the redshift measurements not affected by the lensing and peculiar velocities as well?}\am{We simply assume that $\sigma_z$ encapsulates all the sources of error for EM because it was difficult to find reliable estimates at those magnitudes}. 

\section{\label{sec:data_analysis}Likelihood construction}
%\am{@Lorenzo/Sylvain check that what I wrote here makes sense}
In this section we describe the Bayesian formalism adopted in this study. Suppose we observe $N$ gravitational wave events $\xgw = \{ x_{\rm gw, 1}, \dots, x_{\rm gw, N}\}$ together with their corresponding EM counterparts $\xem = \{ x_{\rm em, 1}, \dots, x_{\rm em, N}\}$. The posterior distribution on the set of cosmological parameters $\theta_c$ based on some cosmological model $\mathcal{H}$ and on the total set of observations $\{ \xgw, \xem \}$ can be expressed as \cite{delpozzo12, Laghi21}
\begin{equation}
p(\theta_c|\xgw \xem \theta_m \hi) = \frac{ p(\xgw \xem|\theta_c \theta_m \hi) }{ p(\xgw \xem| \theta_m \hi)} \pi(\theta_c|\hi) \,,
\end{equation}
where $\theta_m$ collects the parameters describing the astrophysical populations, $\mathcal{I}$ collects all the necessary background information \textcolor{\colorref}{and $\pi$ corresponds to the prior distribution. }%\ls{It seems that we do not really use any of this, so ti might be worth to drop $\hi$. The only place where it matters is in the model assumption of the relation between $z - d_L$} \am{I would like to keep it because it corresponds to the general expression and it's the same notation adopted in Del Pozzo 11}.
The quasi-likelihood $p(\xgw \xem|\theta_c \theta_m \hi)$ can be rewritten as a function of the GW signal and EM counterpart parameters. We define $\thetabin$ as the set of GW signal parameters minus the luminosity distance $d_L$ and $\thetaenv$ as the set of parameters describing the surrounding environment where the EM counterpart is produced minus the redshift $z$. In our case $\thetabin$ corresponds to the two rest-frame BH masses, sky position, inclination, polarization, final phase and time to coalescence and the two spin magnitudes. The set of parameters $\thetaenv$ corresponds to the parameters necessary to produce the EM counterparts as described in M22.

The quasi-likelihood $p(\xgw \xem|\theta_c \theta_m \hi)$ can be expanded as 
% \begin{equation}
% \label{eq:quasi-like}
% \begin{split}
% p(\xgw \xem|\theta_c \theta_m \hi) & = \frac{1}{\alpha(\theta_c \theta_m)} \times \int d \thetabin d \thetaenv d d_L dz \\
% & \times p(\xgw \xem|\thetabin  \thetaenv d_L z \, \theta_c \theta_m \hi) \\
% & \times p(\thetabin  \thetaenv d_L z | \theta_c \theta_m \hi). 
% \end{split}
% \end{equation}
%
%
\begin{eqnarray}
p(\xgw \xem|\theta_c \theta_m \hi) = \frac{p'(\xgw \xem|\theta_c \theta_m \hi)}{\alpha(\theta_c \theta_m)} \label{eq:quasi-like} \\
\begin{split}
    p'(\xgw \xem|& \theta_c \theta_m \hi) = \int \mathrm{d} \thetabin \mathrm{d} \thetaenv \mathrm{d}d_L \mathrm{d}z \\
    & \times p(\xgw \xem|\thetabin  \thetaenv d_L z \, \theta_c \theta_m \hi) \\
    & \times p(\thetabin  \thetaenv d_L z | \theta_c \theta_m \hi).
    \label{eq:quasi-like-noalpha}
\end{split}
\end{eqnarray}
Let's start working out the expression in the integral in Eq.~\ref{eq:quasi-like-noalpha}. The second term in Eq.~\eqref{eq:quasi-like-noalpha} defines how the parameters of the GW and EM events depend on the astrophysical population for a given cosmology. It can be split in the following way
\begin{align}
    p(\thetabin  \thetaenv d_L z | \theta_c \theta_m \hi) =\\ p (d_L|z \theta_c \hi) p(\thetabin  \thetaenv z | \theta_c \theta_m \hi) 
\end{align}
where the second term determines how the parameters of the events depend on the population, whereas the first term $ p (d_L|z \theta_c \hi)$ defines how luminosity distance and redshift are related. In the analysis we assume that $ p(\thetabin  \thetaenv z | \theta_c \theta_m \hi)$ can be treated as a  constant in our inference. We do not fit the population parameters $\theta_m$ and we expect $\thetabin \thetaenv $ to be mostly affected by astrophysical processes, rather than $\theta_c$ or the cosmological and background prior $\hi$. Therefore, we assume that the impact of changing cosmology does not affect the distribution of events. Assessing this assumption would require rerunning the SAM model for different cosmologies and multiple times  which is computationally prohibitive. 
For our purposes, constant quantities can be discharged so only the term $ p (d_L|z \theta_c \hi)$ remains.

% Let's start working out the expression in the integral in Eq.~\ref{eq:quasi-like-noalpha}. First of all, we neglect the contribution coming from the population term $p(\thetabin  \thetaenv d_L z | \theta_c \theta_m \hi)$: we expect that the contributions from different cosmological parameters $\theta_c$ play a minor role in determining $\thetabin$ and  $\thetaenv$ that are mostly affected by astrophysical processes. Moreover, the astrophysical populations  are fixed (no dependence from $\theta_m$)  as well as the cosmological model (no dependence from $\hi$). We note that assessing this assumption would require rerunning the SAM model multiple times which is computationally prohibitive.
% \nt{repetiton!}

% \begin{align}
%     p(\thetabin  \thetaenv d_L z | \theta_c \theta_m \hi) = p(\thetabin  \thetaenv | d_L z \theta_c \theta_m \hi) p (d_L z |  \theta_c \theta_m \hi)
% \end{align}
% }

The first term in the integral of Eq.~\eqref{eq:quasi-like-noalpha} defines how the GW and EM data \{$\xgw,\xem$\} are related to the models that fit the data. 
It can be simplified assuming that the GW and EM measurements are independent. We also suppose that the GW event depends only on the binary parameters, %and the EM counterpart only on the redshift (in this way we can also remove the integral on $\thetaenv$),
leading to 
\begin{equation}
    \begin{split}
        p(\xgw \xem|\thetabin  \thetaenv d_L z \, \theta_c \theta_m \hi)  & =  p(\xgw |\thetabin d_L \, \theta_c \hi) \\
    & \times p(\xem | \thetaenv z \theta_c  \hi).
    \end{split}
\end{equation}
In the likelihood, we assume that the EM observation depends only on the redshift of the source (note, however, that the dependence on the binary environment has already been taken into account while constructing the EMcps catalogues)
and we can write the EM counterpart likelihood as:
\begin{equation}
    \begin{split}
        \int \mathrm{d} \thetaenv p(\xem | \thetaenv \theta_c z \hi)  \textcolor{\colorref}{p( \thetaenv | \theta_c z \hi)} = p(z_{em}|z)
    \end{split}
\end{equation}
\textcolor{\colorref}{where we re-used the assumption on the constant probability of $\thetaenv$ to perform the integral.}

Moreover, the luminosity distance can be expressed as a function of $z$ and the cosmological parameters $\theta_c$ so we can rewrite the integral in Eq.~\ref{eq:quasi-like-noalpha} as 
\begin{equation}
\label{eq:like-simplified}
\begin{split}
    p'(\xgw \xem|& \theta_c \theta_m \hi) = \int \mathrm{d} \thetabin \mathrm{d} d_L \mathrm{d}z \, \\ &  
    \times p(\xgw |\thetabin d_L \, \theta_c \hi) \,
     p(z_{em}|z) \\
     & \times \delta(d_L - d_L^c(z, \theta_c))
\end{split}
\end{equation}
where $d_L^c(z, \theta_c)$ is the luminosity distance according to a specific cosmological model \footnote{To distinguish $d_L^c(z, \theta_c)$ from an integration variable, we use the superscript `$c$'.} as the ones specified in Sec.~\ref{sec:cosmo_intro} and the $\delta(\dots)$ comes from the $ p (d_L|z \theta_c \hi)$ term.
The quantity $p(\xgw |\thetabin d_L \, \theta_c \hi)$ can be expressed as the ratio between the posterior distribution of the binary parameters and the prior, i.e. 
\begin{equation}
    p(\xgw |\thetabin d_L \, \theta_c \hi) = \frac{p(\thetabin d_L| \xgw)}{p(\thetabin d_L)} .
\end{equation}
Since we assumed uniform prior,  Eq.~\ref{eq:like-simplified} becomes  
\begin{equation}
\begin{split}
p'(\xgw & \xem| \theta_c \theta_m \hi)= \int \mathrm{d} d_L \mathrm{d}z \, \\ 
& \times  \, p(d_L|\xgw) \, p(\zem| z) \, \delta(d_L - d_L^c(z, \theta_c)) 
\end{split}
\end{equation}
where we marginalized over $\thetabin$ to obtain $p(d_L|\xgw)$.

The property of the delta function $\delta (f(z)) = \delta (z - z_0) |\partial_z f(z_0)|^{-1}$, where $f(z_0)=0$ allows to rewrite the above equation as follows
\begin{equation}
\begin{split}    p'(\xgw & \xem| \theta_c \theta_m \hi)= \int \mathrm{d} d_L \mathrm{d}z \,p(d_L|\xgw) \, \\ 
& \times   \,  p(\zem| z) \delta(z - z^c(d_L, \theta_c))  \left | \frac{\mathrm{d}d_L(z, \theta_c)}{\mathrm{d}z} \right |^{-1} \, ,
\end{split}
\end{equation}
where we denote $z^c(d_L, \theta_c)$ as the redshift for a given luminosity distance and cosmological parameters, i.e. the inverse of $d_L^c(z, \theta_c)$.
We can now solve the integral in redshift and obtain
\begin{equation}
\begin{split}    p'(\xgw & \xem| \theta_c \theta_m \hi)= \int \mathrm{d} d_L 
    \, p(d_L|\xgw) \, \\ 
& \times   \,  p(\zem| z^c(d_L, \theta_c)) \left | \frac{\mathrm{d}d_L(z^c(d_L, \theta_c), \theta_c)}{\mathrm{d}z} \right |^{-1} \, .
\end{split}
\end{equation}
From the LISA parameter estimation we have the posterior $ p(d_L|\xgw)$ and the associated samples $d^i_L$. This allows to evaluate the integral in a Monte Carlo way:
\begin{equation}
\begin{split}    p'(\xgw & \xem| \theta_c \theta_m \hi)=\\ 
    &\sum_{d_L^i \in p(d_L|\xgw)} p(\zem| z^c(d^i_L, \theta_c)) \left | \frac{\mathrm{d}d_L(z^c(d^i_L, \theta_c), \theta_c)}{\mathrm{d}z} \right |^{-1} \, .
\end{split}
\end{equation}

\begin{comment}
\am{alternative:
Using the delta function to solve the integral for $d_L$ we end up with
\begin{align}
\label{eq:like-intermediate-step}
    p'(\xgw \xem| \theta_c \theta_m \hi) =\\ \int dz \, p(d_L(z, \theta_c)|\xgw) \,  p(\zem| z) \left | \frac{dd_L^c(z, \theta_c)}{dz} \right |.
\end{align}
From the LISA parameter estimation we have the full posterior $p(d_L(z, \theta_c)|\xgw)$ so we can take the luminosity distance samples $d_L^i$ and convert them in redshift samples using the $z^i = z(d_L^i, \theta_c) $ where the conversion depends on the cosmological parameters and model adopted. In this way we can write that $p(d_L(z, \theta_c)|\xgw) = p(z|\xgw \theta_c)$ and Eq.~\ref{eq:like-intermediate-step} becomes
\begin{equation}
\begin{split}
    p'(\xgw & \xem| \theta_c \theta_m \hi) = \int dz \\
    & \times p(z|\xgw \theta_c) \left | \frac{dd_L(z, \theta_c)}{dz} \right | p(\zem| z) \\
    & = \sum_{z_i\in p(z|\xgw \theta_c)} \left | \frac{dd_L(z_i, \theta_c)}{dz} \right | p(\zem| z_i)
\end{split}
\end{equation}
where in the last step we uses the Monte Carlo integration to evaluate the integral and the Jacobian is necessary for the transformation from $z$ to $d_L$.
For the likelihood of the EM counterpart we adopted a Gaussian form as 
\begin{equation}
    p(\zem| z_i) = \frac{1}{\sqrt{2\pi \sigma^2_z}}\exp{\Big[ -\frac{1}{2}\frac{(\zem - z_i)^2}{\sigma^2_z} \Big]}.
\end{equation}
}
\end{comment}

For the likelihood of the EM counterpart we adopted a Gaussian form as 
\begin{equation}
    p(\zem| z) = \frac{1}{\sqrt{2\pi \sigma^2_z}}\exp{\Big[ -\frac{1}{2}\frac{(\zem - z)^2}{\sigma^2_z} \Big]}.
\end{equation}

The quantity $\alpha(\theta_c \theta_m)$ in Eq.~\ref{eq:quasi-like} is the \emph{selection function} and it takes into account that not all the GW events or all the EM counterparts are observed \cite{Mandel17}. %\ls{Maybe it is worth writing the full expression and refer to the checks we did}. 
The fact that we observe only a sub-sample of the entire population might lead to biased estimates if not properly accounted for. On a practical level, selection effects can be understood thinking that, for example, some combinations of $h$ and $\Omega_m$ might move sources outside/inside the GW (or EM) horizon changing the luminosity distance of the source. 
The computation of the selection function requires the integration of the integral in Eq.~\ref{eq:quasi-like-noalpha} over all the possible combinations of $\{ \xgw, \xem \}$ above the respective detection thresholds. Before delving into the calculation of this quantity, we checked how the number of EM counterparts changes for different values of $(h, \Omega_m)$. We picked the median realisation for the Q3d model and the pair of samples $(h, \Omega_m)$ that give the smallest and largest luminosity distance. 
For these two values of $d_L$ we recomputed the EM counterpart for each MBHB in our catalogs and we rescaled the sky localization as in \cite{PhysRevD.93.024003} in order to quantify the number of EMcps that could enter or exit the analysis, varying the cosmological parameters. For Q3d we find a difference of $\sim 0.35$ EMcps in 4 yrs of observation. Since the variation is negligible, we simply assume that $\alpha(\theta_c \theta_m) \sim \rm const.$, and neglect its contribution in the analysis. This assumption is also motivated by the actual results of the cosmological inference, because we do not observe any strong bias coming from selection effects. 
\textcolor{\colorref}{Note that a difference of $\sim 0.35$ EMcps in 4 yrs may not be negligible anymore if the `minimising' model is the correct one, and if LISA operates for only 4 yrs, due to the very small number of EMcps(c.f. with Tab.~\ref{tab:number_of_emcps}). 
However, in a case with such a small number of EMcps,
%we have not run this analysis because 
we anyway do not expect that any meaningful cosmological constraints can be obtained. Therefore, as already mentioned, we always focus on the `maximising' case in the following.}

\section{\label{sec:inference} Cosmological inference analysis}

\begin{table}
\caption{\label{tab:prior_table_local_Universe} Inferred parameters and the corresponding priors for the \hyperref[subsec:local_universe_models]{\color{mycolormodel}`Local Universe'} models. In agreement with \cite{Belgacem19}, for the last model, we adopt 
normal priors as $\mathcal{N}[\mu,\sigma]$ for $h$ and $\Omega_m$, based on CMB+BAO+SNe data; while we adopt truncated normal distributions for $\omega_0, \,\Xi_0$, as given in the table.
%normal priors as $\mathcal{N}[\mu,\sigma]$ (for $\omega_0, \,\Xi_0$ we consider truncated normal distributions). In the last case, the prior on $h$ and $\Omega_m$ are based on CMB+BAO+SNe data. 
}
\begin{ruledtabular}
\begin{tabular}{c|c|c} 
 Model  & Parameter  & Prior  \\
\Xhline{0.1pt}
\multirow{2}{*}{\hyperref[itm:lcdm]{\color{mycolormodel}$(h, \Omega_m)$}}   & $h$  & $\mathcal{U}[0.2,1]$ \\
  & $\Omega_m$ & $\mathcal{U}[0,1]$  \\
\hline
\multirow{4}{*}{\hyperref[itm:cpl]{\color{mycolormodel}$(h, \Omega_m, \omega_0, \omega_a)$}} & $h$  & $\mathcal{U}[0.2,1]$ \\
  & $\Omega_m$ & $\mathcal{U}[0,1]$  \\
& $\omega_0$ & $\mathcal{U}[-3,-0.3]$ \\
& $\omega_a$ & $\mathcal{U}[-2,2]$ \\
  \hline
\multirow{4}{*}{\hyperref[itm:belgacem19]{\color{mycolormodel}$(h, \Omega_m, \omega_0, \Xi_0)$}}
& $h$  & $\mathcal{N}[0.6774,0.012]$ \\
  & $\Omega_m$ & $\mathcal{N}[0.3075,0.0124]$  \\
& $\omega_0$ & truncated $\mathcal{N}[-1,1]$ in $[-3,-0.3]$ \\
& $\Xi_0$ & truncated $\mathcal{N}[1,0.5]$ in $[0,+\infty]$ \\
%\Xhline{0.1pt} 
\end{tabular}
\end{ruledtabular}
\end{table}

\begin{comment}
\begin{table}
\caption{\label{tab:prior_table_local_Universe} Inferred parameters and the corresponding priors for the `Local Universe' models. In agreement with \cite{Belgacem19}, in the last model we adopt normal priors as $\mathcal{N}[\mu,\sigma]$ (for $\omega_0, \,\Xi_0$ we consider truncated normal distributions). In the last case, the prior on $h$ and $\Omega_m$ are based on CMB+BAO+SNe data. \am{add a table with the fiducial values in all cases?} }
\begin{ruledtabular}
\begin{tabular}{c|c|c} 
 Model  & Parameter  & Prior  \\
\Xhline{0.1pt}
\multirow{2}{*}{$(h, \Omega_m)$}   & $h$  & $\mathcal{U}[0.2,1]$ \\
  & $\Omega_m$ & $\mathcal{U}[0,1]$  \\
\hline
\multirow{3}{*}{$(h, \Omega_m, \beta)$}  & $h$  & $\mathcal{U}[0.2,1]$ \\
  & $\Omega_m$ & $\mathcal{U}[0,1]$  \\
  & $\beta$ & $\mathcal{U}[-3,3]$  \\
  \hline
\multirow{4}{*}{$(h, \Omega_m, \omega_0, \omega_a)$} & $h$  & $\mathcal{U}[0.2,1]$ \\
  & $\Omega_m$ & $\mathcal{U}[0,1]$  \\
& $\omega_0$ & $\mathcal{U}[-3,-0.3]$ \\
& $\omega_a$ & $\mathcal{U}[-2,2]$ \\
  \hline
\multirow{4}{*}{$(h, \Omega_m, \omega_0, \Xi_0)$}
& $h$  & $\mathcal{N}[0.6774,0.012]$ \\
  & $\Omega_m$ & $\mathcal{N}[0.3075,0.0124]$  \\
& $\omega_0$ & truncated $\mathcal{N}[-1,1]$ in $[-3,-0.3]$ \\
& $\Xi_0$ & truncated $\mathcal{N}[1,0.5]$ in $[0,+\infty]$ \\
%\Xhline{0.1pt} 
\end{tabular}
\end{ruledtabular}
\end{table}
\end{comment}

\begin{table}
\caption{\label{tab:prior_table_high_Universe} Inferred parameters and the corresponding priors
for the \hyperref[subsec:high-redshift-models]{\color{mycolormodel}`High-redshift Universe'}  models. The first three columns as in Tab.~\ref{tab:prior_table_local_Universe}. The fourth and fifth columns represent the pivot redshift (if applicable) and the corresponding redshift range. In the last column we report the number of EMcps in 10 yr in the redshift range. For most of the parameters we choose uniform priors (more details are given in Sec.~\ref{sec:inference}). The priors on the spline interpolation are given specifically in Sec.~\ref{subsec:inference_splines}.}
\begin{ruledtabular}
\begin{tabular}{c|c|c|c|c|c} 
 Model & Parameter & Prior & $z_p$ & $[z_{\rm min}, z_{\max}]$ & \begin{tabular}{@{}c@{}}EMcps \\ in 10 yr \end{tabular}
  \\
\Xhline{0.1pt}
\multirow{3}{*}{\hyperref[itm:lcdm_beta]{\color{mycolormodel}$(h, \Omega_m, \beta)$}}  & $h$  & $\mathcal{U}[0.2,1]$ & \multirow{3}{*}{-} & \multirow{3}{*}{[$0,+\infty$]} & 16.0 \\
  & $\Omega_m$ & $\mathcal{U}[0,1]$ & & & 37.0 \\
  & $\beta$ & $\mathcal{U}[-3,3]$  & & & 51.7\\
  \hline
\multirow{18}{*}{\hyperref[itm:matteronly]{\color{mycolormodel}\begin{tabular}{@{}c@{}}Matter-only \\ approx. \end{tabular}}} &  \multirow{18}{*}{\begin{tabular}{@{}c@{}}$d_C(z_p)$/Gpc \\ $h(z_p)$  \end{tabular}}  & \multirow{18}{*}{\begin{tabular}{@{}c@{}}$\mathcal{U}[1,\textcolor{\colorref}{10}]$ \\ $\mathcal{U}[0.2,\textcolor{\colorref}{10}]$ \end{tabular}}  & \multirow{3}{*}{2} & \multirow{3}{*}{$[\textcolor{\colorref}{1.2,6}]$} &  \textcolor{\colorref}{12.0} \\
& & & & & \textcolor{\colorref}{29.9} \\
& & & & & \textcolor{\colorref}{40.5} \\ \cline{4-6}
& & &  \multirow{3}{*}{3} & \multirow{3}{*}{$[1.5,\textcolor{\colorref}{10}]$} &  \textcolor{\colorref}{11.4} \\
& & & & & \textcolor{\colorref}{31.0} \\
& & & & & \textcolor{\colorref}{38.9} \\ \cline{4-6}
& & &  \multirow{3}{*}{\textcolor{\colorref}{4}} & \multirow{3}{*}{$[\textcolor{\colorref}{1.7,10}]$} &  \textcolor{\colorref}{10.0} \\
& & & & & \textcolor{\colorref}{30.2} \\
& & & & & \textcolor{\colorref}{37.3} \\ \cline{4-6}
& & &  \multirow{3}{*}{\textcolor{\colorref}{5}} & \multirow{3}{*}{$[\textcolor{\colorref}{1.8,10}]$} &  \textcolor{\colorref}{9.7} \\
& & & & & \textcolor{\colorref}{29.9} \\
& & & & & \textcolor{\colorref}{35.5} \\ \cline{4-6}
& & &  \multirow{3}{*}{\textcolor{\colorref}{6}} & \multirow{3}{*}{$[\textcolor{\colorref}{1.9,10}]$} &  \textcolor{\colorref}{9.3} \\
& & & & & \textcolor{\colorref}{28.5} \\
& & & & & \textcolor{\colorref}{35.2} \\ \cline{4-6}
& & &  \multirow{3}{*}{\textcolor{\colorref}{7}} & \multirow{3}{*}{$[\textcolor{\colorref}{1.9,10}]$} &  \textcolor{\colorref}{9.3} \\
& & & & & \textcolor{\colorref}{28.5} \\
& & & & & \textcolor{\colorref}{35.2} \\

\hline
\multirow{24}{*}{\hyperref[itm:bins]{\color{mycolormodel}\begin{tabular}{@{}c@{}}Redshift \\ bins \end{tabular}}} & \multirow{24}{*}{\begin{tabular}{@{}c@{}}$d_C(z_p)$/Gpc \\ $h(z_p)$  \end{tabular}}  & \multirow{24}{*}{\begin{tabular}{@{}c@{}}$\mathcal{U}[0.1,50]$ \\ $\mathcal{U}[0.1,50]$ \end{tabular}} & \multirow{3}{*}{1} & \multirow{3}{*}{$[\textcolor{\colorref}{0.85,1.2}]$} & \textcolor{\colorref}{1.5} \\
& & & & & \textcolor{\colorref}{2.1} \\
& & & & & \textcolor{\colorref}{4.3} \\ \cline{4-6}
& & &  \multirow{3}{*}{1.5} & \multirow{3}{*}{$[\textcolor{\colorref}{1.25,1.85}]$} &  \textcolor{\colorref}{2.8} \\
& & & & & \textcolor{\colorref}{3.7} \\
& & & & & \textcolor{\colorref}{7.8} \\ \cline{4-6}
& & &  \multirow{3}{*}{2} & \multirow{3}{*}{$[\textcolor{\colorref}{1.7,2.5}]$} &  \textcolor{\colorref}{2.7} \\
& & & & & \textcolor{\colorref}{6.5} \\
& & & & & \textcolor{\colorref}{8.4} \\  \cline{4-6}
& & &  \multirow{3}{*}{2.5} & \multirow{3}{*}{$[\textcolor{\colorref}{2.0,3.2}]$} &  \textcolor{\colorref}{3.8} \\
& & & & & \textcolor{\colorref}{9.4} \\
& & & & & \textcolor{\colorref}{11.8} \\  \cline{4-6}
& & &  \multirow{3}{*}{3} & \multirow{3}{*}{$[\textcolor{\colorref}{2.3,3.9}]$} &  \textcolor{\colorref}{5.0} \\
& & & & & \textcolor{\colorref}{11.7} \\
& & & & & \textcolor{\colorref}{17.4} \\  \cline{4-6}
& & &  \multirow{3}{*}{3.5} & \multirow{3}{*}{$[\textcolor{\colorref}{2.6,4.7}]$} &  \textcolor{\colorref}{5.2} \\
& & & & & \textcolor{\colorref}{13.0} \\
& & & & & \textcolor{\colorref}{18.2} \\  \cline{4-6}
& & &  \multirow{3}{*}{4} & \multirow{3}{*}{$[\textcolor{\colorref}{3.0,5.4}]$} &  \textcolor{\colorref}{4.8}\\
& & & & & \textcolor{\colorref}{14.5} \\
& & & & & \textcolor{\colorref}{18.9} \\  \cline{4-6}
& & &  \multirow{3}{*}{5} & \multirow{3}{*}{$[\textcolor{\colorref}{3.7,6.7}]$} &  \textcolor{\colorref}{3.7} \\
& & & & & \textcolor{\colorref}{13.4} \\
& & & & & \textcolor{\colorref}{15.4} \\
\hline
\multirow{5}{*}{\hyperref[itm:splines]{\color{mycolormodel}\begin{tabular}{@{}c@{}}Splines \\ interp. \end{tabular}}} &  $d_L(z=0.2)$  & \multirow{5}{*}{$d_L^{\Lambda {\rm CDM}}$}  & \multirow{5}{*}{-} & \multirow{5}{*}{[$0,7$]} &  \multirow{5}{*}{\begin{tabular}{@{}c@{}}16.0 \\ 35.8 \\ 51.5 \end{tabular}} \\
&  $d_L(z=0.7)$ & & & \\
&  $d_L(z=2)$ & & & \\
&  $d_L(z=4)$ & & & \\
&  $d_L(z=6)$ & & & \\

\end{tabular}
\end{ruledtabular}
\end{table}

Following the procedure described in Sec.~\ref{sec:cat_construction}, we generate 100 EMcps catalogues, which are realisations of 4 and 10 yrs of LISA observations, for the three astrophysical models, focusing on the `maximising' scenario (see \cref{tab:number_of_emcps}). 
\textcolor{\colorref}{Performing the analysis for multiple realisations of the Universe is necessary to realistically
forecast LISA capabilities when data will be available. 
There are large uncertainties both on the MBHB formation channel, and on the EM emission accompanying MBHB mergers. 
Therefore, depending on our `luck', 
LISA might observe more or less EMcps, leading to better or worse cosmological constraints, respectively. 
Indeed, we observe that the results of the cosmological inference on the  cosmological models which we analyse (see \cref{subsec:local_universe_models} and \cref{subsec:high-redshift-models}) do depend on the Universe realisation: even within the same MBHB formation model, and for the same mission duration, some realisations provide better constraints and some worst.
Therefore, in \cref{sec:results}, we present our results on the forecast errors on the cosmological parameters as cumulative distributions to help the reader identify the fraction of realisations providing that particular error value 
When we quote errors on a given cosmological parameter, we take as reference the value which is provided by at least 50\% of the realisations.}

\textcolor{\colorref}{As will be clear later, there are cosmological models for which the fluctuations are larger than others, and consequently, the results depend on the realisations heavily than for others. 
For example,
%later discussed, we found that 
for the \hyperref[itm:bins]{\color{mycolormodel}redshift bins} approach, we found that not all realisations 
%might 
provide useful results. 
Indeed, dividing the redshift span of EMcps catalogues in redshift bins renders the effect of the EMcps number fluctuations more severe.
If there are not enough EMcps in a given bin, the Hubble parameter and the comoving distance at that pivot redshift 
%are 
turn out to be unconstrained. 
We do not, however, discard this model.
%Since t
The current work attempts at forecasting LISA abilities to constrain cosmological parameters with different approaches and techniques, and in some realisation cases the redshift bin approach does indeed provide useful constraints, interesting since they are model independent. However, if we are not `lucky' and the Universe corresponds to a realisation in which
%we believe that it is important to state that this approach might 
not enough EMcps are present, this approach simply will not be %performed 
adopted in the future LISA data analysis
%. We refer the interested reader to the 
(see discussion in Sec.~\ref{subsec:bins}). }

Concerning the parameter estimation, we run the MCMC for 2500 iterations with $n \times 16$ walkers where $n$ corresponds to the number of parameters in the model.
In Tab.~\ref{tab:prior_table_local_Universe} we report the prior range for the parameters over which we performed the inference for the  \hyperref[subsec:local_universe_models]{\color{mycolormodel}`Local Universe'} models. In the first two cases we assume uniform priors, while for the  \hyperref[itm:belgacem19]{\color{mycolormodel}$(h, \Omega_m, \omega_0, \Xi_0)$} model, we follow the same approach of \cite{Belgacem19}: since the proposed modification can be measured only with GWs, we assume CMB+BAO+SNe priors for $h$ and $\Omega_m$.

In the case of the  \hyperref[subsec:high-redshift-models]{\color{mycolormodel}`High-redshift Universe'} scenarios,
\textcolor{\colorref}{for the \hyperref[itm:lcdm_beta]{\color{mycolormodel}$(h, \Omega_m, \beta)$} model we adopt uniform priors, reported in Tab.~\ref{tab:prior_table_high_Universe}.}
Additional considerations are necessary to clarify our procedure for the other models, to which are dedicated the next three sections.
\textcolor{\colorref}{As a common feature, it is important to point out that
%First of all, 
the \hyperref[itm:matteronly]{\color{mycolormodel}matter-only}, \hyperref[itm:bins]{\color{mycolormodel}redshift bins}, and   \hyperref[itm:splines]{\color{mycolormodel}spline interpolation} %following approaches 
models all have built-in luminosity distance relations that differ from the $\Lambda$CDM one. 
Therefore, to provide relialable estimates during the analysis, it is important to quantify the model's accuracy with respect to 
$\Lambda$CDM: this must always be smaller than the uncertainties on the model parameters provided by the  inference analysis.
The accuracy is defined as 
\begin{equation}
\label{eq:accuracy}
    \mathcal{A}(z) = \frac{d^{\rm model}(z) - d^{\Lambda \rm CDM}(z)}{d^{\Lambda \rm CDM}(z)}
\end{equation}
where $d^{\rm model}$ ($d^{\Lambda \rm CDM}$) is the luminosity or comoving distance for the particular model ($\Lambda$CDM).
We will comment more on this aspect in the following on a case by case basis.}

\subsection{Matter-only approximation}
% \am{cite somewhere the LISA low-frequency document. How I can cite it? is it published somewhere? or in the Atrium?}
% \nt{It has not been published (yet). I would not cite it.}
\label{sec:matter_sub}

The  \hyperref[itm:matteronly]{\color{mycolormodel}matter-only} model 
necessitates to choose the pivot redshift around which to expand the comoving distance-redshift relation.
Since we assume that the Universe is matter-dominated, we have to choose values of $z_p$ which are sufficiently large. 
\textcolor{\colorref}{In the following, we consider six values, equally spaced from $z_p=2$ to $z_p=7$.
By construction, the matter-only approximation diverges from \lcdm at low redshift and, to a lower extent, at high redshift too.
As a consequence, in order to avoid significant biases in the reconstructed parameters, one needs to remove the EMcps where the model is not enough accurate. 
This is a clear disadvantage of the model, since it causes a reduction in the number of EMcps useful for the analysis. }

\textcolor{\colorref}{For a given pivot redshift $z_p$, one must then find suitable boundaries $z_{\rm min}$ and $z_{\rm max}$ such that the accuracy of the model over the redshift interval always remains smaller than the uncertainty on the comoving distance $d_c(z_p)$, i.e.~the inferred  model parameter.\footnote{To clarify with respect to Eq.~\ref{eq:accuracy}, here the accuracy is  defined as the difference between the comoving distance in the matter-only approximation and in \lcdm at a given redshift divided by $d_C$ of \lcdm.} 
The choice for the final values of $z_{\rm min}$ and $z_{\rm max}$ for each $z_p$ has been determined by a process of trial-and-error: we performed a first analysis with a large interval and, if the uncertainties on the comoving distance were smaller than the model's accuracy, we rerunned the analysis.
We repeated the process until the reported uncertainties were larger than the model's accuracy. 
For example, for $z_p=2$, the matter-only approximation is accurate at $\sim 0.9\%$ at $z=1.2$, so we removed all the EMcps at $z<1.2$ and set $z_{\rm min}=1.2$. 
Similarly, to keep an accuracy smaller than $1\%$, we also set $z_{\rm max}=6$. 
The threshold of 1\% has been chosen based on the results of the relative uncertainties on the cosmological parameters (see \cref{fig:matter_only_uncer_new}).
A similar procedure could also be performed with real data, i.e. we could add or remove a portion of the EMcps until the uncertainties on the cosmological parameters are larger than the model's accuracy. However, in the case that the Universe do not follow \lcdm (or any other known model), it would be impossible to evaluate the accuracy of the matter-only approximation.} 

\textcolor{\colorref}{The priors for $(h(z_p), d_C(z_p))$, the pivot redshifts and the values of $[z_{\rm min},z_{\rm max}] $ are reported in Tab.~\ref{tab:prior_table_high_Universe}.}

%two values: $z_p=2$ and $z_p=3$. In the former case, the matter-only approximation is accurate \footnote{Here the accuracy is  defined as the difference between the comoving distance in the matter-only approximation and in \lcdm at a given redshift divided over the $d_c$ from \lcdm} at $\sim 2-1\%$ in the redshift range $1<z<7$, while in the latter we have an accuracy of $\sim 4\%$ already at $z=1$. As a consequence, in order to avoid significant biases in the reconstructed parameters, we chose to remove the EMcps at low redshift. In particular, in the case $z_p=2$, we remove all the EMcps at $z<1$ while for $z_p=3$ we remove all systems at $z<1.5$ \textcolor{\colorref}{where we have an accuracy of 0.9\%}. We do not apply any cut at high redshift where deviations are below $\sim 1\%$ at $z=10$.

\textcolor{\colorref}{We highlight that there is nothing special in the pivot redshifts that we choose for this approach. As the model is based on the assumption that the Universe is matter-dominated, any sufficiently large redshift (i.e. $z\gtrsim 1$) can be adopted as pivot redshift. However, as we go to larger $z_p$, we have to remove more EMcps because the approximation diverges quickly from $\Lambda$CDM for $z<z_p$. For example, setting $z_p=4$, we have an accuracy of $\sim 1.1\%$ already at $z=1.7$, so we have to remove all the EMcps at $z<1.7$, leading to fewer and fewer EMcps as we move to higher $z_p$. %The priors and the redshift ranges adopted for this approach are reported in the second row of Tab.~\ref{tab:prior_table_high_Universe}
}

\subsection{\label{subsec:bins}Redshift bins approach}

\begin{figure}
    \includegraphics[width=0.5\textwidth]{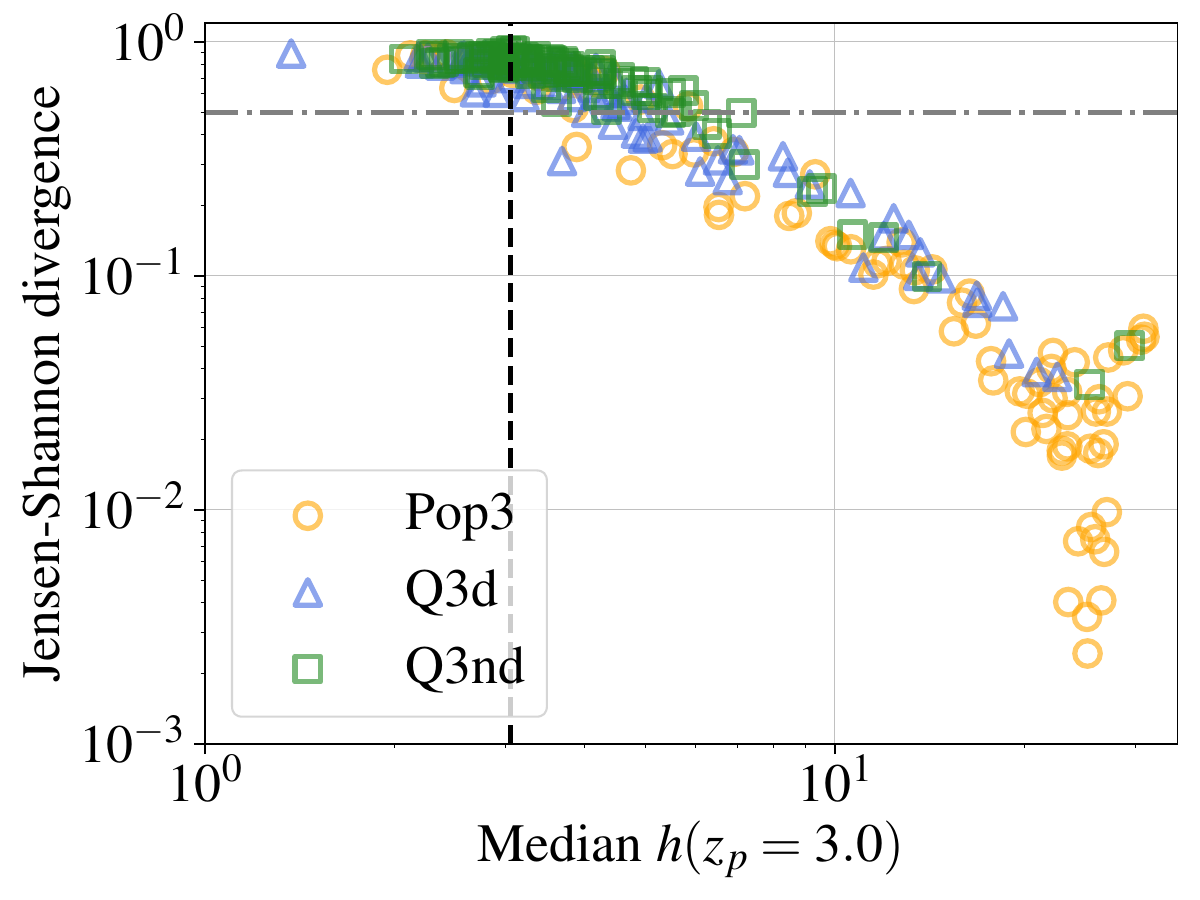} 
    \caption{\textcolor{\colorref}{The JS divergence as a function of the median of the posterior on the parameter $h(z_p)$, for all the EMcps catalogues simulated for the three MBHB formation scenarios, and taking as example the redshift bin centred on $z_p=3$. The JS divergence is
    computed between the posterior distribution of $h(z_p=3)$ and its uniform prior, which is extended to the interval %between 
    $[0.1,50]$ to distinguish the \emph{uninformative} realisations.% versus the inferred median value of $h(z_p=3)$. 
    Each point corresponds to a single realisation for each astrophysical model, according to the legend. The horizontal grey dotted-dashed line corresponds to the arbitrary cut-off of 0.5 which we impose
    on the value of the JS divergence to select the informative realisations (more details in the text). The black dashed line represents the \lcdm true value, $h(z_p=3)= 3.06$.
    %, according to $\Lambda$CDM. 
    %We consider as \emph{informative} realisations only the ones above the grey line. 
    Uninformative realisations show inferred median values 
    %at 
    far from the \lcdm one and, specifically, decaying towards $ h(z_p=3)\sim25$, corresponding to the midpoint of the prior range. Furthermore, the JS divergence decays towards 0, %i.e. 
    meaning that the posterior is similar to the prior. Informative realisations, on the other hand, cluster around the \lcdm value, and have $\mathrm{JS}\gtrsim 0.5$.}}
    \label{fig:js_divergence_vs_hvalue}
\end{figure}

\begin{figure}
    \includegraphics[width=0.5\textwidth]{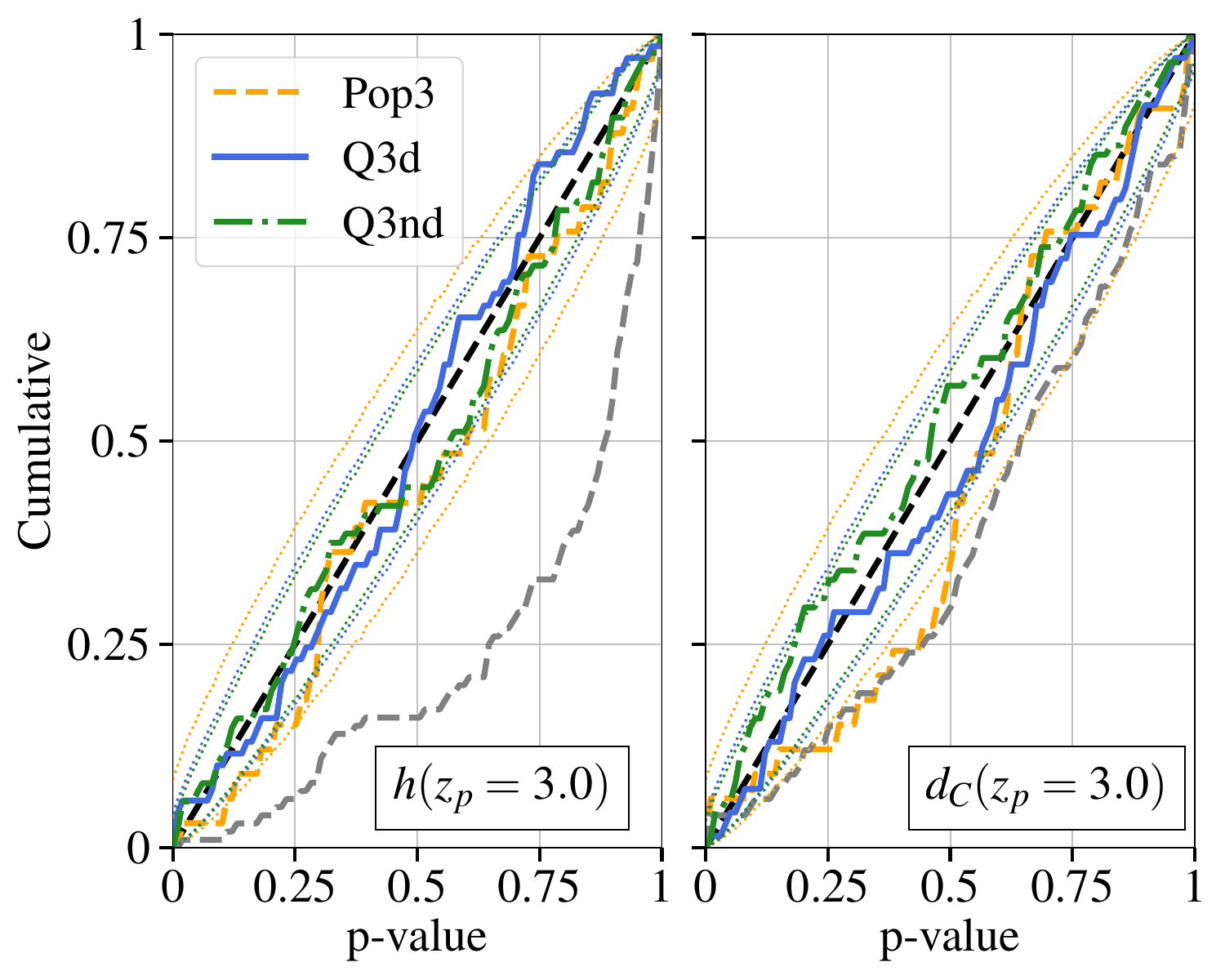}
    \caption{Cumulative distribution of events with p-value smaller or equal to the abscissa. 
    %Colors and line styles 
   The solid, dashed and dot-dashed lines represent the \emph{informative} realisations only, for the different MBHB formation scenarios 
    according to the legend. The dotted lines represent the 90\% uncertainty region expected from the finite number of realisations. The dashed black line represents the expected value for unbiased results. Grey dashed line: same as the orange one for Pop3, but including all the realisations, even the uninformative ones (according to the JS criterion (see the meain text). It is clear that selecting only the informative realisations provide results consistent with the expected values.  %\am{need inputs from Sylvain} \am{making the plot square is a bit too hard because each panel is too small or too big}
     \label{fig:ppplot_for_hz3}
    }
\end{figure}

In the  \hyperref[itm:bins]{\color{mycolormodel}redshift bins} approach, we approximate the comoving distance as in Eq.~\ref{eq:linear_fit}, in eight redshift intervals. 
For each redshift interval,
%we consider 
only the EMcps that fall in that particular bin are used for the analysis.\footnote{In order to select the systems in a given redshift bin, we use the value of $z_{\rm em}$ from Eq.~\ref{eq:z_infer}.}
%and discharge all the others. 
%One can easily see that  there are t
Two competing effects enter in the implementation of this model: on one hand, one would like to increase the size of the redshift bin as much as possible, in order to include more EMcps; on the other hand, extending the redshift range leads to an inaccurate representation of the $d_C(z)$ relation (it can no longer be approximated as a straight line), which introduces significant biases in the recovered parmaters $(h(z_p), d_C(z_p))$. 
This approach might also lead to biased results for the comoving distance by construction: 
if, by chance, the EMcps redshifts lie only close to the bin edges, the inference recovers a comoving distance that is slightly smaller than the expected value at the pivot redshift, while the slope is still the same. 
Therefore, the lower and upper limit of the redshift bin play an important role in the inference: we choose them in order to have the maximum number of EMcps in a redshift bin, without compromising the recovered parameters or the accuracy between the linear approximation and  $\Lambda$CDM.

\textcolor{\colorref}{We report the pivot redshifts and the corresponding redshift bins in Tab.~\ref{tab:prior_table_high_Universe}. Similarly to the \hyperref[itm:matteronly]{\color{mycolormodel}matter-only} model, the chosen pivot redshifts have no particular physical meaning. The analysis can be performed at any pivot redshift (as long as there are enough EMcps in the corresponding redshift bin), and each bin is treated independently. In other words, we do not need a consistent cosmological model across %cosmic time 
redshift,
and that's the strength of the approach. For example, if the analysis of actual LISA+EM facilities data shows with enough credibility that the EMcps observations are consistent with \lcdm at $z_p=2$ but not at $z_p=4$, it could be the evidence of new underlying physics.} 

\textcolor{\colorref}{Due to the two aforementioned competing effects, the possibility to perform the redshift bins approach %will 
strongly depends on the number of EMcps 
%around a certain $z_p$ 
that LISA+EM facilities will observe in each redshift bin. 
If we are unlucky and not enough EMcps are present (this occurs e.g.~in the Pop3 MBHB formation scenario), this approach cannot be implemented. For this reason,
we perform the analysis assuming that LISA lasts for 10 years. 
We find that, if the mission lasts 4yrs only, LISA+EM facilities won't observe a sufficient number of EMcps for the redshift bin approach to be effective.
%will not be performed. 
%Indeed, this is the result we find in our study. 
%As discussed later in this section, 
However, even assuming 10yrs as the mission duration, we still find that the outcome of the analysis is very much realisation-dependent.
Some realisations of EMcps catalogues, especially within the Pop3 astrophysical model, do not contain enough EMcps to
%were not able to 
provide useful constraints on the model parameters $((h(z_p),d_C(z_p))$, within some of the redshift bins, while others do. However, instead of disregarding the approach altogether, we investigated the issue more carefully, and came up with the following procedure to evaluate the 
number of cases where we can obtain useful cosmological constraints within this approach.} 

Depending on the astrophysical models and on the redshift bin, 
%we might 
there are realisations that do not have 
%many 
enough EMcps (this occurs both at low and at high pivot redshift) or for which the EMcps in the bin have too large $d_L$ and $z$ uncertainties (this occurs specifically at high redshift). In these cases, 
%we have 
one has to be extra cautious, because 
%if 
whenever the data are not sufficiently informative, the priors play a pivotal role.

For both $h(z_p)$ and $d_C(z_p)$ we impose uniform priors between $[0.1,50]$ (in the corresponding units, see \cref{tab:prior_table_high_Universe}). 
However, in standard $\Lambda$CDM, one expects $h(z)$ to span from $h(z=0) \sim 0.7$ to $h(z=7) \sim 8.5$ . 
Considering these values, a prior extending up to 50 might seem too broad to be realistic, given our current and future `degree of belief' on $h(z)$. 
%Here we justify our choice.
Indeed, originally we had chosen smaller priors for $h(z_p)$ and $d_C(z_p)$, symmetric and centered on the \lcdm values. 
%assuming $\Lambda$CDM. 
However, after a deeper inspection of the results, we noticed that some Universe realisations, i.e.~some EMcps catalogues, were not informative, since the posteriors on $h(z_p)$ and $d_C(z_p)$ inferred from the EMcps data provided by these realisations were identical to the priors. 
As expected, this happened if the data were not sufficiently informative, i.e.~whenever there was a too small number of EMcps in a given bin, or the $d_L$ and $z$ uncertainties were too large.
%Without accounting for these \emph{uninformative realisations}, 
We therefore realised that it was imperative to account for the presence of these \emph{uninformative} realisations, otherwise
our forecasts would have been \emph{prior dominated}. 
%In other words, LISA might not be able to provide always data-driven results in all the redshift bins and we need to quantify this information.

\textcolor{\colorref}{In order to account for the {uninformative realisations}, we decided to set
%the 
unphysically large priors on $h(z_p)$ and $d_C(z_p)$, which allow us to %are necessary to 
distinguish  between \emph{informative} and \emph{uninformative} realisations. 
%Here the idea is that, i
Indeed, if a realisation is \emph{uninformative}, the parameter inference on that realisation returns a
%the 
median of the posterior distribution 
%will be 
that is close to the median of the prior one, and 
%that the 
a 90\% error 
%will be 
that is similar to the prior range. 
On the other hand, if the 
realisation %posterior distribution 
is \emph{informative}, the result of the data analysis procedure is a %will give us additional information and the 
posterior distribution 
%will be 
that is significantly different from the prior, and a median
%will be 
that is close to the true value. 
Therefore, it is intuitively clear that by expanding the prior range it becomes easier to distinguish the {uninformative} from the the informative realisations, by inspection of their mean value.}

Still, we need a more rigorous and concrete criterion to compare the posterior distributions. In the following, we have adopted
the Jensen-Shannon (JS) divergence \cite{JS-divergence} as a way to quantify the degree of similarity between two distributions. 
In our case, we apply the JS divergence to the posterior and the prior distributions for $h(z_p)$ and $d_C(z_p)$.
If the realisation is \emph{informative}, the posterior distribution will be significantly different from the prior, and the JS divergence will be close to 1. If the realisation is \emph{uninformative}, the posterior will be instead similar to the prior and the JS will be closer to 0.
In Fig.~\ref{fig:js_divergence_vs_hvalue} we show an example of the JS divergence computed between the posterior in $h(z_p)$ and a flat uniform prior in $[0.1,50]$ for all the EMcps catalogues that we have simulated, across the three MBHB formation scenarios and fixing 10 yrs of LISA mission. 
For this example, we have chosen the redshift bin centred on $z_p=3$.
On the x-axis we report the inferred median value of $h(z_p=3)$, and on the y-axis the JS divergence. 
It is clear that there is a fraction of realisations with median value $h(z_p=3)\sim 25$ (the midpoint of the extended prior range) and low JS divergence, indicating that these realisations do not provide any constraint on $h(z_p=3)$. 
As expected, it is the Pop3 formation scenario that show the largest number of uninformative realisations, due to the smaller average number of EMcps and larger errors on luminosity distance and redshift. 
\textcolor{\colorref}{However, %we also note that 
%there are 
Fig.~\ref{fig:js_divergence_vs_hvalue} also shows that the majority of realisations %that 
correctly recover the median value $h(z_p=3)\sim 3$, i.e.~close to the true \lcdm value of $h(z_p=3)= 3.06$.%according to $\Lambda$CDM). 
%Therefore, it's clear that s
Therefore, from the analysis of the JS divergence, summarised in Fig.~\ref{fig:js_divergence_vs_hvalue} for the bin at $z_p=3$, we conclude that some Universe realisations %might tell us something on 
do contain enough EMcps to enable the redshift bins approach and in turns to provide meaningful constraints on
the expansion of the Universe at high redshift, while others do not.} 
%\am{FIX: In order to get rid of the uninformative cases}, 

In the following, we have decided to select as \emph{informative} realisations %only 
the ones with JS divergence $>0.5$. 
This value is somewhat arbitrary, but \textcolor{\colorref}{it is motivated by the results of Fig.~\ref{fig:js_divergence_vs_hvalue}. Indeed, setting a threshold of $\mathrm{JS}=0.5$ effectively eliminates all the realisations with median $h(z_p=3)\sim 25$. 
We provide more details on this choice in Appendix~\ref{sec:js_convergence}.}
Note that, to distinguish the informative realisations, we make use of the JS criterion exclusively on
%apply this criterion only on 
the posterior distributions of $h(z_p)$, because it's the parameter we are mostly interested in in this approach. %, because the posterior in $d_C(z_p)$ is typically significant different from the prior distribution. Indeed, the comoving distance represent just the intercept of the line at the corresponding pivot redsfhit and it's therefore easier to constrain than the slope. 
%We decide to 
% \am{maybe to remove or move to appendix} 
Moreover, in order to further support the choice of 
%this value
the JS threshold, we present a probability-probability (PP) plot of the p-value following the approach in \cite{2021arXiv211206861T} and considering only the \emph{informative} realisations. 
For each informative realisation, we compute the quantile in which  the true value is contained,  and we assume $n=$p-value$\times \rm N$ and $\rm N = \rm N_{info-real} $
where $\rm N_{info-real}$ corresponds to the number of informative realisations.
We show an example of the PP plot for $z_p=3$ in Fig.~\ref{fig:ppplot_for_hz3}. For $h(z_p)$, the pp-plots follow the diagonal line and they are compatible with the corresponding errors. The results for $d_C(z_p)$ are a bit worse but still compatible with the overall errors. This is due to the nature of the bin approach, which tends to produce slightly biased results for the comoving distance. 
It is also clear that, if we include 
%all the realisations 
also the realisations classified as uninformative
($\rm N=100$), there is a significant deviation in the cumulative distribution of the p-value from the expected one, as showed by the grey line.
Note that the JS divergence and the PP plot convey different information: the former quantifies the difference between the posterior and the prior distribution without %telling us 
accounting for where the posterior peaks; the latter 
%tell us 
shows the fraction of realisations that contain the true value in a given percentile interval.

In all the results for the redshift bins approach given in \cref{sec:results}, we applied the JS divergence and the PP-plot to 
%assess the number of 
select the informative realisations, and performed the cosmological inference only on the latter. 
However, we stress again that our choice of the JS threshold to classify the informative realisations is arbitrary and
%, as a consequence, 
that the number of informative realisations depends on this criterion.
%should be taken with a pinch of salt.
Among all the cosmological model approaches tested in this work, the redshift bins is the one showing the strongest realisation-dependence (see \cref{sec:results}): the fraction of realisations providing errors on the parameters $h(z_p)$ and $d_C(z_p)$ that are less than 100\% can be very marginal, depending on the redshift bin and on the MBHB formation scenario. This means that it is very  improbable that the Universe will correspond to a realisation for which the LISA+EM facilities will allow us to implement the redshift bins approach.
We extended %this 
the JS+PP-plot analysis also to the other cosmological models tested in this work, but we found no particular issues with them. Indeed, these tests are necessary within the redshift bisn approach because of its particular nature:
%of the redshift bins approach: 
most of the other models adopt a functional form for the $d_L(z)$ relation that is more complicated than a simple straight line, and covers a wider range in redshift. Therefore, even with a smaller number of EMcps, these models can provide reasonable constraints because we are adding an `a priori' additional information on the Universe model. The only exception is represented by the \hyperref[itm:splines]{\color{mycolormodel}spline interpolation} model. However, in this case, the absence of the binning allows one to exploit all EMcps for the inference, as opposed to the redshift bins case.

%Finally, since the bin approach is the most sensitive to the number of EMcps, we perform this analysis assuming only the scenario of 10 years of time mission. If LISA will provide data for only 4 yr, this analysis could not be performed. 

\subsection{\label{subsec:inference_splines}Spline interpolation}
\label{sec:splines_sub}

\textcolor{\colorref}{In the field of cosmology, model-independent techniques have become increasingly popular in the recent years \cite{2009PhRvD..80l1302S,2010PhRvD..82j3502H,2012JCAP...06..036S,2020ApJS..246...13W,2022JCAP...02..023D,2024ApJ...960...61M}, as they do not require to impose a physical model to interpret the data. In the present work, we 
%rely on 
adopt a model-independent approach based on a spline interpolation of the $d_L(z)$ relation}.

For the \hyperref[itm:splines]{\color{mycolormodel}spline interpolation} model, we approximate the luminosity distance with cubic polynomials. We fix the knots at $z=[0.2,0.7,2,4,6]$ with the additional information that the luminosity distance $d_L(z=0)=0$.
\textcolor{\colorref}{The choice on the number and position of the knots is determined by the fact that the model's accuracy on $d_L$ should be smaller than the uncertainties inferred on the parameters, i.e.~ the values of $d_L$ at the knots (reported in the results section). 
In particular, the two smallest values are crucial for capturing the initial rapid increase in luminosity distance, while the three final points are simply equally spaced. We also tested different configurations of the number of knots and their position but the one we adopted yielded the best accuracy for $z>2$. We detail our choice in Appendix~\ref{sec:accuracy_spline}.}
In order to avoid interpolation issues at high redshift due to the lack of EMcps, we remove all the systems at $z>7$. 

The result of the inference is a 5-dimensional posterior distribution of the $d_L$ values at the aforementioned knots. The $d_L$ posteriors can then be easily converted into $d_C$ posteriors at the knot redshifts and we can evaluate the slope of the spline to obtain information on $h(z)$ at any redshift.
Contrary to the \hyperref[itm:bins]{\color{mycolormodel}redshift bins} model, approximating the luminosity distance with splines allows us to use the entire population of EMcps. 
To construct the prior on the luminosity distance at a given knot, we take samples from the uniform priors in $(h, \Omega_m)$ and convert them in $d_L$ assuming $\Lambda$CDM, i.e. according to Eq.~\ref{eq:hubble_rate_h0_omegam}. The resulting $d_L$ distribution is assumed as the prior and we repeat this process at each of the knots. %In appendix \am{define the appendix}, we show that this choice do not affect our conclusion.
We note that we use $\Lambda$CDM only to fix the priors for the luminosity distance: otherwise, the spline approach is model-independent, and we have tested that the choice of priors does not impact our results.
%\nt{From these last sentences it is not clear whether the splines approach is a LCDM independent test or not...}

\section{\label{sec:results} Results}

In this section we present the results of the inference analysis applied to the cosmological models listed in Sec.~\ref{sec:cosmo_intro}. 
\textcolor{\colorref}{Since the results depend on the Universe realisation, the relative uncertainties on the model parameters are presented in plots showing the fraction of realisations for which they are obtained. 
When we quote the relative uncertainty on a given cosmological parameter, we take as reference the value which is provided by at least 50\% of the realisations, i.e.~the median of the error distribution (the value for which the error distribution cross 0.5 on the y axis of the plots). This value is
reported for the Q3d (Pop3) \{Q3nd\} model following this bracket convention. }
%\textcolor{\colorref}{and we always refer to the uncertainty of the median realisation , unless stated otherwise.}

%\nt{Added this to avoid repeating it everywhere. If you don't like it the old text is in the .tex file.}

\subsection{Results for the local Universe models}

% \begin{figure*}
%     \subfigure{\includegraphics[width=1\textwidth]{figures/H0_and_omega.pdf}} \\ 
%     \vspace{-0.6cm}
%     \subfigure{\includegraphics[width=1\textwidth]{figures/H0_and_omega_uncer.pdf}} 
%     \caption{\am{add median to the x-axis of upper panels}Upper panels: $H_0$ and $\Omega_m$ median distributions for different astrophysical populations, according to legend. The black vertical lines represent the fiducial values. Lower panels: cumulative distributions of the 90\% relative uncertainty for $H_0$ and $\Omega_m$, respectively. Overall we are able to constrain $H_0$ with a relative precision of few percents in $\sim 50\%$ of the cases but the \uncer distributions spread for approximately one order of magnitude. $\Omega_m$ is poorly constrained in all scenarios \am{ The upper panel is necessary to show that results are not biased}\am{should I change the label for $H_0$ to $h$ for consistency with the main text?} \am{add the grey area in the uncertainty at $>1$}}
%     \label{fig:H0_and_omegam}
% \end{figure*}

\begin{figure*}
    \includegraphics[width=1\textwidth]{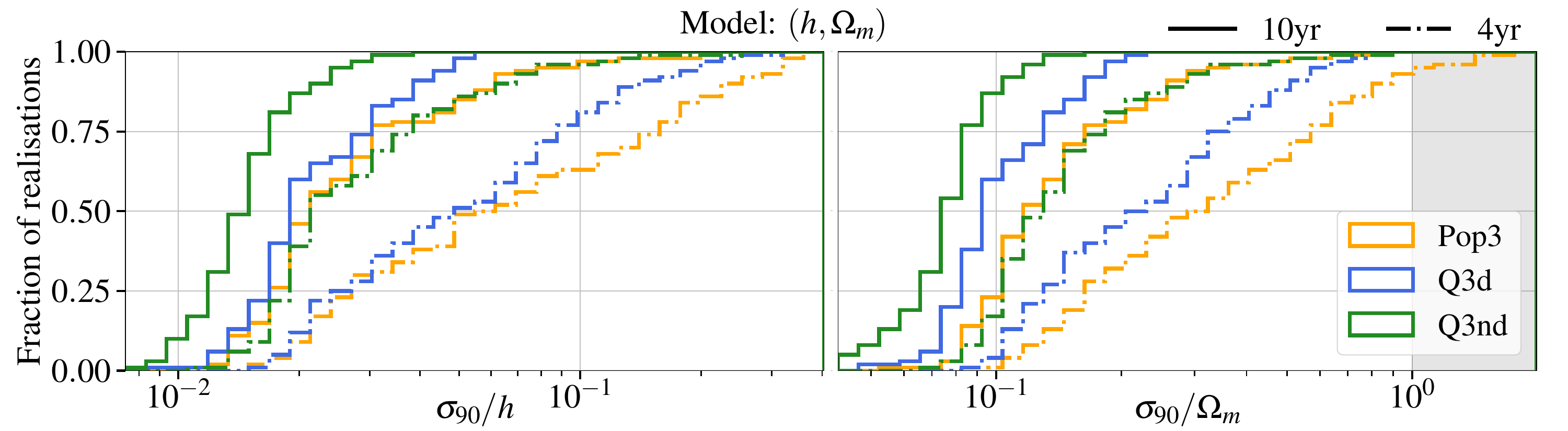}
    \caption{Cumulative distributions over realisations of the relative uncertainties for $h$ (left panel) and $\Omega_m$ (right panel) in the  \hyperref[itm:lcdm]{\color{mycolormodel}$(h, \Omega_m)$} model, namely $\Lambda$CDM. Solid (dashed) lines correspond to 10 (4) yrs of observations with LISA. Colors represent different astrophysical models as described in the legend and the grey area represents uncertainties larger than $100\%$. 
    We expect relative errors on $h$ of $\lesssim 5\%$ in 4 yrs and $\lesssim 2\%$ in 10 yrs for at least $50\%$ of the realisations. The relative errors on $\Omega_m$ descend below $\lesssim 10\%$ for at least $50\%$ of the realisations only in the 10 yrs LISA scenario.}
    \label{fig:H0_and_omegam}
\end{figure*}

\begin{figure}
    \includegraphics[width=0.5\textwidth]{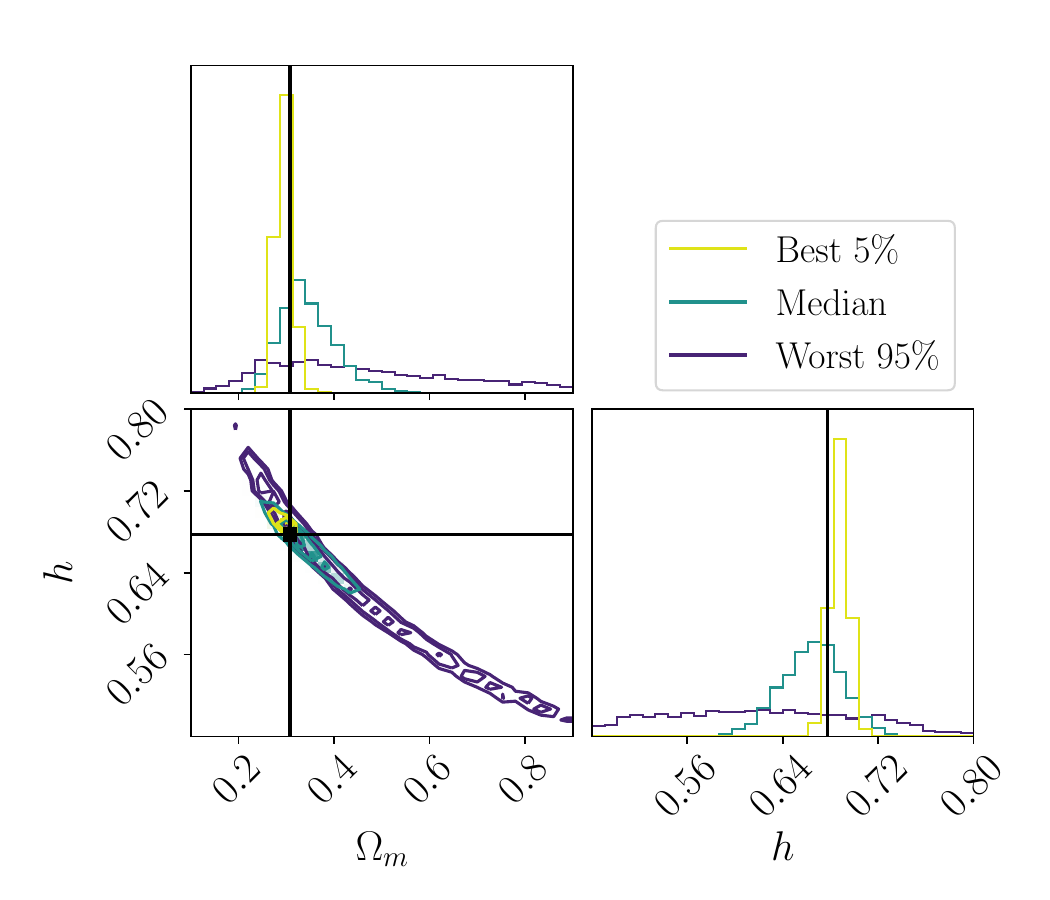} 
    \caption{Corner plot for $(h, \Omega_m)$ for the %\hyperref[itm:lcdm]{\color{mycolormodel}$(h, \Omega_m)$} 
    \lcdm model in 4 yrs of LISA mission, for the average, best and worst realisations of the Q3d MBHBs formation scenario, i.e.~those corresponding to the median, 5th and 95th percentile of the distribution pf the realisations
    %of Q3d 
    (more details in the text). Colors according to legend.}\label{fig:corner}
\end{figure}

% \begin{figure*}
%     \includegraphics[width=1\textwidth]{figures/dl_errors_90perc.pdf}
%     \caption{Luminosity distance uncertainty at 90\% confidence interval obtained from the $(h, \Omega_m)$ samples. Blue line corresponds to the median, while blue and green regions are the 68 and 95 percentile. To be compared with right panel in Fig.2 of \citealt{Bull21}.}
%     \label{fig:dl_errors}
% \end{figure*}

% \begin{figure}
%     \includegraphics[width=0.5\textwidth]{figures/hz_resampling.pdf} 
%     \caption{tbd}\label{fig:h_of_z_resampling}
% \end{figure}

\begin{figure*}
    \includegraphics[width=1\textwidth]{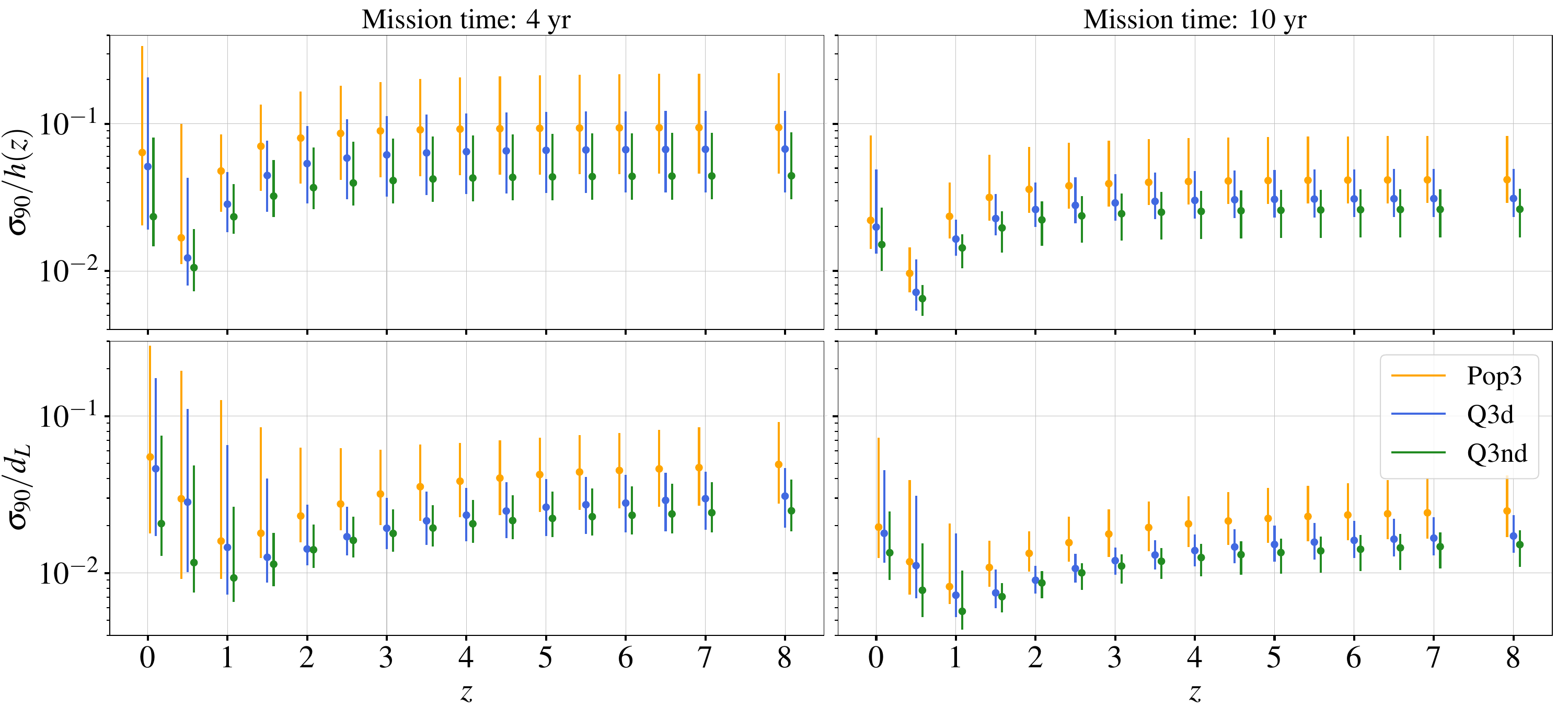}
    \caption{Relative errors at 90\% for $h(z)$ (upper panels) and $d_L(z)$ (lower panels) as a function of redshift, derived using~\cref{eq:hubble_rate_h0_omegam} and \cref{eq:dl} from the posterior samples of the  \hyperref[itm:lcdm]{\color{mycolormodel}$(h, \Omega_m)$} model, shown in \cref{fig:H0_and_omegam}.
    %, i.e. assuming $\Lambda$CDM. 
    Error bars %also 
    correspond to the $90\%$ uncertainty coming from the distribution over realisations. The three colors correspond to different astrophysical models, according to the legend. %Uncertainties are obtained from the $(h, \Omega_m)$ samples, assuming Eq.~\ref{eq:hubble_rate_h0_omegam} . 
     To avoid null values in the lower panels, the first point for $d_L$ is at $z=0.1$.}
    \label{fig:hz_and_dl_errors}
\end{figure*}

% \begin{figure*}
% \includegraphics[width=1\textwidth]{figures/dl_errors_68perc_bars_10yr_test.pdf} 
% \end{figure*}

\subsubsection{\lcdm $(h, \Omega_m)$}

We start with the result of the analysis on \lcdm i.e.~the two-parameter model \hyperref[itm:lcdm]{\color{mycolormodel}$(h, \Omega_m)$}.
In Fig.~\ref{fig:H0_and_omegam} we report the 
relative errors at 90\% confidence level for $h$ and $\Omega_m$
%as in the (aka $\Lambda$CDM,) 
for 4 yrs and 10 yrs of LISA mission duration. 
The median of the error distribution provides a relative uncertainty on $h$
%, we predict a median uncertainty  
of $5.1\%$ ($6.3\%$) \{$2.3\%$\}
%for Q3d (Pop3) \{Q3nd\}
in 4 years. These figures improve to 
$2.0\%$ ($2.2\%$) \{$1.5\%$\} in 10 years.

Overall Q3nd provides the best estimates, because it  has on average $\sim 3$ times more EMcps than Pop3. However, moving from 4 yrs to 10 yrs of observations, the uncertainty for the Pop3 model reduces by a factor $\sim 3$, while for Q3nd it reduces only by a factor $\sim 1.5$. This is due to the fact that the estimate on the cosmological parameters improves as $\sim N_\mathrm{EMcps}^{-1/2}$: therefore, increasing the mission time has a bigger effect on the constraints coming from models for which the initial
%providing larger improvements when the 
number of events is small.
%Our estimates are affected by large uncertainties: 
The variability of the relative error due to the realisation-dependence is large:
for the Q3d model and in 4 yrs, for example, the relative uncertainty varies from a few percent up to $\sim 20\%$. 
%As we move to 10 yr, 
Increasing the mission duration reduces the
%the uncertainties 
variability: 
%decreases because we have more 
since a larger number of EMcps is available overall to perform the inference, the latter is
%and we are 
less sensitive to the fluctuations due to single events in each realisation.

For $\Omega_m$, we 
%expect 
find relative \uncer of $23\%$ ($32\%$) \{$13\%$\}%for Q3d (Pop3) \{Q3nd\}
, in 4 yrs and of $10\%$ ($12\%$) \{$8\%$\} in 10 yrs. %We note that 
$h$ is typically determined better than $\Omega_m$, since the former corresponds to the first derivative of $d_L(z)$ at $z=0$, and $d_L(z=0)=0$ is fixed. 
This means that a single precise measurement at low $z$ can constrain $h$ to a tight value (similarly to the CMB). 
To constrain $\Omega_m$ one needs to probe the curvature (e.g.~the second derivative) of $d_L(z)$, meaning that one needs multiple precise measurements at low-redshift, i.e.~where the curvature of $d_L(z)$ is more pronounced, to get a good precision on $\Omega_m$. %Low redshift data do not give good measurements of $\Omega_m$ because the $d_L(z)$ function is too ``linear''.

%\am{Possible explanation (let me know if it sounds solid to you):} From Fig.~(1) in M22 it's clear that the majority of the EMcps are located at high redshift $z\gtrsim1.5$. However these high-redshift sources are characterised by large errors on $d_L$ and redshift. Even if the Universe can be considered as matter-dominated at high redshift, we obtain better estimates on $h$ thanks to the few well-constrained low-redshift sources.

In Fig.~\ref{fig:corner} we 
show the correlation between $(h, \Omega_m)$ for three representative realisations of Q3d for 4 yrs of LISA duration. 
I order to rank the realisations, we have decided to use
%ranked all the realisations according to 
the area covered by the posterior samples. 
This quantity keeps track of the correlation between the two parameters of the model, and it is more representative than the marginalised error in each of them. It is also simpler to use, as it is a single number.
%Following this procedure, 
We have therefore calculated
%e obtain 
the areas of the 2-dimensional parameter posteriors of all the 100 realisations in Q3d, and thereby constructed their distribution.
In Fig.~\ref{fig:corner}, we plot the average, best and worst realisations according to our classification, i.e.~the three realisations providing error
%with the 
area closest to the median, the 5th and the 95th percentiles of the areas distribution.
The parameters $(h, \Omega_m)$ are negatively correlated because small values of $h$ require large value for $\Omega_m$ to compensate in the $d_L-z$ relation.
We highlight this feature as other cosmological probes might have different orientations for the correlation. 
Among standard sirens, for example, EMRIs are expected to have more negative correlation between $h$ and $\Omega_m$ \cite{Laghi21}. 
Combining different categories of GW sources and sirens methods, one
would be able to %further 
break degeneracies between cosmological parameters \cite{Tamanini:2016uin,Laghi_paper, 2014A&A...568A..22B}.

Within the \lcdm model, it is also possible to use the posterior samples on $(h, \Omega_m)$ to get the forecast errors from MBHBs bright sirens on $h(z)$ or $d_L(z)$ at higher redshift.
Following Eq.~\ref{eq:hubble_rate_h0_omegam} and Eq.~\ref{eq:dl}, we can convert each pair of 
$(h, \Omega_m)$ in the posterior sample in a %sample for $h(z)$ and for the luminosity distance, at any redshift. 
couple $(h(z), d_L(z))$ at any redshift, thereby reconstructing a posterior for the two parameters at any chosen redshift.
%We note that the forecasts obtained in this case are based on the assumption of $\Lambda$CDM.
By doing this for each realisation, we also obtain the error distribution across realisations for $(h(z), d_L(z))$.
In Fig.~\ref{fig:hz_and_dl_errors} we report the relative errors at $90\%$ on $h(z)$ and $d_L(z)$ as a function of redshift, together with the 90\% error-bars coming from the error distribution over realisations.
The best constraints on $h(z)$ are achieved at $z\sim 0.5$: as demonstrated in Appendix~\ref{sec:pivot_parameters}, this is indeed the redshift at which the correlation between $h(z)$ and $\Omega_m$ is zero.
%because 
%At low redshift we expect better constraints on $h$ than on $\Omega_m$ and vice-versa at high redshift. \cc{COMMENT: ARE WE SURE ABOUT THIS? IT DOESN'T SEEM TO BE THE CASE?}\am{I don't remember why I put this sentence here but we can remove it. It's not connected with the rest of the text}
%Moreover, we also 
%We find that $z\sim 0.5$ is where the correlation between $h(z)$ and $\Omega_m$ is zero (more details in Appendix~\ref{sec:pivot_parameters}). 
%Starting from 4 yr of time mission, 
MBHBs bright sirens with LISA can therefore constrain the Hubble parameter $h(z)$
at $z\sim 0.5$ 
%we expect 
with a relative uncertainty 
%on $h(z)$ 
of $1.2\%$ ($1.7\%$) \{$1.0\%$\}, within the \lcdm model.
%for Q3d (Pop3) \{Q3nd\}.
Above $z\gtrsim 2$, the relative errors flatten around $7\%$ ($9\%$) \{$4\%$\}
%, respectively for each model, 
while at $z=0$ we recover the %previous 
aforementioned results. 
%If we will have 
With 10 years of data, the relative uncertainties on $h(z)$ at $z\sim 0.5$ improve to $0.7\%$ ($0.9\%$) \{$0.6\%$\} while at $z\gtrsim 2$ they  %uncertainties 
are between $2\%$ and $5\%$, depending on the MBHB population model.

For $d_L(z)$,  we find the same trends of $h(z)$ but the best constraints are obtained at $z\sim1$. % due to the effect of the integral.
%\am{is it sufficient? It's a bit difficult to prove by `` words'' but there is really no other difference between h(z) and $d_L(z)$}\nt{It's hard to pin down the exact reason for this. I would simply report the fact without mentioning any explanation.}
In particular, in 4 yr at $z\sim 1$ we predict $d_L(z)$ relative errors of $1.4\%$ ($1.6\%$) \{$0.9\%$\}. While in 10 yr, we expect the errors to be between $0.5\%$ and $0.8\%$ at $z\sim 1$ and at $\sim 1-2\%$  for  $z\gtrsim 4$.

%The results from MBHBs can be compared to other cosmological probes. 
It is interesting to compare the high-redshift constraints on $(h(z), d_L(z))$ obtained with MBHBs bright sirens at LISA with other cosmological probes. 
%In particular, we can compare 
For example, Ref.~\cite{Bull21} analyses the potential of
forthcoming galaxy surveys and 21 cm intensity mapping to constrain the universe expansion at high redshift. Their results for $(h(z), d_L(z))$ are summarised in their Fig.~2, to be compared with Fig.~\ref{fig:hz_and_dl_errors} of the present paper
%Fig.~\ref{fig:hz_and_dl_errors} with Fig.~2 of \cite{Bull21}: 
(note that we are plotting the $90\%$ error while they are reporting the $1\sigma$ uncertainty so the values from Fig.~\ref{fig:hz_and_dl_errors}  should be divided by a factor $\sim 1.6$ for a proper comparison). 
At low redshift, the 
%results for 
error on $h(z)$ forecast using
MBHBs standard sirens is
%are 
slightly worse than what 
%we might be able to do with other cosmological probes: 
can be achieved with baryon acoustic oscillations and/or 21cm intensity mapping: indeed,
in 4 yrs, at $z\sim1$ the median relative error for Q3d is $\sim4$ times larger than the forecast for HIRAX, and one order of magnitude larger than the one from DESI. 
However, from the results presented in Fig.~2 of \cite{Bull21} it also appears that MBHBs provide comparable results with a high-redshift version of HIRAX at $z>5$.
Concerning the luminosity distance, at $z\sim1$, the relative error from Q3d model is $<2$ times larger than the forecasts from DESI, but at the same level of HIRAX. 
As we move to higher redshift, our forecasts tend to flatten, thanks to the fact that MBHBs bright sirens extend to high redshift. On the other hand,
%the high-redshift sources, while 
the predictions from EM probes degrade quickly: for example, MBHBs bright sirens provide better constraints than a high-redshift version of HIRAX at $z\sim4$ and than a stage 2 intensity mapping experiment at $z\sim 6$. 
%\am{this is a bit strange: why do we have worse estimate for h(z) at almost all redshift but for $d_L$ we have better estimates? I think it's due to the correlation between $(h_0, \Omega_m)$ but it's difficult to test without samples from other probes (i.e. I would need the samples for  $(h_0, \Omega_m)$ to redo the same analysis with the conversion of the samples)}

%\am{You can look at Fig.~\ref{fig:pp_plot_H0_and_omega} and Fig.~ \ref{fig:error_bars_H0} in  Appendix~\ref{sec:random_plot} for more plots on $(h, \Omega_m)$. Not sure I want to put them in the main text.}

\begin{figure*}
    \includegraphics[width=1\textwidth]{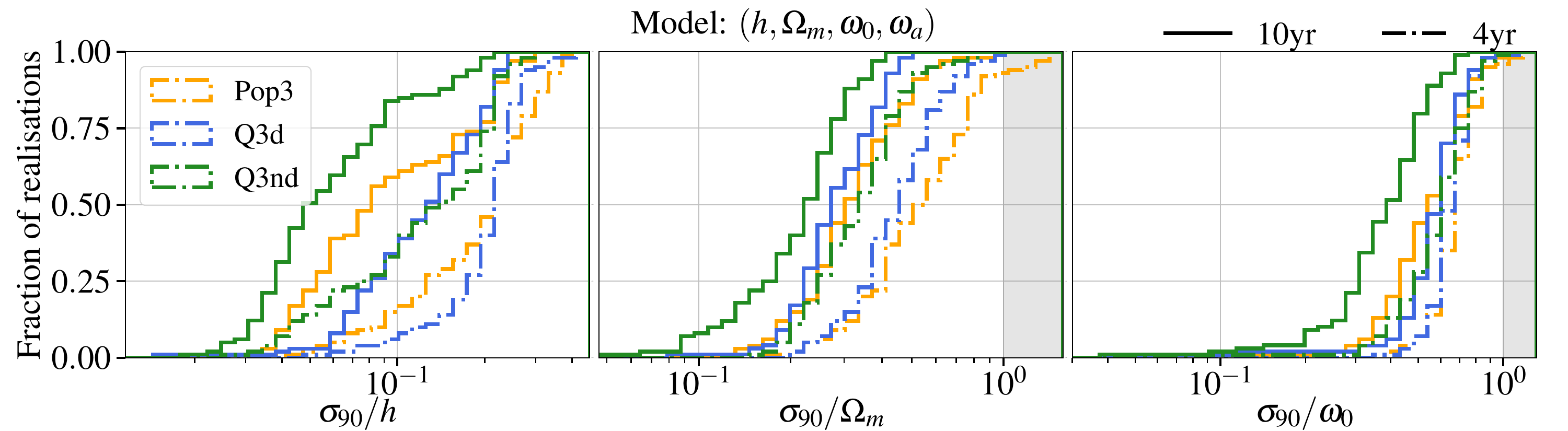}
    \caption{Same as Fig.~\ref{fig:H0_and_omegam} but with $h, \Omega_m$ and $\omega_0$ from the \hyperref[itm:cpl]{\color{mycolormodel}$(h, \Omega_m, \omega_0, \omega_a)$} model. $\omega_a$ is unconstrained and not reported. The addition of two parameters worsen the estimates on $(h, \Omega_m)$.}
    \label{fig:H0_and_omegam_and_cpl}
\end{figure*}

\begin{figure*}
    \includegraphics[width=1\textwidth]{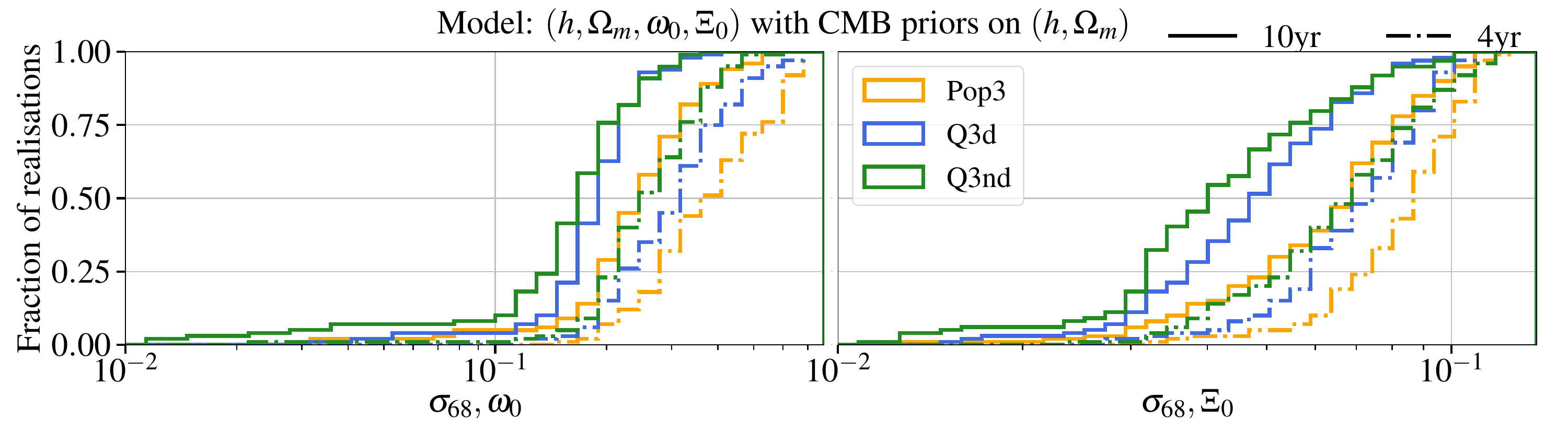}
    \caption{Same as Fig.~\ref{fig:H0_and_omegam} but for $\omega_0$ and $\Xi_0$ for the  \hyperref[itm:belgacem19]{\color{mycolormodel}$(h, \Omega_m, \omega_0, \Xi_0)$} model. Note that we report the absolute $1\sigma$ uncertainties for comparison with Tab.2 in \cite{Belgacem19}. The uncertainties on $h$ and $\Omega_m$ %are not reported because they 
    coincide with the CMB priors. We can constrain $\Xi_0$ to $<10\%$ using only MBHBs bright sirens.}
\label{fig:Xi0_omega0_uncer_belgacem19}
\end{figure*}

\subsubsection{Dark energy $(h, \Omega_m, \omega_0, \omega_a)$}

Moving to the dark energy parametrisation \hyperref[itm:cpl]{\color{mycolormodel}$(h, \Omega_m, \omega_0, \omega_a)$}, we report in Fig.~\ref{fig:H0_and_omegam_and_cpl} the  distribution of the relative uncertainties on $h$, $\Omega_m$ and $\omega_0$ over the fraction of realisations. 
As expected, the presence of two additional parameters worsen the estimates on $h$ and $\Omega_m$ with respect to the \lcdm case. In 10 yrs, $h$ is constrained to $\sim 10\%$ accuracy and $\Omega_m$ to $\sim 20-30\%$ for at least 50\% of the realisations. We also find that $\omega_0$ is poorly constrained, with uncertainties $>30\%$ in all cases, while $\omega_a$ is unconstrained. As showed by previous work \cite{Tamanini16, Belgacem19}, 
%we conclude saying that 
in this scenario LISA could hardly provide %information. 
tany constraint, due to the lack of MBHBs bright sirens at low redshift, compared to other cosmological probes such as SNe.

\subsubsection{Modified gravity $(h, \Omega_m, \omega_0, \Xi_0)$}

In Fig.~\ref{fig:Xi0_omega0_uncer_belgacem19} we report the absolute uncertainties on $\Xi_0$ and $\omega_0$ for the \hyperref[itm:belgacem19]{\color{mycolormodel}$(h, \Omega_m, \omega_0, \Xi_0)$} model. 
For this scenario we adopt the same priors of \cite{Belgacem19} on $h$ and $\Omega_m$ derived from the CMB+BAO+SNe in order to assess LISA ability to constrain $ \Xi_0$, and we report the $1\sigma$ absolute uncertainty to compare with their results. 
%In comparison to \cite{Belgacem19}, w
We obtain uncertainties on $\Xi_0$ approximately 2-3 times larger for at least 50\% of the realisations. This is due to the fact that, in \cite{Belgacem19}, the authors also included the 
%information from 
CMB, BAO, and SNe data in their analysis, leading to a better estimate of $\omega_0$ and, consequently, of $\Xi_0$ (this can be appreciated from their Fig.~17 and Fig.~18). 
In 4 yrs, the median relative errors on $\Xi_0$ are $7.6\%$ ($8.9\%$) \{$7.0\%$\}.
% for Q3d (Pop3) \{Q3nd\}. 
Assuming 10 years of observation with LISA, the estimates improve to $4.9\%$ ($7.1\%$) \{$4.1\%$\}, respectively.
Due to the choice of priors on $h$ and $\Omega_m$, the uncertainties on these parameters are comparable with the prior, i.e.~\textcolor{\colorref}{the simulated LISA data do not constrain these parameters more strongly than CMB data, included via the prior.}

We can also compare our results with the forecasts for EMRIs \cite{Liu:2023onj}, although, in this case, the comparison is not straightforward due to the different analysis setups. In its fiducial model and assuming only $\Xi_0$ as free parameter, Ref.~\cite{Liu:2023onj} reports an error of $8.5\%$ at 90\% C.I. This value is slightly better than our results in 4 yrs. However, when more parameters are left free to vary, the reported errors on $\Xi_0$ in \cite{Liu:2023onj} are larger than %ours. 
those given by MBHBs bright sirens.
Taking into account the uncertainties from the different priors adopted, we expect MBHBs and EMRIs to provide similar constraints on $\Xi_0$.

\subsection{High-redshift Universe results}

\begin{figure*}
    \includegraphics[width=1\textwidth]{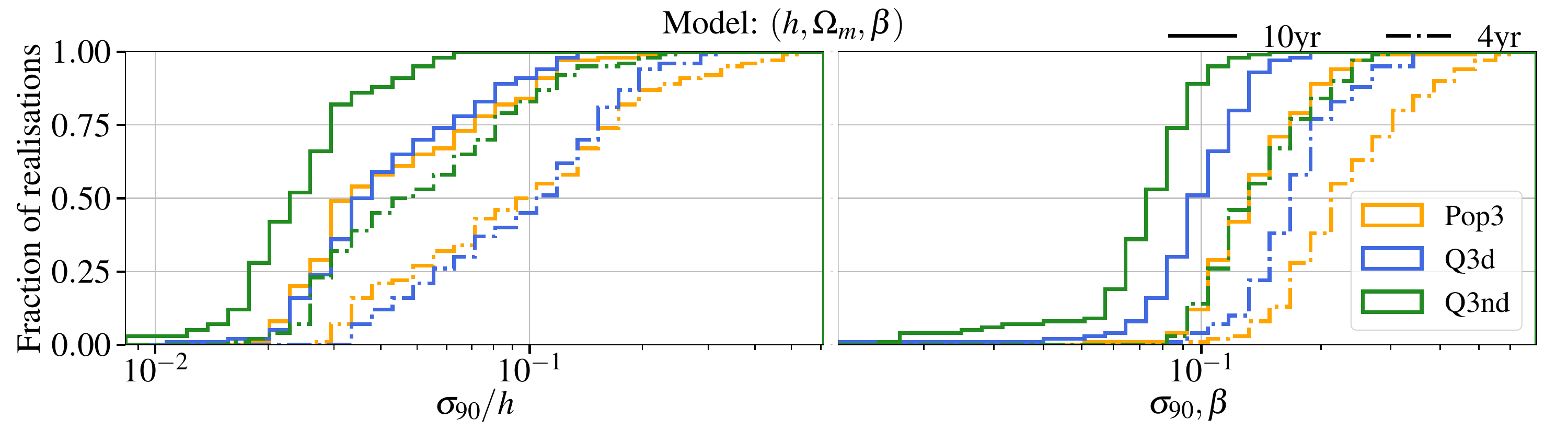}
    \caption{Same as Fig.~\ref{fig:H0_and_omegam} but  for $h$ and $\beta$ for the  \hyperref[itm:lcdm_beta]{\color{mycolormodel}$(h, \Omega_m, \beta)$} model. The parameter $\beta$ models a possible deviation from zero in the matter equation of state, i.e. $\omega_m = \beta$. We forecast a relative accuracy on $\beta$ of $<20\%$ with 4 yrs of LISA observations and $\lesssim 10\%$ with 10 yrs for at least 50\% of the realisations. }
    \label{fig:H0_and_omegam_and_beta}
\end{figure*}

\begin{figure*}
    \includegraphics[width=1\textwidth]{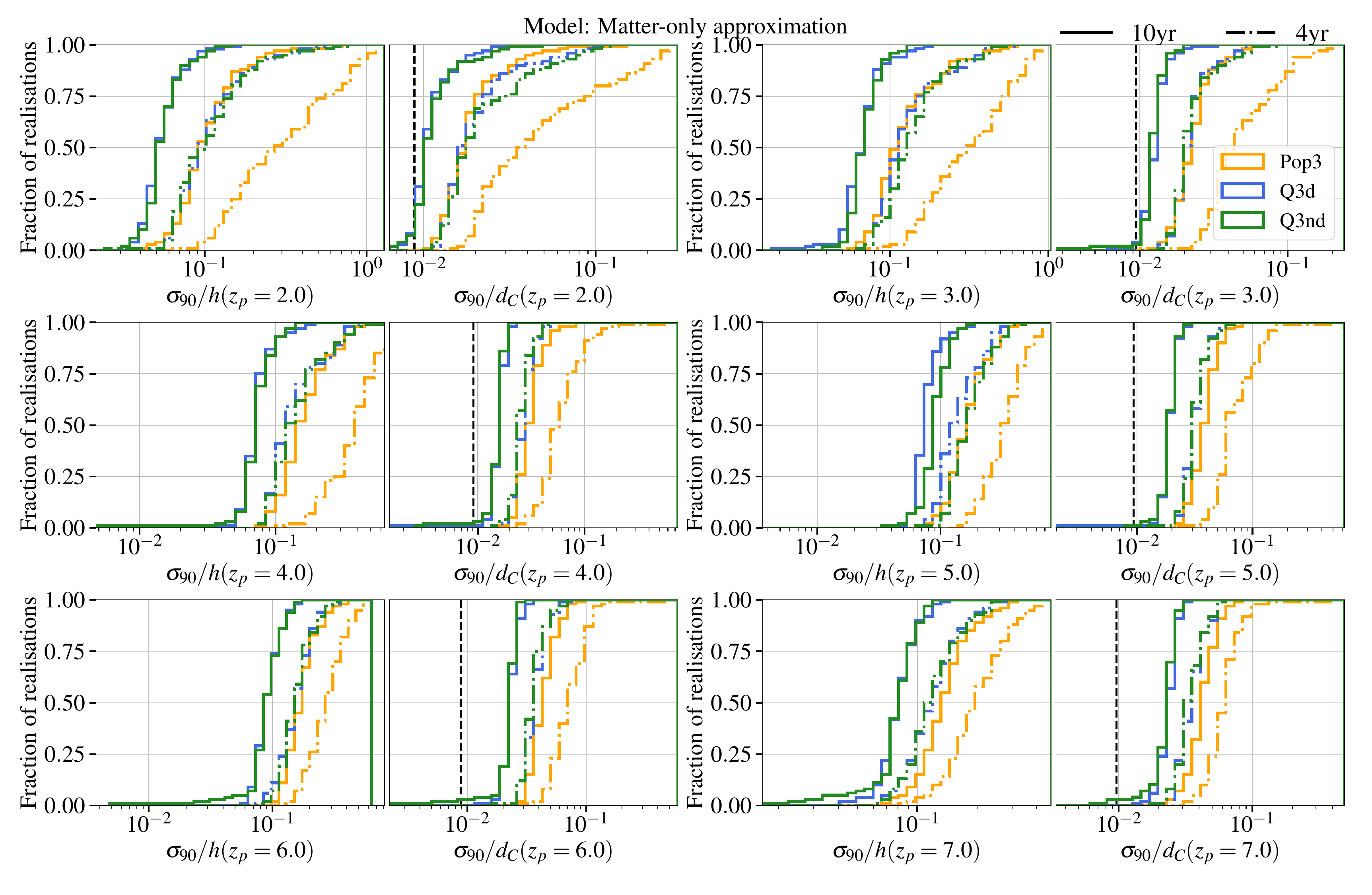}
    \caption{Same as Fig.~\ref{fig:H0_and_omegam} but for $h(z_p)$ and $d_C(z_p)$ from the   \hyperref[itm:matteronly]{\color{mycolormodel}matter-only} approximation at different pivot redshift, %according to 
    as specified in the x-axis labels. \textcolor{\colorref}{Colors represent different astrophysical models as reported in the legend. The vertical dashed black line represents the model's accuracy.
    In 10yrs of observation, we expect constraints on $h(z_p=2)$ at $3-5\%$. At higher redshift, estimates get slightly worse but we still have relative errors on $h(z_p)$ of $10\%$ at $z_p=7$.}}
    \label{fig:matter_only_uncer_new}
\end{figure*}

\subsubsection{Deviation to the matter equation of state}

In Fig.~\ref{fig:H0_and_omegam_and_beta} we report the results of the analysis for  the \hyperref[itm:lcdm_beta]{\color{mycolormodel}$(h, \Omega_m, \beta)$} model, i.e.~assuming $\omega_m = \beta$ as the matter equation of state. 
As expected, the addition of $\beta$ worsen the constraining power on $h$ with respect to \lcdm.
In 4 yrs of LISA mission, $h$ is constrained within this model at $11\%$ ($10\%$) \{$4.9\%$\}%for Q3d (Pop3) and \{Q3nd\} 
for at least 50\% of the realisations,
while in 10 yrs the estimates improve to $3.8\%$ ($3.4\%$) \{$2.5\%$\}.
%, respectively. 
%We note that the Q3d model loses part of its constraining power on $h$, leading to the same estimates from the Pop3 model.

Concerning the matter part, $\Omega_m$ and $\beta$ are degenerate: if $\beta$ decreases, $\Omega_m$ increases to compensate. 
We find that in 4 yrs $\Omega_m$ is unconstrained and, for this reason, we do not plot it.
In 10 yrs, $\Omega_m$ can be constrained with large uncertainty of $\sim 30-40 \%$.
For $\beta$, we expect constraints of $18\%$ ($23\%$) \{$14\%$\} and $10\%$ ($14\%$) \{$8.0\%$\} in 4 yrs and 10 yrs, respectively, for at least 50\% of the realisations.

The fact that $\beta$ is constrained while $\Omega_m$ is not can be understood 
%looking at how they appear in 
from the role they play in Eq.~\ref{eq:hubble_rate_h0_omegam_and_beta}. While $\Omega_m$ acts a multiplicative factor, $\beta$ is an exponential. Therefore a small variation in $\beta$ can lead to a large difference in the expected luminosity distance. Or, in other words, if $\beta$ varies, $\Omega_m$ has to vary even more in order to compensate and reproduce the expected $d_L-z$ relation.

\subsubsection{Matter-only approximation}

The \hyperref[itm:matteronly]{\color{mycolormodel} matter-only} approximation 
entails fitting for $h(z_p)$ and $d_C(z_p)$ at a high enough pivot redshift $z_p$ (see \cref{sec:matter_sub}).
In Fig.~\ref{fig:matter_only_uncer_new} we report the relative uncertainties at $90\%$ on $h(z_p)$ and $d_C(z_p)$ \textcolor{\colorref}{at different pivot redshift.  Starting from $h(z_p)$, 
in 4 yrs we expect median errors of $10.0\%$ ($28.9\%$) \{$10.2\%$\} at $z_p=2$ and $11.9\%$ ($33.0\%$) \{$13.5\%$\} at $z_p=3$. Moving to higher pivot redshift, the estimated errors remains similar: even at $z_p=7$, we forecast median relative errors of $12.5\%$ ($19.8\%$) \{$11.9\%$\}. We stress that even a relative uncertainty on $h(z_p)$ at $z_p\sim 2-7$ is a remarkable results that could be achieved with only a handful of EMcps available, i.e. even the Pop3 model could give interesting results in only 4 yr of observations.
For 10 yr, we estimate
median errors of $5.4\%$ ($9.9\%$) \{$5.4\%$\} at $z_p=2$ and $7.1\%$ ($11.4\%$) \{$7.0\%$\} at $z_p=3$. At higher redshift, we forecast median relative errors of $8.6\%$ ($13.7\%$) \{$8.3\%$\}  at $z_p=7$. 
Two effects come into play: 
%This is due to the fact that, 
on one side, for larger $z_p$, the value of the lower redshift of the bins $z_{\rm min}$ increases (as reported in Tab.~\ref{tab:prior_table_high_Universe}), leading to fewer EMcps. 
On the other side, the matter-only approximation gets better and better: at $z_p>4$, %we are fully in the matter-only approximation 
it is very accurate, so that $z_\mathrm{min}$ increases slowly (see Tab.~\ref{tab:prior_table_high_Universe})
and correspondingly, the number of EMcps decreases slowly.}

%for $z_p=2$ and $z_p=3$. Starting from $h(z_p)$, we find overall better results at $z_p=2$ than $z_p=3$. For example, in 10 yr we predict median errors of $3.8\%$ ($6.0\%$) \{$3.6\%$\} at $z_p=2$ and $6.8\%$ ($11\%$) \{$6.8\%$\} at $z_p=3$.
%for Q3d (Pop3) \{Q3nd\} respectively. 
%We recall that for the case at $z_p=2$ ($z_p=3$), we removed all the systems at $z<1$ ($z<1.5$). As a consequence, the case $z_p=2$ contains overall more standard sirens than the one at $z_p=3$, leading to better estimates.

\textcolor{\colorref}{
%Moving to 
For the comoving distance $d_C(z_p)$, still in 10 yrs, we forecast relative uncertainties at $1.0\%$ ($1.8\%$) \{$1.1\%$\} at $z_p=2$, and 
$1.3\%$ ($2.3\%$) \{$1.3\%$\} at $z_p=3$. 
%Similarly to the case 
As for $h(z_p)$, the relative uncertainties remain similar at high redshift: the relative error is  $2.5\%$ ($4.5\%$) \{$2.4\%$\} at $z_p=7$ in 10 yrs.}

\textcolor{\colorref}{As explained in \cref{sec:matter_sub}, we need to ensure that the matter-only approximation in the each redshift bin around $z_p$ is more accurate than the forecast relative errors in the model's parameters.
The matter-only model's accuracy is represented by a dashed vertical black line for each $z_p$ in the panels for $d_C$.
It can be appreciated that most of the realisations predict uncertainties above the corresponding accuracy in each redshift bin. However, for some $z_p$, there are few realisations whose errors is below the model's accuracy: in this case, the model's accuracy itself is the smallest constraint we can put on $h(z_p)$ or $d_C(z_p)$. }

It is interesting to compare these results with the uncertainties reported in Fig.~\ref{fig:hz_and_dl_errors} \textcolor{\colorref}{(in both cases, we report relative distances at fixed redshift so the comparison can be done looking directly at the plot)}. As it can be appreciated, the two approaches predict similar uncertainties, even at high redshift since the number of EMcps is similar.
%\sout{We think that the explanation is different for these two cases. In Fig.~\ref{fig:hz_and_dl_errors}, we propagate the uncertainties on $(h, \Omega_m)$ to the other redshifts using the \lcdm relation. In \sout{this} \am{that}  case, we expect that only EMcps at low redshift contribute significantly to the determination of $(h, \Omega_m)$. Instead, in the matter-only approximation, the EMcps that dominate the inference are the one close to the pivot redshift. For example, even if we had kept the EMcps at $z<1.9$ for $z_p=7$, they would not have  significantly contributed to its measurements. }
%} \cc{COMMENT: I'm not sure I understand...}\am{re-reading it, I don't know what I intended to say here. I'll remove it and replace the motivation}
%The estimates on $h(z_p=2)$ in the matter-only approximation are marginally worse while at $z=3$ the difference increases to a factor of $\sim 2-3$. This is expected because in the matter-only approximation we removed the low-redshift sources so there are effectively less EMcps.
%The results for the distance uncertainties are more similar between the two approaches: we think that the explanation resides in the fact that the comoving distance appears in Eq.~\ref{eq:matter-only} as a simple normalizing factor for the whole expression, leading to better estimates even with less EMcps. %\am{I didn't want to make the text too heavy so I chose not to report the results for the distance. Let me know what you think.} 

\subsubsection{Redshift bins approach}

\begin{figure*}
    \includegraphics[width=1\textwidth]{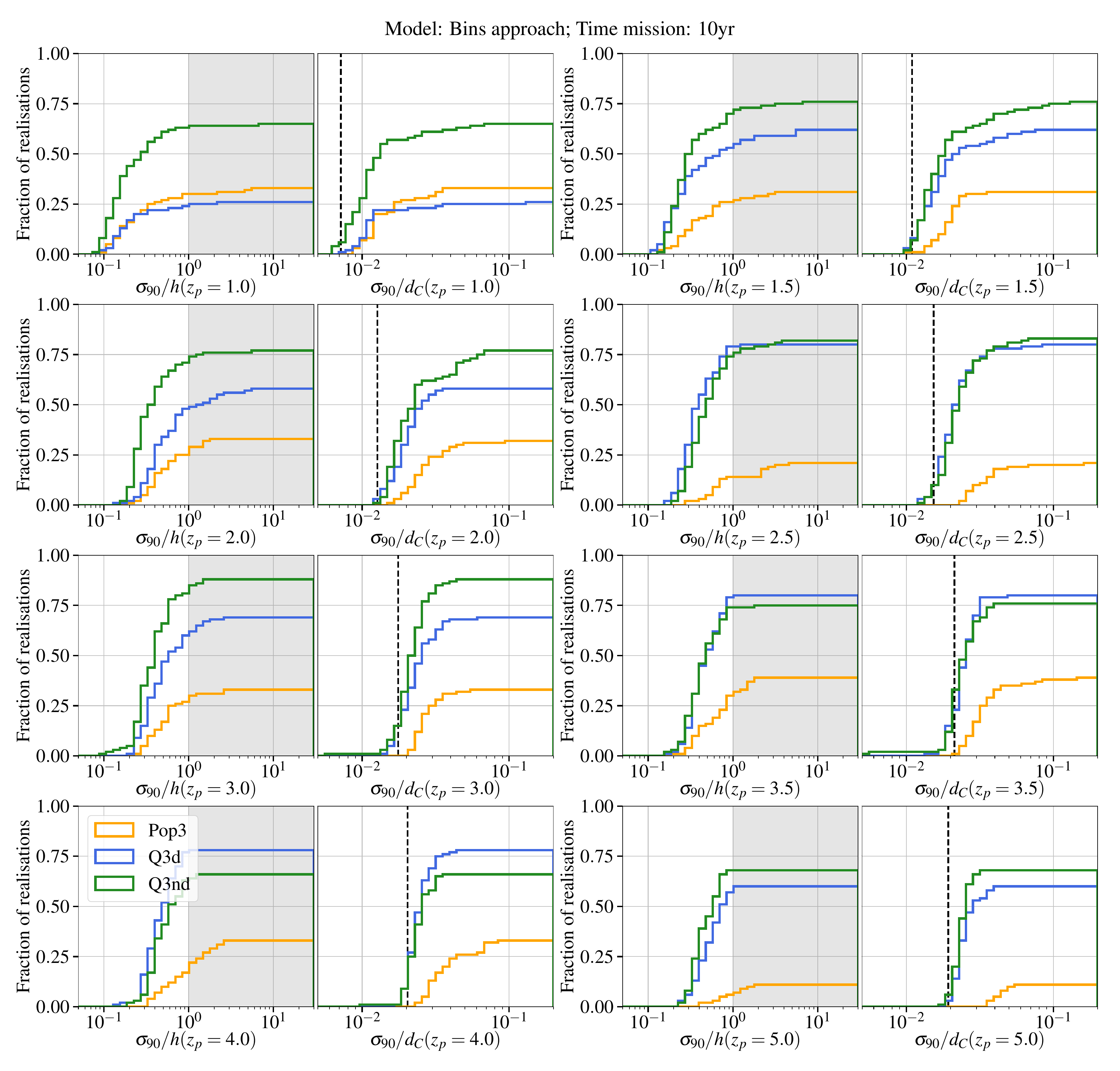}
    \caption{Same as Fig.~\ref{fig:H0_and_omegam} but for $h(z_p)$ and $d_C(z_p)$ from the   \hyperref[itm:bins]{\color{mycolormodel}redshift bins} approach at different pivot redshift, as specified in the x-axis labels. Colors represent different astrophysical models as reported in the legend. \textcolor{\colorref}{The vertical dashed black line represents the model's accuracy.} The cumulative distributions reach the fraction of \emph{informative} realisations (see Sec.~\ref{subsec:bins}): for example at $z_p=3.5$, only $\sim75\%$ of the Q3d and Q3nd realisations provide useful constraints.}
    \label{fig:uncer_bin_approach}
\end{figure*}

In Fig.~\ref{fig:uncer_bin_approach} we report the uncertainties on $h(z_p$) and $d_C(z_p)$ from the \hyperref[itm:bins]{\color{mycolormodel}redshift bins} approach, assuming 10 yrs of observation only, since for 4 yrs the number of EMcps is always too low for the method to provide any meaningful constraint. 
As discussed in Sec.~\ref{subsec:bins}, even for 10yrs of LISA mission duration, there are realisations for which the method is not applicable due to the low number of EMcps. We therefore present the results of the analysis only for the \emph{informative} realisations. As a consequence,
%Following the discussion in , 
the cumulative distributions of the relative uncertainties in \cref{fig:uncer_bin_approach} do not asymptote to 
%the value of 
1, i.e.~providing the relative error within 100\% of the distributions, but to the fraction of informative realisation in that particular redshift bin. 
For example at $z_p=1$, we have \textcolor{\colorref}{$26$ ($33$) \{$65$\} } informative realisations out of 100 in total. \textcolor{\colorref}{In other words, we expect that once LISA will fly, there is only $26\%$ ($33\%$) \{$65\%$\} chance (depending on the MBHBs formation scenario) that it will be possible to perform the
the redshift bins approach %can be performed  
at $z_p=1$.
%only in  $26\%$ ($33\%$) \{$65\%$\} of the cases. 
} 
The number of informative realisation is small at low redshift ($z_p=1$), it increases, reaching its maximum at $z_p\sim3$ (this value changes slightly for the three MBHBs formation astrophysical models) and then it decreases again at higher redshift. 
The largest fraction of informative realisations at $z_p\sim2.5-3.5$ reflects the redshift distribution of merging MBHBs (c.f. solid green line in Fig.1 of M22). At $z<1.5$, we do not expect many EMcps, so the number of uninformative realisations increases. The same argument can be applied at high redshift $z>4$, where we also expect larger errors on the luminosity distance and redshift of the source.

\textcolor{\colorref}{From Fig.~\ref{fig:uncer_bin_approach} it appears clearly that
we 
%do not expect to 
won't be able to test this approach at any redshift if the Pop3 model is the correct one because at best only $39\%$ of the realisations are informative (at $z_p=3.5$). 
The results improve if we consider the Q3 models. In these scenarios, 
%cases, 
it will still be challenging to apply 
this technique at $z\lesssim2$ and $z\gtrsim 5$ because the number of informative realisations is below 75\%; however, we might be able to obtain some constraints for $2.5 \lesssim z \lesssim$ 4. For example, at $z_p=3.5$, 50\% of the realisations predict a relative error on $h(z_p)$ smaller than 50\%. We note that at low redshift %we have 
the forecast uncertainties are better than at high redshift: for instance, the best 25\% realisations of Q3nd have smaller uncertainty at $z_p=1$ than at any other redshift. However also the number of EMcps is small and so is the number of informative realizations. 
At higher redshift $(z_p\gtrsim 5)$ the lack of EMcps and the larger errors on luminosity distance and redshift lead naturally to a decrease of the informative realisations and to worse constraints.}

%A part from $z_p=1$ where the number of informative realisation is below $50\%$ even for Q3d, we have the best constraints on $h(z_p)$ at $z_p=3$ with $75\%$ of the realisation predicting a relative error on $h(z_p)$ smaller than $30\%$ for the massive models, while this fraction decreases to only $25\%$ for Pop3. 

Comparing these results with those reported in Fig.~\ref{fig:hz_and_dl_errors}, it is clear that the redshift bins approach is much less performing. Even if the prospect of a model-independent test is appealing, the small number of EMcps in each bin makes it feasible only between \textcolor{\colorref}{$2.5<z<4$} if the Q3 models are the correct ones \textcolor{\colorref}{ \emph{and} if we will be lucky with the real realisation of the Universe}. %\am{I don't think there is more to report about bins}

\subsubsection{Spline interpolation}

\begin{figure*}
    \includegraphics[width=1\textwidth]{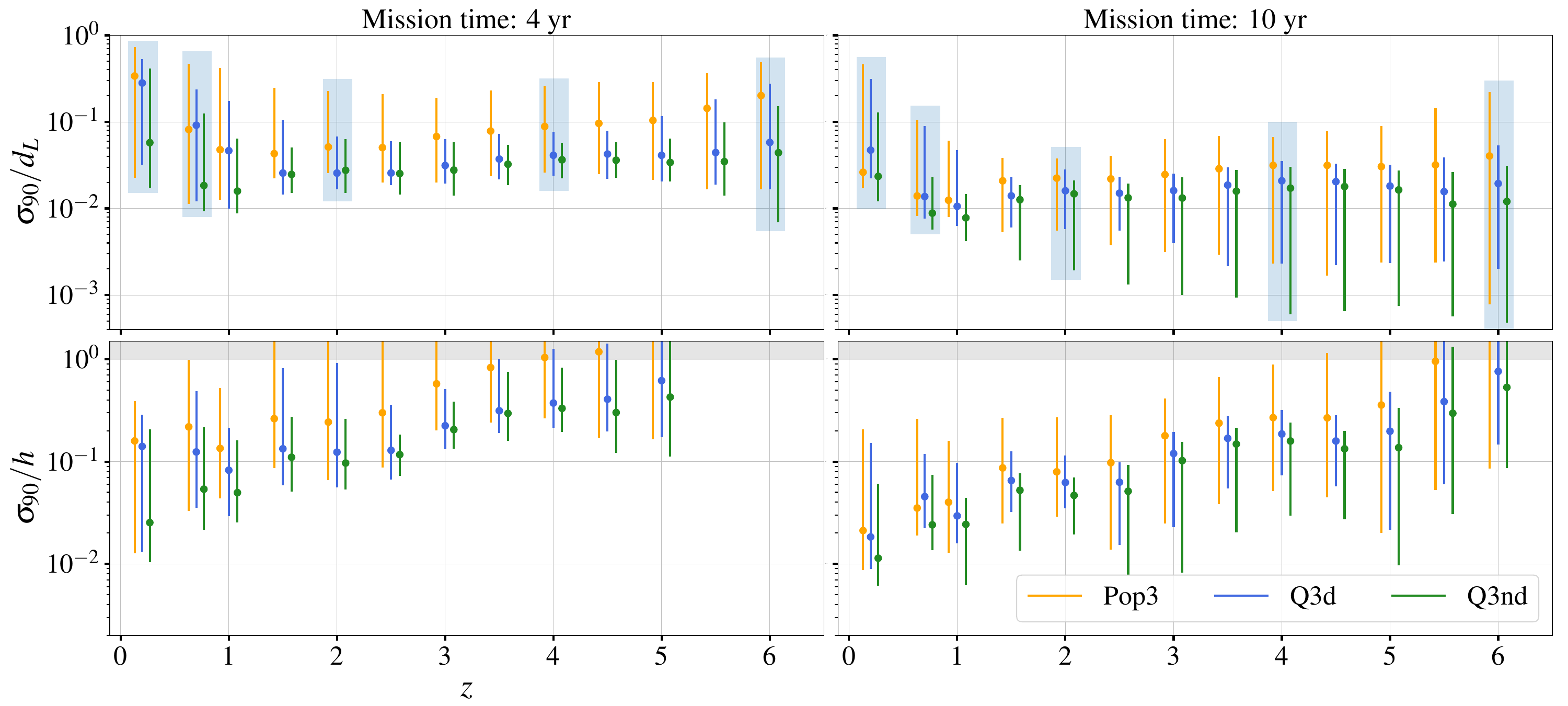}
    \caption{Relative errors at 90\% for $d_L(z)$  (upper panels) and $h(z)$ (lower panels) as a function of redshift from the \hyperref[itm:splines]{\color{mycolormodel}spline interpolation} model. Error bars correspond to the $90\%$ uncertainty on the relative errors from the distribution of realisations. Light blue boxes highlight the uncertainties at the knot redshifts.
    Splines can constrain the luminosity distance to $<10\%$ in 4 yrs at $1<z<4$ and up to $z\sim6$ in 10 yrs of observations. }
\label{fig:hz_and_dl_errors_splines}
\end{figure*}

Motivated by the model-independent technique adopted with the redshift bins, we searched for another model-independent method that could allow us to use the entire set of EMcps, thereby improving its constraining capability. 
%The results of this search is 
The best option turns out to be
the \hyperref[itm:splines]{\color{mycolormodel}spline interpolation} model (see \cref{sec:splines_sub}), whose results are reported in Fig.~\ref{fig:hz_and_dl_errors_splines}. 
%Starting from 
The outcome of the parameter analysis within the spline interpolation model are the posterior distributions on $d_L$ at the redshift knots.
We forecast relative errors of $<10\%$ between $0.7<z<4$ in 4 yrs of observations, taking as reference the result provided by 50\% of the realisations. The errors bars in Fig.~\ref{fig:hz_and_dl_errors_splines} denote the 90\% uncertainty on the relative error, due to the distribution over realisations.
For 10 yrs of observation, it is possible to reach $1-2\%$ precision in between $0.7<z<3$, while at $z>4$ the uncertainty is on few percent, with wider error bars. 
We compare our results with those reported in Fig.2 of \cite{Bull21} (note that they report the uncertainties at  $1\sigma$ so our values must be divided by a factor $\sim 1.6$ for proper comparison), and \textcolor{\colorref}{this is the only correction to apply since we both report relative uncertainties at fixed redshift)}.  
Spline can constrain the luminosity distance to $<10\%$ in 4 yr at $1<z<4$ and up to $z\sim6$ in 10 yr of observations. One can appreciate that LISA EMcps with the spline method can provide estimates that are competitive with future EM observations. The fact that we obtain better estimate at $z\sim6$ than at $z\sim2$ is due to the nature of the spline that do not possess a rigid model by construction. For example, for the Q3nd model in 10 yr, the relative error on $d_L$ at $z=5.5$ is $\sim 1\%$ while at $z=2$ is $\sim 2\%$. However we also notice that the error bars increase moving from low to high redshifts.

%Moving to 
The relative uncertainty on $h(z)$ is obtained from the one on $d_L$ using \cref{eq:H_of_z}.
The spline model can provide constraints $<10\%$ up to $z\lesssim 3$ in 10 yrs of observations. 
Note that, the within the spline model, the best estimate on $h(z)$ is not at $z\sim 0.5$ as for \lcdm (see Fig.~\ref{fig:hz_and_dl_errors}): indeed, this was due to the  correlation
%With the spline we break the correlation 
between $h$ and $\Omega_m$, which is absent when using splines.
%because the cosmological model is not anymore fixed and, as a consequence, the best estimates are not anymore at $z\sim 0.5$ as in Fig.~\ref{fig:hz_and_dl_errors}.%\am{anything else I forgot to mention?}

\section{Other tested models}

During the realisation of this manuscript, we also tested additional cosmological models, which we decided not to include in the final analysis. 
Most of them predict a $d_L-z$ relation \textcolor{\colorref}{that cannot be reconciled with $\Lambda$CDM for an optimal choice of the parameters}. Since our data are constructed according to the $\Lambda$CDM model, we can %make predictions 
obtain meaningful constraints
only assuming this cosmology or any extension of it: \textcolor{\colorref}{for example the \hyperref[itm:belgacem19]{\color{mycolormodel}$(h, \Omega_m, \omega_0, \Xi_0)$} recovers $\Lambda$CDM for $\omega_0=-1$ and $\Xi_0=1$. The only exception among the models presented in this work is the \hyperref[itm:matteronly]{\color{mycolormodel}matter-only} approximation that recover $\Lambda$CDM only to a certain precision (as previously discussed in the text). } 
Any alternative cosmology that cannot be reduced to \lcdm leads to systematic biases in the recovered parameters.
In particular, we tested:
\begin{enumerate}
    \item A phenomenological expression where the luminosity distance is approximated by a third-order polynomial expansion  \cite{Risaliti19}. The expected $d_L-z$ relation is close to $\Lambda$CDM at small redshift but deviates from it at $z>2$ \cite{2021PhLB..81836366B,Speri21}, resulting in systematic biases in the recovered $h$.
    For this case, we also attempted to expand the luminosity distance in redshift (scale factor) around a general pivot redshift (scale factor) but with no success. 
    \item A phenomenological tracker model \cite{Bull21} where the dark energy equation of state undergoes a smooth transition, parameterised by four additional parameters. This model aroused our interest because the transition might happen in the matter-dominated era, at 
    $1<z<7$. However, the large number of parameters (6 = 4 + $h$ + $\Omega_m$) and the limited number of EMcps makes it challenging to test this model with MBHBs.
    \item A dark energy model, in which the cosmological constant switches sign at a certain $z_\dagger$ due to a transition from an anti-de Sitter to a de Sitter Universe \cite{2023arXiv230710899A}. Assuming $\Lambda$CDM, $z_\dagger \rightarrow +\infty$, so we were only able to establish lower limits on this parameter. Since it was not particularly informative, we decided to remove it.
    \item A vacuum metamorphosis model \cite{PhysRevD.62.083503,PhysRevD.73.023513} where the Ricci scalar $R$ evolves during cosmic history until a certain $z_\star$ where it freezes to $R = m^2$, being $m$ the mass of the scalar field. The interest in this model lies in the fact that it is not a phenomenological description but rather a consequence of first principles theory. However, the predicted $d_L-z$ relation by this model differs significantly from the $\Lambda$CDM one, leading to strong biases in our estimates.
\end{enumerate}

\section{\label{sec:discu_and_concl}Discussion and conclusions}
In this paper we provide forecasts on the ability of LISA to constrain the expansion of the Universe, combining the luminosity distance information from the GW signal of MBHBs with the redshift information obtained from the identification of their host galaxy. 
We built this paper following the results of M22 \cite{Mangiagli22}, where we constructed synthetic catalogues
of EMcps, i.e.~systems with a sufficiently good sky localization from LISA and a detectable EM counterpart, and estimated the number of these objects that LISA together with several electromagnetic facilities could detect. 
Since EMcps provide independent estimates of $d_L$ and $z$, they can be considered as \emph{standard sirens}, perfect tools for cosmological tests. The advantage of MBHBs as standard sirens is that we expect to detect their EM counterpart up $z\sim 7-8$, which means that we can use these systems to test the expansion of the Universe at intermediate redshifts $2\lesssim z \lesssim 8$.
This is the possibility we have explored in this work.

Starting with 90 years of simulated data obtained with SAMs, we generated 100 realisations of the Universe for three astrophysical models of MBHBs formation, assuming two scenarios for the duration of LISA, of 4 years and 10 years. 
For each EMcp, we convolved the $d_L$ posterior distributions from the LISA data analysis process with the expected errors from lensing and peculiar velocities. For the redshift error, we assumed Gaussian distributions with a standard deviation depending on the particular electromagnetic detection technique and device.
We divide the cosmological analysis in \hyperref[subsec:local_universe_models]{\color{mycolormodel}`Local Universe'} models, where we focused on local measurements as the Hubble constant at $z=0$, and \hyperref[subsec:high-redshift-models]{\color{mycolormodel}`High-redshift Universe'} models, where we explored the ability of EMcps with LISA to put constraints on $h(z)$ with $z\gtrsim 2$. For each cosmological model, we performed  the Bayesian inference of the corresponding cosmological parameters for each of the %and we combined the 
100 realisations, and presented the relative error on the cosmological parameters in relation with the fraction of realisations that would provide an error equal or smaller than that particular value. Indeed, the number of EMcps fluctuates from realisation to realisation, rendering the errors realisation dependent. When we qoted the forecast error on a specific cosmological parameter, we always referred to the one provided by at least 50\% of the realisations.
%to provide realistic forecasts of LISA capabilities.

As a general trend, we find that LISA will likely not provide constraints on the \hyperref[subsec:local_universe_models]{\color{mycolormodel}`Local Universe'} models
%$h$ and $\Omega_m$ 
competitive with future EM measurements, due to the limited number of expected EMcps. 
For instance, assuming $\Lambda$CDM, LISA will constrain $h$ with a relative error of less than $5\%$ in 4 years and less than $2\%$ in 10 years, while $\Omega_m$ is constrained with an accuracy of only $10\%$ in 10 years.
%\ls{Since this is the most accurate analysis, we can compare to previous papers such as \cite{Speri21} where we found 5\% precision with 15 events}. 
In the (unlikely) scenario that
%If 
the Hubble tension remains unresolved by around 2040, LISA observations of MBHBs can 
%potentially shed light on the true value 
provide a measurement of $h$ %, addressing 
and help disentangling one of cosmology's long-standing puzzles.
LISA will also not be particularly 
sensitive to deviations in the dark energy equation of state: assuming the standard CPL formalism to describe dark energy, we found constraints on $\omega_0$ greater than $30-40\%$ in almost all cases and no constraining power on $\omega_a$.
However, LISA can test alternative gravity theories where GWs do not propagate as photons even if they have the same speed : within the parametrisation proposed by \cite{Belgacem18}, we found that MBHBs-only observations can constrain $\Xi_0$ to $<10\%$ in just 4 yr. 

On the other hand,
%The detection of the 
EM counterparts from MBHBs can be detected up to $z\lesssim 8$: therefore, EMcps systems detected by LISA in combination with electromagnetic observatories offer the opportunity
%gives us the possibility 
to test the expansion of the Universe in a still unmapped redshift range. 
As the GW emission from MBHBs can only be observed with LISA, providing a very precise measurement of the luminosity distance and a relatively good sky localisation, this represents a unique science case for the mission.
To fully assess LISA capabilities, we investigated four \hyperref[subsec:high-redshift-models]{\color{mycolormodel}`High-redshift Universe'} models, within which one can constrain the matter equation of state or the value of the Hubble parameter and of the luminosity distance at certain pivot redshifts. 
This includes two agnostic models, for which one does not need to assume a particular cosmology.

In the \hyperref[itm:lcdm_beta]{\color{mycolormodel}$(h, \Omega_m, \beta)$} model we tested LISA's ability to determine the matter equation of state, assuming $\omega_m=\beta$. LISA constrains $\beta$ within $10\%$ ($20\%$) in 10 yrs (4 yrs) of observations. However, we also found that LISA has no constraining power on $\Omega_m$ due to its degeneracy with $\beta$.

In the \hyperref[itm:matteronly]{\color{mycolormodel}matter-only} approximation we assumed that the Universe is matter-dominated and we constrained the $d_L-z$ relation under this assumption, featuring $h(z_p)$ and $d_C(z_p)$ as unknown parameters (where $z_p$ denotes a pivot redshift). We found that $h(z_p=2)$ can be constrained at $3-5\%$ in 10 yrs, \textcolor{\colorref}{but we also obtained useful constraints up to $z_p=7$: $h(z_p=7)$ is determined within a $\sim 10\%$ accuracy} in 10yrs. For the comoving distance we expect few percent precision at almost all pivot redshift.
%in almost all cases.

The \hyperref[itm:bins]{\color{mycolormodel}redshift bins} model is the first of our two agnostic models, in which $d_L(z)$ is Taylor expanded as a straight line around a pivot redshift $z_p$. The advantage of this model is that we do not assume a functional form for $h(z)$ (the only assumption at the level of the luminosity distance is that the Universe is flat). However, the disadvantage of this model is that only the EMcps pertaining to a given redshift bin can be accounted for in the analysis, and reducing the number of EMcps reduces the constraining power of the measurement.
\textcolor{\colorref}{ The small amount of EMcps available for the analysis in each bin increases the dependence of the results on the particular Universe realisation: the results are heavily influenced by small fluctuations in the number of EMcps, such that two realisations with perfectly equivalent physical conditions can provide results which are meaningful in one case, but prior dominated in the other. This
required a more thoughtful approach to understand if and when this approach could be adopted in the future.}  
We have therefore decided to run the analysis only on the \emph{informative} realisations, under the understanding that the fraction of realisations giving meaningful results provides an indication on whether the method is viable and can be used, or not.
In order to identify the \emph{informative} realisations, we computed the JS divergence between the posterior and the prior distributions of $h(z_p)$ and we retained only the realisations with $ \rm JS>0.5$ .
\textcolor{\colorref}{Following this procedure, the largest number of informative realisations is achieved at $z_p\sim 2.5-4$ because it is where the distribution of the EMcps peaks in redshift. However, we found that the PopIII MBHB formation scenario never provides enough EMcps for this method to be useful; furthermore, only %the best 
50\% of the realisations within the Q3 models predict relative uncertainties on $h(z_p)$ which are smaller than $\sim 50\%$ for $z_p>1.5$, an error that might be too large given the cosmological landscape expected in the late 2030s. 
Consequently, we conclude that the \hyperref[itm:bins]{\color{mycolormodel}redshift bins} is not a promising approach for cosmological inference with LISA, even though
it might still be performed if a sufficient number of EMcps is detected. 
Note that, even if a realisation
turns out to be not informative under the redshift bins approach, in most cases it will still be informative if we apply another technique/approximation, as the matter-only or spline interpolation.}

%However, the constraints on $h(z_p=3)$ are only at $\sim 20\%$. At lower redshift we still have a similar precision in the recovered values of $h(z_p)$ because $d_L$ and $z$ errors are smaller but the number of informative realisations decrease due to the lack of sources. At higher redshifts $(z_p>3.5)$ the lack of EMcps and the larger errors lead naturally to a decrease of the informative realisations and to worse constraints on $h(z_p)$. \am{dire qui che abbiamo anche alcune realizzazioni che non sono informative per i bins ma lo sono per matter-only o spline}

The key results of this work are obtained within the second of our agnostic models, namely the \hyperref[itm:splines]{\color{mycolormodel}spline interpolation} model, where we fit the $d_L-z$ relation with cubic spline polynomials from $z=0$ up to $z=7$. The outcomes of the analysis are the $d_L$ posterior distributions at 5 knot redshifts, which can be used directly to determine the luminosity distance or, if the derivative is computed, the Hubble parameter at any redshift.
In this model, we recover the luminosity distance with an error of 
$<10\%$ up to $z\sim 6$ with 10 yrs of data, a forecast competitive with future cosmological, EM-only observations.

In this work, we restrict our analyses to MBHBs. However LISA will most probably also observe EMRIs and, at higher frequencies, the early inspiral stellar-mass binary black holes. Both of these populations can in principle be used as dark (or spectral) sirens~\cite{2018MNRAS.475.3485D,Kyutoku:2016zxn,Muttoni22,MacLeod:2007jd,Laghi21,Liu:2023onj}.
We expect LISA cosmological constraints to significantly improve when EMRIs and MBHBs are analyzed together \cite{Tamanini:2016uin, Laghi_paper}. The former can efficiently probe the low-redshift portion of the Universe, while the latter will test the intermediate redshift range, as demonstrated in this work.
This interplay between the redshift range probed by different LISA source populations is expected to help breaking degeneracies in the $d_L(z)$ relation, substantially improving the constraints we found in the present study.

\textcolor{\colorref}{
Finally, in this work we assume that we will be able to identify the galaxies hosting the MBHB mergers, thanks to the strategy developed in M22 \cite{Mangiagli22}. We focus on the `maximising' case, whereby the assumptions are such that the number of EMcps is maximal. Even if there are preliminary study on this topic (\cite{DalCanton19, Lops_paper}), the identification of the host galaxies is one of the major challenges of multimessenger astrophysics involving GWs. 
Assessing the validity of this assumption amounts to realistically simulate the conditions of the strategy put forward in M22, and would require %realistic 
the use of galaxy catalogues to populate LISA error boxes, a model for the EM counterpart emitted by the other accreting AGNs and statistical tools to %manage 
sensibly model the EM observations by several EM facilities. 
This is beyond the scope of the present work, which is therefore entirely based on the assumption that the strategy put forward in M22 is feasible. However, in the construction of the EMcps catalogues of M22, a severe criterion on the localisation of the MBHBs is implemented, which limits the number of potential candidates at high redshift but at the same time vastly increases the chances for a coincident EM detection.
}

\begin{acknowledgments}
We wish to thank Walter Del Pozzo, Jonathan Gair, Danny Laghi, Alexandre Toubiana and Marta Volonteri for fruitful discussions. We also thank the anonymous referee for their comments and suggestions. AM acknowledges support from the postdoctoral fellowships of IN2P3 (CNRS). This project has received funding from the European Union’s Horizon 2020 research and innovation programme
under the Marie Skłodowska-Curie grant agreement No. 101066346 (MASSIVEBAYES). A.M. and C.C. acknowledge funding from the French National Research Agency (grant ANR-21-CE31-0026, project MBH\_waves). 
CC is supported by the Swiss National Science Foundation (SNSF Project Funding grant \href{https://data.snf.ch/grants/grant/212125}{212125}).
S.M.~and N.T.~acknowledge support form the French space agency CNES in the framework of LISA.
This project has received financial support from the CNRS through the MITI interdisciplinary programs. Numerical computations were performed on the DANTE platform, APC, France. We gratefully acknowledge support from the CNRS/IN2P3 Computing Center (Lyon - France) for providing computing and data-processing resources needed for this work. The data underlying this article will be shared on reasonable request to the corresponding author.

\end{acknowledgments}

\bibliography{bibliography.bib}% Produces the bibliography via BibTeX.

\clearpage
\appendix
\section{\label{sec:app_convergence_realisation} Convergence of the realisations}

\begin{figure*}
    \includegraphics[width=1\textwidth]{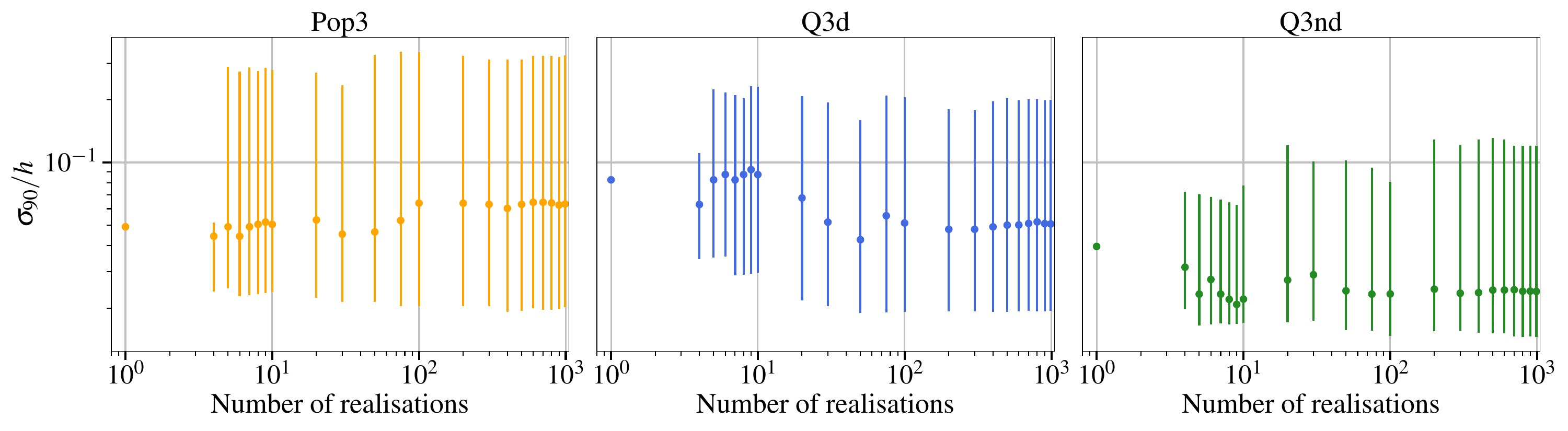}
    \caption{Median relative uncertainty on $h$ as a function of the number of realisations. Errors bars represent the 90 percentile. For all astrophysical models, 100 realisations are sufficient to construct a statistically representative sample.}
\label{fig:conver_error}
\end{figure*}

%In this section we present the results of the tests we performed to assess the validity of our methodology.
Since we adopted mock catalogues of MBHB mergers, one important question is how many realisations are necessary in order to provide reliable estimates on the cosmological parameters. If the number of realisations is too small, the results are strongly affected by the Poisson noise, i.e.  we might have better or worse estimates with respect to the median value, depending on the MBHBs we randomly select. However, we expect that after a sufficiently large number of realisations the average results will not be  sensitive anymore to the particular realisation. To answer this question, we generated 900 additional realisations with respect to the 100 that were used for the results in the main text, for a total of 1000 realisations. For each of these additional realisations, we performed the inference only in the \hyperref[itm:lcdm]{\color{mycolormodel}$(h, \Omega_m)$} model. In Fig.~\ref{fig:conver_error} we show the median relative uncertainty in $h$ and the range of the 90 percentile for the three astrophysical models as a function of the number of realisations. 
As expected, if we consider $<10$ realisations, the median and 90 percentile values fluctuate by a factor of $\sim 2$. After 100 realisations, the average results stabilise, justifying the choice of 100 realisations that we adopted throughout the entire analysis.

\section{\label{sec:gaussianity_dl} 
Gaussianity of the luminosity distance posteriors}
\begin{figure}
    \includegraphics[width=0.5\textwidth]{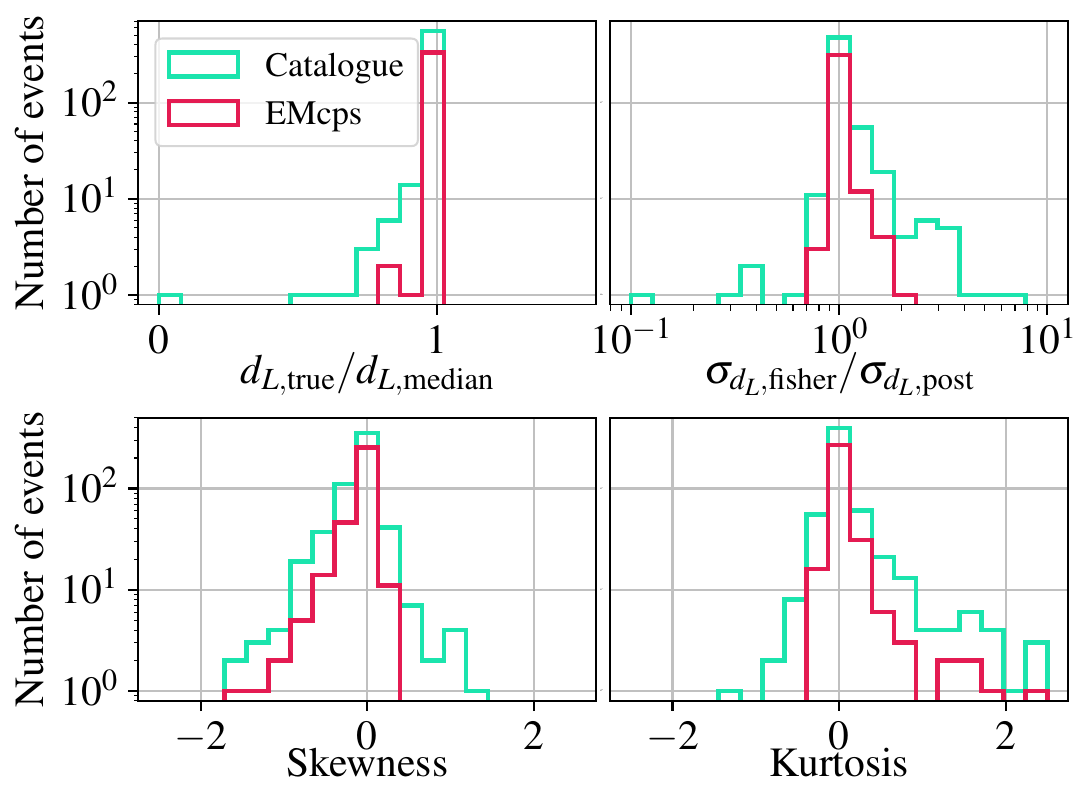} 
    \caption{Upper left panel: ratio between the true value of the luminosity distance and the median value from the $d_L$ posterior distributions. Upper right panel: ratio between the $1\sigma$ uncertainty from fisher analysis and from $d_L$ posterior distribution. Lower left panel: skewness of the $d_L$ posteriors. Lower right panel: same as the left one but for the kurtosis. Aquamarine (crimson) lines correspond to the distribution for the entire MBHBs catalogues (the subset of EMcps). This plot is only for the Q3d model. Overall, the $d_L$ posterior distributions for the EMcps can be considered as Gaussian distributions.}
    \label{fig:test_gauss_dl}
\end{figure}

%In this appendix, we discuss the gaussianity of the luminosity distance posterior distributions. This approximation has been extensively used in the past, in all the analysis performed with a fisher matrix (by construction, the fisher matrix `assumes' that the posterior is a Gaussian distribution). Here we check the validity of this approximation.

\textcolor{\colorref}{As reported in Sec.~\ref{sec:data_analysis}, we compute the likelihood for the inference of the cosmological parameters using the luminosity distance posterior distributions, $p(d_L, \xgw)$, and the associated samples $d_L^i$. The posteriors distributions are the results of the inference process and, consequently, they may have a generic shape and not follow a specific distribution.
However, in the past, most data analysis studies have been performed within the Fisher matrix formalism. In this formalism, we assume that the posterior distribution is a Gaussian centered at the true value with a standard deviation given by the inverse of the Fisher matrix. This approximation is valid only in the limit of large SNR \cite{PhysRevD.77.042001}, and we expect that this condition might not be satisfied by all the MBHBs that LISA will observe (cf. Fig.~1 in M22).\\ 
In this section, we aim to estimate the "gaussianity" of  $p(d_L, \xgw)$ by comparing the luminosity distance estimates obtained via a Fisher matrix with those obtained via the MCMC posterior distributions. The data analysis code reports both the posterior distributions from the full MCMC analysis and the covariance matrix (which is a spurious product and has never been used in the analysis).}

%the likelihood can be treated as a multivariate Gaussian with covariance given by the inverse of the Fisher matrix 

In Fig.~\ref{fig:test_gauss_dl}, we report:
\begin{enumerate}
    \item the ratio between the true luminosity distance and the median of the posterior distributions;
    \item the ratio between the variance from the fisher and from the posterior distributions;
    \item the skewness of the luminosity distance posterior distributions;
    \item the kurtosis of the luminosity distance posterior distributions;
\end{enumerate}
for the entire Q3d catalogue and for the subset of EMcps. If we consider the entire population of MBHBs, it is evident that the results from the fisher analysis are not fully compatible with the Bayesian results. However if we consider the sub-population of EMcps, the posterior distributions appear more Gaussian 
with median values closer to the real luminosity distance and the ratio of the variances peaking at 1. Similarly, also the skewness and the kurtosis are more centered around zero, as expected from a normal distribution. 
We obtain similar results for Pop3 and Q3nd.
This is expected, since the EMcps are a subset of the MBHBs for which the parameter estimation is particularly accurate, in particular for the sky localisation.

\section{\label{sec:js_convergence} Assessment of the JS convergence statistic}
\begin{figure*}
    \includegraphics[width=1\textwidth]{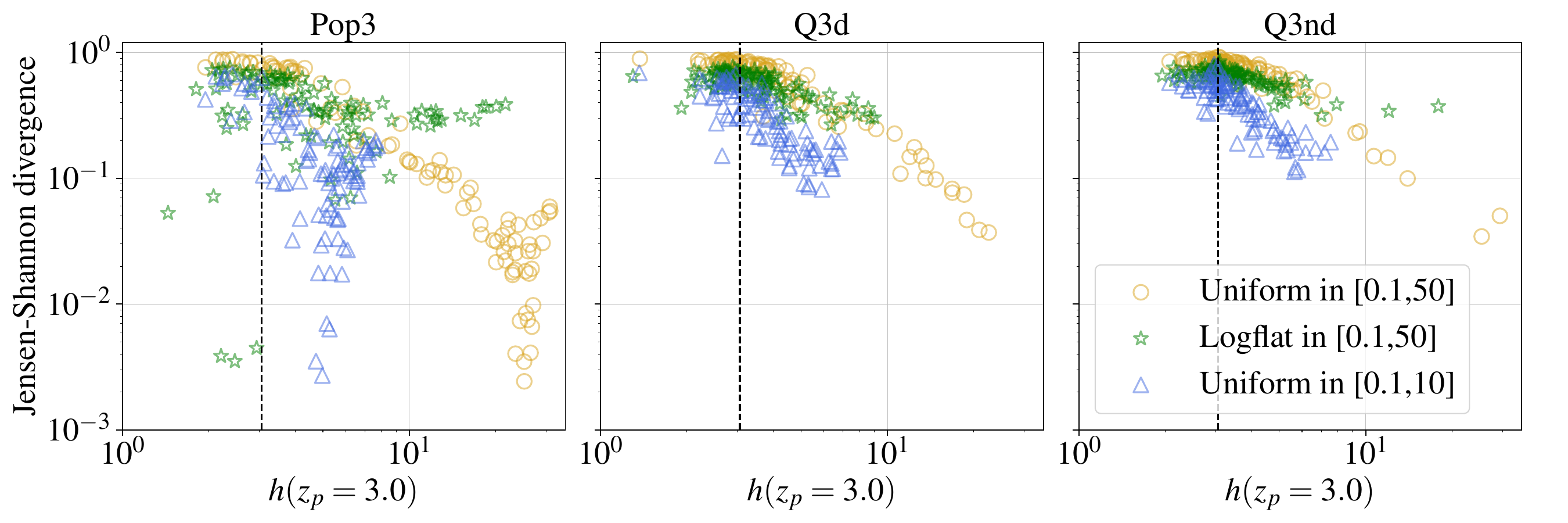}
    \caption{Scatter plot of JS divergence computed between the posterior distribution of $h(z_p=3)$ and different priors (as specified in the legend) versus the inferred median value of $h(z_p=3)$. The dashed black line represents the value of $h(z_p=3)\sim3.06$, according to $\Lambda$CDM.}
\label{fig:js_divergence_for_different_prior}
\end{figure*}

\begin{figure*}
    \includegraphics[width=1\textwidth]{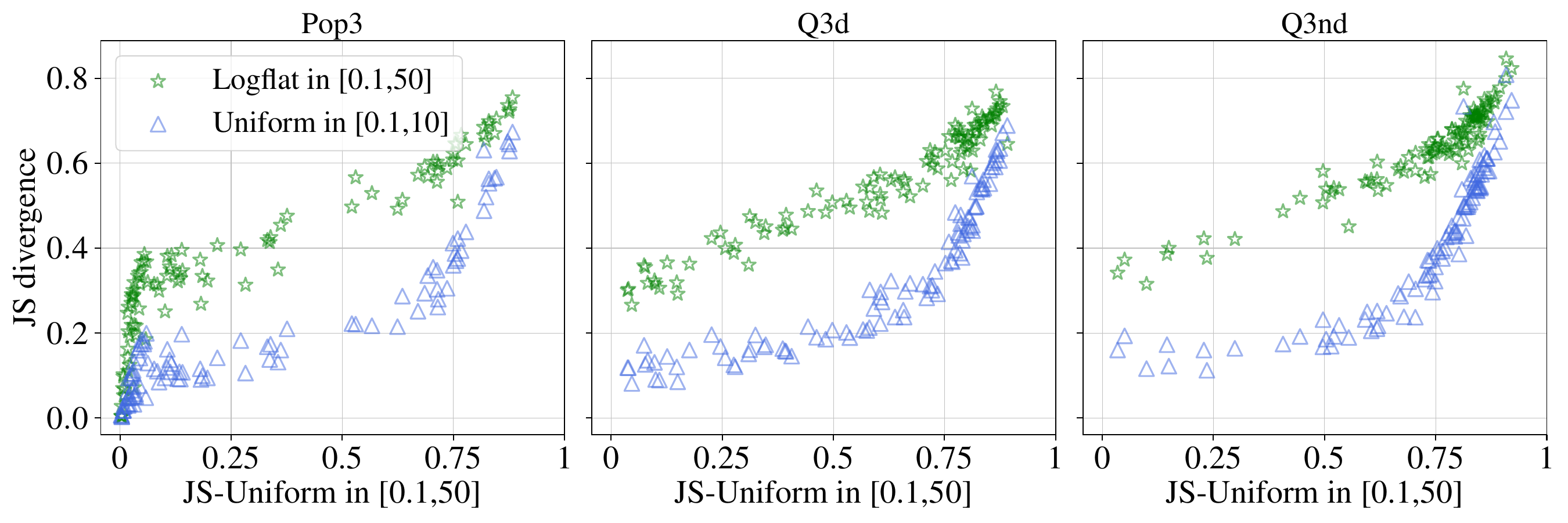}
    \caption{Comparison between JS divergence values for $h(z_p=3)$ for different priors, as stated in the legend. On the x-axis we report the JS divergence adopted in the main text, i.e. with uniform prior in [0.1,50]. }
\label{fig:comparison_js_divergence}
\end{figure*}

\begin{figure}
    \includegraphics[width=0.5\textwidth]{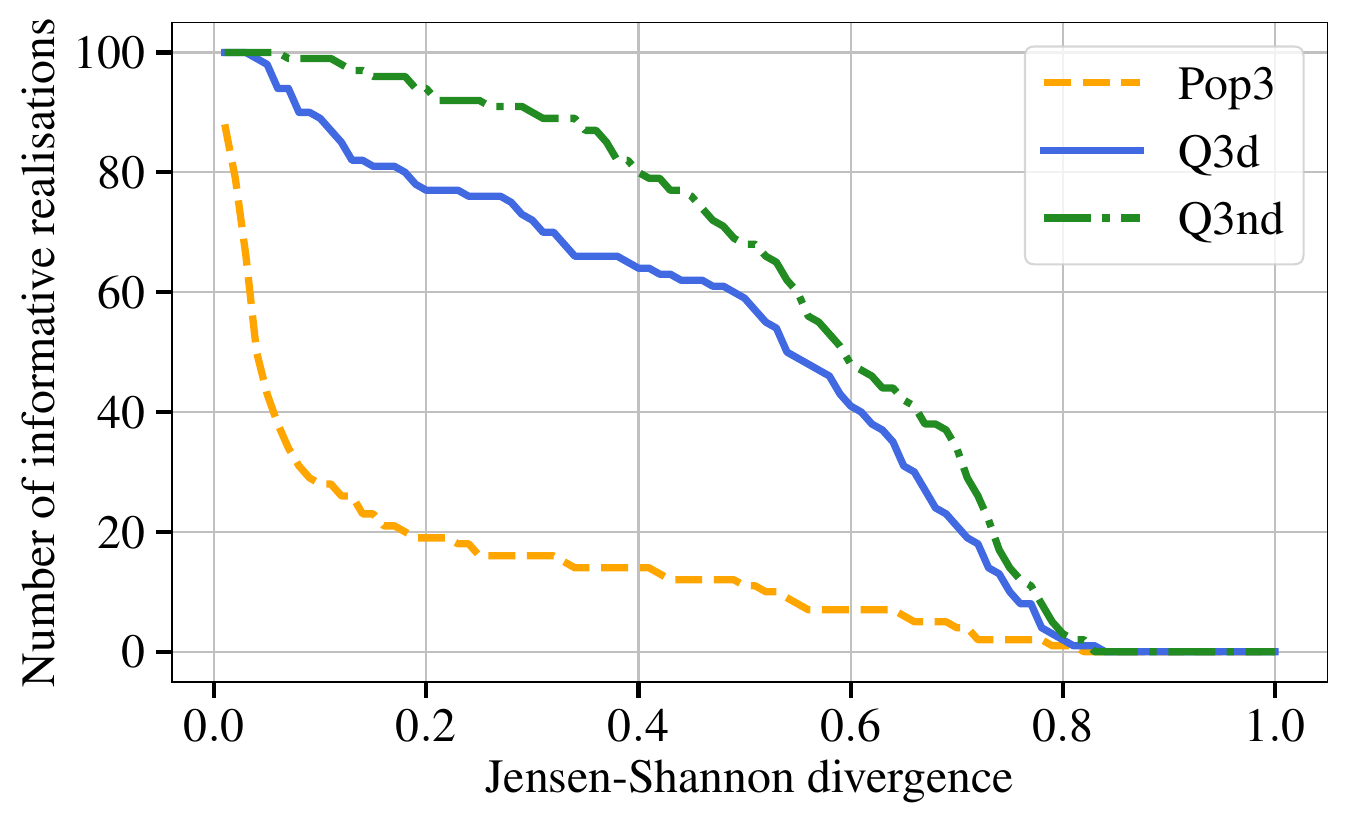} 
    \caption{Number of informative realisations as a function of the JS divergence for the three astrophysical models, as in the legend. The JS divergence is computed for $h(z_p=3)$ with the prior adopted in the main text. }\label{fig:number_of_info_real_vs_threshold}
\end{figure}

\begin{comment}
\begin{figure}
    \includegraphics[width=0.5\textwidth]{comparison_posterior_hzp.pdf} 
    \caption{Comparison between the posterior distributions on $h(z_p)$ and the corresponding priors, according to legend. Here we select a particular event realisation of Q3d \am{for future note: I selected the realisation 1}}\label{fig:comparison_posterior_hzp}
\end{figure}
\end{comment}

In this appendix, we present further details on the JS divergence adopted to distinguish between informative and uninformative realisations in the case of the \hyperref[itm:bins]{\color{mycolormodel}redshift bins} approach. 
As reported in the main text, the value of the JS divergence depends on the choice of the prior adopted. In Fig.~\ref{fig:js_divergence_for_different_prior} we compare the JS divergence for the redshift bin centered at $z_p=3$ (as in Tab.~\ref{tab:prior_table_high_Universe}) with the median value obtained for $h(z_p=3)$ for three different  priors: 
\begin{enumerate}
    \item Uniform in $[0.1,50]$;
    \item Logflat in $[0.1,50]$;
    \item Uniform in $[0.1,10]$.
\end{enumerate}
The first one corresponds to the prior adopted in the analysis in the main text. The other two correspond to two priors we tested during the preparation of this paper. As expected, different priors lead to different JS divergence values. 
In particular, both the logflat prior and the uniform prior in [0.1,10] present smaller JS divergences. This stems from the fact that in the logflat case we give more weight to small values of $h(z_p)$, where the true value is. In the uniform case between [0.1,10], the JS values are smaller because the range of the prior is narrower and, therefore, the posteriors look  more similar to the priors.
%\am{I think that this can be easily understood just looking at Fig.~\ref{fig:comparison_posterior_hzp}  but I'm not sure I want to put this figure in the paper.}
%Moreover i
In the uniform case in [0.1,10], we can observe some realisations with low JS divergence and median value $\sim 5$, the midpoint of this prior range. As expected, this subset of systems corresponds to the uninformative realisations.
From Fig.~\ref{fig:js_divergence_for_different_prior}, it is clear that the choice of the threshold for the JS divergence depends on the adopted prior. As discussed in the main text, this value is somewhat arbitrary, and we aim to understand if there is a way to justify or support our choice of 0.5.

In Fig.~\ref{fig:comparison_js_divergence} we present the scatter plot of the JS divergence in the case of logflat prior or in the case of a uniform prior in $[0.1,10]$ versus the JS values from the uniform prior in $[0.1,50]$ adopted in the main text. Both cases exhibit a monotonic behavior (with only a few exceptions for Pop3 at small JS values), granting us the possibility to change priors, keeping the same informative realisations. 
For example, a JS divergence threshold of $0.5$ for the uniform prior in $[0.1,50]$ roughly corresponds  to the same value for a logflat prior in $[0.1,50]$. However, if we adopt a uniform prior in $[0.1,10]$, we need to lower the threshold to $\sim 0.2$ to maintain consistency in the analysis and select the same realisations.
%systems.
While this comparison allows us to keep the same number of informative realisations, independently from the choice of the prior, the open question remains on how to set the ``first'' threshold. In the main analysis with the uniform prior in $[0.1,50]$, we fixed the JS divergence threshold at $0.5$ in order to remove the spurious points with the inferred median values of $h(z_p) \sim 25$. 

Performing the cosmological analysis with different priors and fixing the thresholds following the scaling relations in Fig.~\ref{fig:comparison_js_divergence}, we found that the uninformative realisations were always the same,  irrespective of the adopted priors. 
Finally, in Fig.~\ref{fig:number_of_info_real_vs_threshold} we report the number of informative realisations as a function of the JS threshold for the uniform prior in $[0.1,50]$. As expected, setting a value close to 0 renders all realisations informative, while a value close to 1 makes none of them informative. However, it's worth noting that there isn't a `plateau' around the value of $0.5$ we adopted, so the presented figures slightly change if we increase/decrease the threshold.

\section{\label{sec:pivot_parameters} Pivot parameter for $(h(z), \Omega_m)$}
A set of 
%variables 
parameters
might present some degree of correlation. Therefore, a natural approach is to construct a new set of 
%variables 
parameters
where the correlation is minimised.
%zero. 
For example, in the context of dark energy, this approach has been adopted by several studies in the past (see \cite{1999ApJ...518....2E} for the original idea and \cite{2006astro.ph..9591A} for an application). Assuming the same dark energy expression in Eq.~\ref{eq:cpl_dark_energy}, one can search for a pivot redshift $z_p$ (or corresponding scale factor $a_p$) where the error on $\omega(a)$ is minimized.
It can be demonstrated that this redshift corresponds to
\begin{equation}
     1-a_p = - \frac{\sigma_{\omega_0 \omega_a}}{\sigma_{\omega_a}^2}
\end{equation}
where $\sigma_{\omega_0 \omega_a}$ corresponds to the correlation between $\omega_0$ and  $\omega_a$ and $\sigma_{\omega_a}^2$ corresponds to the variance for $\omega_a$.

In order to explain the fact that we have the best constraints on $h(z)$ at $z\sim 0.5$ in Fig.~\ref{fig:hz_and_dl_errors}, we want to find a redshift for which the correlation between $h(z)$ and $\Omega_m$ is minimum.
We start defining the variance-covariance matrix between $h$ and $\Omega_m$ as
\begin{equation}
    \Sigma_{h, \Omega_m} = 
    \begin{pmatrix}
    \sigma_{h}^2 & \sigma_{h m} \\
    \sigma_{h m} &  \sigma_{m}^2 \\
    \end{pmatrix}
\end{equation}
where $\sigma_{h}^2$ ($\sigma_{m}^2$) corresponds to the variance for $h$ ($\Omega_m$) and $\sigma_{h m}$ is the correlation term. We want to go from the old set of variables $(h,\Omega_m)$ to the new set formed by $(h(z),\Omega_m)$ with $z$ that will be specified imposing the minimum correlation. 
Following the rules for the  propagation of errors, we can write the variance-covariance matrix for $(h(z),\Omega_m)$ as
\begin{equation}
    \Sigma_{h(z), \Omega_m} = \rm{J} \, \Sigma_{h, \Omega_m} \, \rm{J}^{\rm T}.
\end{equation}
where $\rm J$ is the Jacobian of the transformation defined as 
\begin{equation}
    \rm{J} = \begin{pmatrix}
    \frac{\partial h(z)}{\partial h} & \frac{\partial h(z)}{\partial \Omega_m} \\
    \frac{\partial \Omega_m}{\partial h} & \frac{\partial \Omega_m}{\partial \Omega_m} \\
    \end{pmatrix}.
\end{equation}
Performing the calculation, we get
\begin{widetext}
\begin{align}
 \Sigma_{h(z), \Omega_m}^{11} & = \frac{h z \left(z^2+3 z+3\right) \left(h \sigma_m^2 z \left(z^2+3 z+3\right)+4 \sigma_{hm} \left(\Omega_m z \left(z^2+3 z+3\right)+1\right)\right)+4 \left(\Omega_m \sigma_h z \left(z^2+3 z+3\right)+\sigma_h\right)^2}{4 \Omega_m z \left(z^2+3 z+3\right)+4} \\ 
\Sigma_{h(z), \Omega_m}^{12} & = \frac{h  \sigma_m^2 z \left(z^2+3 z+3\right)+2 \sigma_{hm}\left(\Omega_m z \left(z^2+3 z+3\right)+1\right)}{2 \sqrt{\Omega_m z \left(z^2+3 z+3\right)+1}} \\ 
\Sigma_{h(z), \Omega_m}^{22} & =\sigma_m^2
\end{align}
\end{widetext}
We highlight three points:
\begin{enumerate}
    \item The element $\Sigma_{h(z), \Omega_m}^{22}$ is left unchanged from the transformation as expected because $\Omega_m$ remains the same;
    \item If we set $z=0$ in $\Sigma_{h(z), \Omega_m}^{11}$, we recover the original value of $\sigma_h^2$;
    \item For the term $\Sigma_{h(z), \Omega_m}^{12}$, if we insert the value of variance-covariance matrix from the median realisation, we find that the expression has a zero at $z\sim 0.6$, compatible with the results in Fig.~\ref{fig:hz_and_dl_errors}. 
\end{enumerate}

\section{\label{sec:accuracy_spline} Accuracy of spline interpolation}

\begin{figure*}
    \includegraphics[width=1\textwidth]{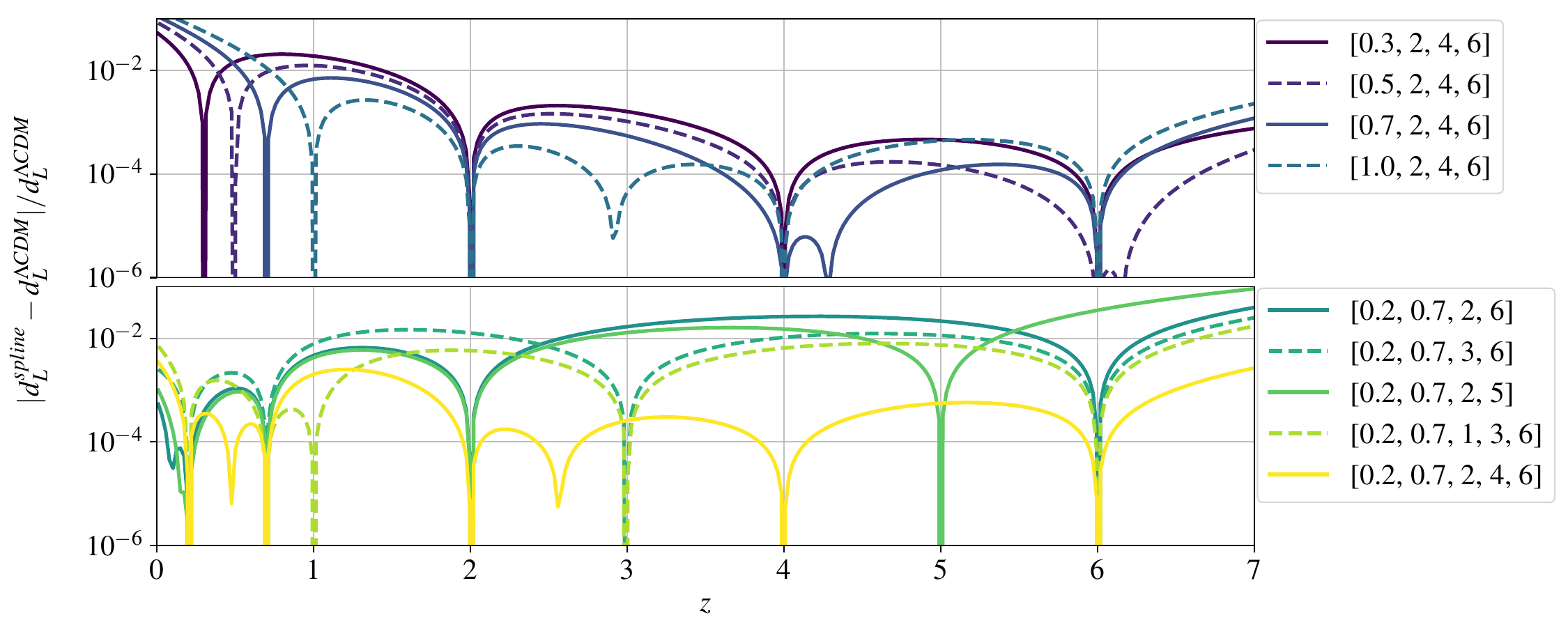}
    \caption{\textcolor{\colorref}{Difference between the  luminosity distance from spline interpolation $d_L^{spline}$ and the $\Lambda$CDM luminosity distance $d_L^{ \Lambda CDM}$, normalised over $d_L^{ \Lambda CDM}$ and for different choices of knot redshifts as reported in the legends. 
    Upper and lower panels report the same quantity: we split the results and use solid and dashed lines only to facilitate the visual comparison. At the values of the knots, the spline is forced to pass through $\Lambda$CDM values and the difference goes to zero.}}
\label{fig:spline_accuracy}
\end{figure*}

\color{\colorref}
As reported in Sec.~\ref{subsec:inference_splines}, for the spline interpolation we set the knot redshifts at $z=[0.2,0.7,2,4,6]$. This choice is determined by the fact that we wish the model accuracy to be smaller than the uncertainties reported in Fig.~\ref{fig:hz_and_dl_errors_splines}. To justify this choice more quantitatively, in Fig.~\ref{fig:spline_accuracy} we report the difference between the luminosity distance from the spline interpolation $d_L^{spline}$ (assuming the true \lcdm values) and the $\Lambda$CDM luminosity distance $d_L^{ \Lambda CDM}$ , for different choices of the knots redshift.
We start the investigation replacing the two smallest knots (i.e. $z=[0.2,0.7]$) with only one, testing values from $z=0.3$ to $z=1$. As reported in the upper panel, removing the first two knots leads to larger differences between $d_L^{ \Lambda CDM}$ and $d_L^{spline}$ at $z<1$. For example, if we set the knots at $z=[0.5,2,4,6]$, we have an accuracy of $1\%$ at $z\sim0.7$. Similar considerations apply also if we remove one of the three knots at $z=[2,4,6]$. We also tested another choice for the position of the knot redshifts, keeping the same number as in the main text. Namely, we tested the case where the knots are at $z=[0.2,0.7,1,3,6]$. However, we find an accuracy larger than one order of magnitude at $z>2$ with respect to our original choice. Therefore, the original values adopted in Sec.~\ref{subsec:inference_splines} represent a good choice for the spline interpolation. We do not increase the number of knots to avoid over-fitting. We finally note that the choice of knots redshift is also affected by the choice of cubic spline as interpolating function.

\end{document}